\documentclass[prd, 
superscriptaddress,preprintnumbers,tightenlines,nofootinbib, eqsecnum]{revtex4-2}

\usepackage{amsmath}
\usepackage{amsfonts}
\usepackage{amssymb}
\usepackage{bm}
\usepackage{hyperref}
\usepackage{mathrsfs}
\usepackage{graphicx}
\usepackage{subcaption} 
\usepackage{makecell, booktabs} 
\usepackage{empheq}
\usepackage{ulem}
\usepackage[usenames]{color}
\usepackage{ytableau}

\usepackage{ragged2e} 
\allowdisplaybreaks

\DeclareSymbolFontAlphabet{\mathrsfs}{rsfs}
\DeclareMathAlphabet{\mathcal}{OMS}{cmsy}{m}{n}

\newcommand{\nn}{\nonumber}
\newcommand\calO{\mathcal{O}}
\newcommand{\dd}{\mathrm{d}}
\newcommand{\di}{\mathrm{i}} 
\newcommand{\de}{\mathrm{e}}

\newcommand{\be}{\begin{equation}}
\newcommand{\ee}{\end{equation}}
\newcommand{\bse}{\begin{subequations}}
\newcommand{\ese}{\end{subequations}}

\newcommand{\g}{g^{\rm 2PN}}

\newcommand{\n}{\bm{n}}
\newcommand{\la}{\bm{\lambda}}
\newcommand{\el}{\bm{\ell}}
\newcommand{\E}{\tilde{E}}
\newcommand{\Lp}{\tilde{L}}
\newcommand{\Ve}{V_{\rm eff}}

\definecolor{darkgreen}{rgb}{0,0.5,0}

\hypersetup{
 unicode=false,          
 pdftoolbar=true,        
 pdfmenubar=true,        
 pdffitwindow=false,     
 pdfstartview={FitH},    
 pdftitle={My title},    
 pdfauthor={Author},     
 pdfsubject={Subject},   
 pdfcreator={Creator},   
 pdfproducer={Producer}, 
 pdfkeywords={keyword1} {key2} {key3}, 
 pdfnewwindow=true,      
 colorlinks=true,       
 linkcolor=red,          
 citecolor=cyan,        
 filecolor=magenta,      
 urlcolor=darkgreen,           
 linktocpage=true
}

\makeatletter
\g@addto@macro\bfseries{\boldmath}
\makeatother

\allowdisplaybreaks

\begin{document}

\title{Innermost stable circular orbit (ISCO) of arbitrary-mass compact binaries \\with spins at the fourth post-Newtonian order}
    
\author{Luc \textsc{Blanchet}}\email{luc.blanchet@iap.fr}
\affiliation{$\mathcal{G}\mathbb{R}\varepsilon{\mathbb{C}}\calO$, 
	Institut d'Astrophysique de Paris, \\UMR 7095, CNRS, Sorbonne Universit{\'e},
	98\textsuperscript{bis} boulevard Arago, 75014 Paris, France}

\author{David \textsc{Langlois}}\email{david.langlois@apc.in2p3.fr}
\affiliation{Université Paris Cité, CNRS, Astroparticule et Cosmologie, 75013 Paris, France}

\author{Etienne \textsc{Ligout}}\email{etienne.ligout@apc.in2p3.fr}
\affiliation{$\mathcal{G}\mathbb{R}\varepsilon{\mathbb{C}}\calO$, 
	Institut d'Astrophysique de Paris, \\UMR 7095, CNRS, Sorbonne Universit{\'e},
	98\textsuperscript{bis} boulevard Arago, 75014 Paris, France}
\affiliation{Université Paris Cité, CNRS, Astroparticule et Cosmologie, 75013 Paris, France}

\begin{abstract}
We compute, up to 4th post-Newtonian (4PN) order, the gauge-invariant stability criterion determining the innermost stable circular orbit (ISCO) for arbitrary-mass compact binaries with spins aligned with the orbital angular momentum. Our calculation, a perturbation analysis,  incorporates spin-orbit and spin-spin contributions, both up to next-to-next-to-leading order, as well as leading-order cubic and quartic spin contributions, all calculated in prior works within the effective-field-theory (EFT) Hamiltonian formalism. In the test-mass limit, we recover the Kerr ISCO truncated at 4PN order. Estimating the ISCO shift at first order in the mass ratio and  comparing our results with numerical gravitational self-force (GSF) calculations, we find good agreement for retrograde spin (the smallest discrepancy being obtained for maximal retrograde spin) and moderate prograde spin. For nearly maximal prograde spin, the PN approach fails to determine the ISCO, for it is close to the black hole (BH) horizon. We also estimate the ISCO shift  due to the test-particle spin  and compare our result with the exact prediction from the Mathisson-Papapetrou-Dixon (MPD) equations for a spinning particle orbiting a Kerr black hole. Finally, we  discuss the stability and the ISCO of corotating black-hole binaries.
\end{abstract}


\maketitle


\section{Introduction}
\label{sec:intro}

Recent efforts to investigate the dynamical stability of circular orbits in the post-Newtonian (PN) approximation have established a well-defined, gauge-invariant notion of the innermost stable circular orbit (ISCO), valid for arbitrary-mass (in particular, comparable-mass) compact binary systems. This result relies on a stability criterion $C(x)$ defined in terms of  the dimensionless parameter $x\equiv\left({G m \Omega}/{c^3}\right){}^{2/3}$ constructed from the invariant orbital frequency $\Omega$ measured at infinity from the system (and where $m$ is the total mass). The circular orbit is stable whenever $C(x)>0$, so the ISCO corresponds to the solution $x_\text{ISCO}$  of the marginal case $C(x)=0$. The stability criterion can be obtained by analysing the linear perturbation at the level of the equations of motion (EoM) around a circular orbit, as put forward in~\cite{KWWisco}. 

The gauge-invariant criterion was derived at 3PN order in~\cite{BI03CM} and found to exactly reproduce the Schwarzschild ISCO in the test-mass limit (at all known PN order). Furthermore, as shown by Favata~\cite{F11a}, it provides a good approximation of the ISCO shift due to the finite-mass correction in the gravitational self-force (GSF) approach, as computed numerically in~\cite{BarackS09,LBB12} (see also~\cite{BarackS07,BarackS11,Akcay12}). These interesting features motivated us to extend the criterion to 4PN order in our previous work~\cite{BLL25}. We found a significant improvement over the 3PN order in the criterion's ability to locate the ISCO and determine the ISCO shift. The 4PN criterion was obtained by perturbing both the 4PN EoM in harmonic coordinates and, with equivalent result, the 4PN Hamiltonian in ADM coordinates. Note that at 4PN order, the non-local tail term~\cite{BD88} is crucial and must be consistently perturbed. 

In this paper, we generalize the 4PN criterion to spinning compact binaries. Favata~\cite{F11b} already included the spin contribution in the criterion up to 2.5PN, showing that, at this order, it yields the Kerr ISCO in the test-mass limit. Here, we extend this analysis to 4PN order, by including the leading-order (LO), next-to-leading order (NLO) and next-to-next-to-leading order (NNLO) spin-orbit terms; the LO, NLO and NNLO spin-spin terms; the LO cubic spin terms and the LO quartic spin terms. We adopt the convention where  spin variables are defined as $S \equiv c S_\text{physical} \equiv G m^2 \chi$, $S_\text{physical}$ having the dimension of an angular momentum and the dimensionless spin $\chi$ satisfying $-1\leqslant\chi\leqslant 1$ for black holes (BHs). For maximally rotating compact objects the spin variable scales as $S\sim G m^2$, while for slowly rotating objects (with ``surface'' velocity $v_\text{spin} \ll c$) we have $S\sim G m^2 v_\text{spin}/c \ll G m^2$. 

With this convention, the LO, NLO and NNLO spin-orbit terms respectively enter at 1.5PN, 2.5PN and 3.5PN orders; the LO, NLO and NNLO spin-spin terms at 2PN, 3PN and 4PN orders; the LO cubic spin terms at 3.5PN order and the LO quartic spin terms at 4PN order (any other spin term is at least of order 4.5PN). Pioneering works on the LO and NLO spin-orbit and spin-spin interactions in the EoM include Refs.~\cite{BOC75, BOC79, KWW93, K95, TOO01, FBB06, DJSspin}. Further contributions along these lines (in particular, using harmonic coordinates), such as the inclusion of spin effects in the gravitational-wave (GW) field, can be found in Refs.~\cite{BBF06,MBFB13, BMFB13, BFMP15, M15, MBBB13}.

In the present work, we rely on the effective field theory (EFT) for spinning objects, developed at the Hamiltonian level  up to 4PN order by Levi and Steinhoff~\cite{LS14,LS15a,LS15b,LS16a,LS16b,LS21}. As is well known, from 2PN order onward, spin-dependent finite-size effects must be taken into account. The leading-order contribution comes from the spin-squared interaction,   first derived in~\cite{P98} and later computed to NLO in~\cite{PoR08b,SHS08c,HergtS08a,HSS10}. These finite-size effects introduce coefficients that  depend on the internal structure of the bodies and arise from the spin-induced multipole moments of the compact objects; they can in principle be computed numerically. For BHs, all such coefficients are equal to unity.

After computing the gauge-invariant stability criterion \textit{via} perturbation of the 4PN EFT Hamiltonian~\cite{LS14,LS15a,LS15b,LS16a,LS16b,LS21}, we rederive it — as an important consistency check — by perturbing the 4PN EFT EoM, obtained  from the Hamiltonian (hence in the same EFT coordinates, which differ from both the harmonic coordinates and ADM coordinates used in our previous work~\cite{BLL25}).\footnote{In Appendix~\ref{app:hamiltonianNS} we provide the expression of the spinless part of the EFT Hamiltonian in a general frame up to 4PN order, as it does not seem to have been given previously in the literature.} A  further check consists in recovering the criterion up to 3.5PN order using the EoM in harmonic coordinates, because all spin interactions relevant at 3.5PN order are directly available in this gauge~\cite{MBFB13, BMFB13, BFMP15, M15} (contrary to the 4PN contributions).  

For a non-spinning particle around a Kerr BH, our criterion reproduces the exact Kerr stability criterion in the test-mass limit, up to  4PN order. We then compute the shift of the ISCO due to the finite mass ratio between the particle and the Kerr BH. When compared with (conservative) GSF numerical computations~\cite{IBDLNSTW14}, our results show  very good agreement for retrograde orbits, whereas  the estimate becomes increasingly inaccurate for prograde orbits as the BH spin increases. This is due to the fact that for a prograde spin approaching maximality, the Kerr ISCO gets closer to the horizon of the BH, where the post-Newtonian approximation is expected to break down.

We then move on to two applications. Firstly, we compute the ISCO shift due to the spin of the test-particle and compare it with the exact result derived from the Mathisson-Papapetrou-Dixon (MPD) equations~\cite{Mathisson37,Papa51spin,Dixon79} applied to the case of a spinning particle around a Kerr BH. Secondly, we examine the special case of corotating BHs. In this configuration, the two spins are no longer free parameters but are fixed by the orbital frequency of the binary system. This yields a specific criterion which depends solely on the parameter $x$ and can thus be compared to the stability criterion for spinless bodies. We find that corotating systems are more stable than their spinless counterparts.  

The plan of this paper is as follows. In section~\ref{sec:Equations of motion}, we present the general structure of the EoM and introduce an appropriate moving basis to decompose them. We proceed similarly for the Hamiltonian formulation. We then linearise the EoM (or the Hamiltonian equations) and analyse the stability of the resulting system, essentially extending the analysis of our previous work~\cite{BLL25}. Focusing on the perturbation of circular orbits with spins aligned with the orbital angular momentum, we derive  our stability criterion (using spin formulas detailed  in Appendices~\ref{app:spinEFT} and~\ref{app:EoM formulae}). In Sec.~\ref{sec:criterion}, we discuss the criterion specialised to the case of two BHs, comparing it to the exact Kerr stability criterion when one of the bodies is a test mass. We compute the ISCO shift at first order in the mass ratio and compare our results with GSF numerical calculations. We discuss the particular case of a spinning test particle around the Kerr BH. Additionally, we consider the stability of corotating BH binaries. Sec.~\ref{sec:discussion} is devoted to a discussion and our conclusions. The paper also includes  appendices with  all  relevant details: the full criterion for aligned spins with arbitrary spin-induced finite-size deformability coefficients (App.~\ref{app:ISCOcriterion}); the analytical expressions for the first-order corrections to the ISCO parameters in the MPD approach (App.~\ref{app:MPD}); the spinless part of the EFT Hamiltonian up to 4PN order (App.~\ref{app:hamiltonianNS}); the spin contributions to the EFT Hamiltonian up to 4PN (App.~\ref{app:spinEFT}); and, finally, the spin contributions to the EFT EoM up to 4PN (App.~\ref{app:EoM formulae}).

\section{Equations of motion and their perturbations}
\label{sec:Equations of motion}

\subsection{Equations of motion}

We consider a compact binary system modeled by two point-like spinning particles with respective masses $m_a$, positions $\bm{x}_a$, velocities $\bm{v}_a=\dd\bm{x}_a/\dd t$ and spins $\bm{S}_a$ $(a=1,2)$. We adopt the constant-magnitude spins defined, \textit{e.g.}, in Sec.~II.A of~\cite{BMFB13}. In the center-of-mass (CM) frame, the evolution of the relative position $\bm{x}\equiv \bm{x}_{1}-\bm{x}_{2}$ is governed by an EoM of the form
\begin{equation}
	\label{acceleration}
	\frac{\dd \bm{v}}{\dd t} = \bm{a}\bigl[\bm{x}(t),\bm{v}(t),\bm{S}_a(t)\bigr]\,,
\end{equation}
where $\bm{v}\equiv \bm{v}_{1}-\bm{v}_{2} = \dd \bm{x}/\dd t$. The right-hand side should be seen as a  \emph{functional} of its argument, due to the presence of the non-local tail term at 4PN order, see~\eqref{eq:tail acceleration} in Appendix~\ref{app:EoM formulae}. The evolution of the spins is determined by precession equations of the form 
\begin{equation}
\label{precession}
	\frac{\dd \bm{S}_a}{\dd t}=\bm{\Omega}_a \times \bm{S}_a\,.
\end{equation}
At 4PN order, the right-hand side of \eqref{acceleration} and the spin precession vectors  $\bm{\Omega}_a$ in \eqref{precession} are very long expressions, which include spinless and spin contributions. Most expressions are given in the Appendix~\ref{app:EoM formulae} and all the necessary formulas are systematically given in the ancillary file~\cite{ancillaryfile}.

When working in the CM frame, instead of the vectors $\bm{S}_a$, it is convenient to work with the linear combinations~\cite{K95}
\begin{equation}
\label{definition spin variables S and Sigma}
        \bm{S} \equiv \bm{S}_1+\bm{S}_2\,, \qquad  \bm{\Sigma} \equiv m \left( \frac{\bm{S}_2}{m_2}-\frac{\bm{S}_1}{m_1}\right)\,,
\end{equation}
where $m\equiv m_1+m_2$, and to rewrite the precession equations~\eqref{precession} as evolution equations for $\bm{S}$ and $\bm{\Sigma}$. This yields
\begin{subequations}
\label{EOM_S_Sigma}
\begin{align}
\frac{\dd \bm{S}}{\dd t} &= \left( \frac{1 + \delta}{2}\, \bm{\Omega}_1 + \frac{1 - \delta}{2}\, \bm{\Omega}_2 \right) \times \bm{S} + \nu (\bm{\Omega}_2 - \bm{\Omega}_1) \times \bm{\Sigma}\,,\\
 \frac{\dd \bm{\Sigma}}{\dd t}&= (\bm{\Omega}_2 - \bm{\Omega}_1) \times \bm{S} + \left( \frac{1 - \delta}{2}\, \bm{\Omega}_1 + \frac{1 + \delta}{2}\, \bm{\Omega}_2 \right) \times \bm{\Sigma}\,,
\end{align}
\end{subequations}
where we have introduced the symmetric mass ratio $\nu \equiv  m_1 m_2/m^2$ and the relative mass difference $\delta \equiv (m_1-m_2)/m$ (such that $|\delta|=\sqrt{1-4\nu}$).

In practice, it is also convenient to introduce a moving orthonormal basis $(\n, \la, \el)$, adapted to the relative motion, such that $\bm{n}=\bm{x}/r$ where $r\equiv\vert\bm{x}\vert$, $\la$ is defined by the decomposition of the velocity as $\bm{v}= u \,\bm{n}+r \Omega\,\bm{\lambda}$, 
where $u\equiv \Dot{r}$ and $\Omega \equiv \Dot{\varphi}$ is the orbital frequency, and finally $\bm{\ell} = \bm{n} \times \bm{\lambda}$. The time derivatives of the basis vectors are given by
\begin{equation}
\label{basis_dot}
    \frac{\mathrm{d}\bm{n}}{\mathrm{d}t}=\Omega \,\bm{\lambda}\,,\qquad \frac{\mathrm{d}\bm{\lambda}}{\mathrm{d}t}=-\Omega\,\bm{n}+\varpi\,\bm{\ell}\,,\qquad\frac{\mathrm{d}\bm{\ell}}{\mathrm{d}t}=-\varpi\,\bm{\lambda}\,,
\end{equation}
where $\varpi \equiv -\bm{\lambda} \cdot \Dot{\bm{\ell}}$ is the precession frequency  of the orbital plane. As a consequence, the time derivative of any vector $\bm{T}$, for instance $\bm{S}$ or $\bm{\Sigma}$, decomposes as
\begin{equation}
\label{dotX}
\frac{\dd\bm{T}}{\dd t} = \left(\Dot{T}_n -\Omega\,T_{\lambda}\right)\n +\left(
    \Dot{T}_\lambda +\Omega\,T_{n}- \varpi\, T_{\ell}\right)\la+\left(\Dot{T}_\ell +\varpi\, T_{\lambda}\right)\el\,,
\end{equation}
denoting $T_n$, $T_\lambda$ and $T_\ell$ the components of $\bm{T}$ on this basis. For the velocity,  the above general formula reduces to 
\begin{align}  
\frac{\dd\bm{v}}{\dd t} &= \bigl( \Dot{u}-r \Omega^2\bigr)\bm{n}+\Bigl( 2 u\,\Omega+r \Dot{\Omega}\Bigr)\bm{\lambda} +r\Omega\,\varpi\,\bm{\ell}\,.
\end{align}
Substituting this decomposition into~\eqref{acceleration}, we obtain the equations 
\begin{equation}\label{equations Dot r, Dot u, Dot omega}
\Dot{u} = r \Omega^2 + a_n\,, \qquad       \Dot{\Omega} = -\frac{2 u \Omega}{r} + \frac{a_\lambda}{r}\,,
\end{equation}
while the component of the acceleration along $\el$ yields $\varpi =a_\ell/(\Omega r)$. Similarly, inserting~\eqref{dotX} into~\eqref{EOM_S_Sigma}, one gets the expressions for the time derivatives of the six components of $\bm{S}$ and $\bm{\Sigma}$. Completing with the equation $\dot r=u$, we finally end up with a first-order differential system for the following nine variables: 
%
\begin{equation}\label{defXalpha}
    X^\alpha = \bigl\{r, u, \Omega, S_n, S_\lambda, S_\ell, \Sigma_n, \Sigma_\lambda, \Sigma_\ell  \bigr\}\,.
\end{equation}

An equivalent set of differential equations is obtained in the Hamiltonian formalism, where we have a Hamiltonian functional of the form
\begin{equation}
    H\equiv H[\bm{x}(t),\bm{p}(t),\bm{S}_a(t)]\,,
\end{equation}
with the Poisson brackets structure $\{x^{i},p^{j}\}= \delta^{ij}$ and $\{S_a^{i},S_b^{j}\}=c \,\delta_{ab}\,\varepsilon^{ijk}\,S_a^{k}$ (the other brackets being zero). The Hamilton equations for $\bm{x}$ and $\bm{p}$ read
\begin{equation}
\frac{\dd \bm{x}}{\dd t} = \frac{\delta H}{\delta \bm{p}}\,,\qquad \frac{\dd \bm{p}}{\dd t} = - \frac{\delta H}{\delta \bm{x}} \,, 
\end{equation}
where we use functional derivatives in the right side to account for the fact that the Hamiltonian is nonlocal. The Hamiltonian equations for the spins give the precession equations
\begin{align}\label{precequation}
\frac{\dd S_a^i}{\dd t} = \varepsilon^{ijk} \,\Omega_a^j \,S_a^k\,,\quad  \text{where}\quad \Omega_{a}^{i} \equiv  c\,\frac{\partial H}{\partial S_{a}^{i}}\,,
\end{align}
which can be rewritten in terms of the variables $\bm{S}$ and $\bm{\Sigma}$ defined in \eqref{definition spin variables S and Sigma}: 
\begin{subequations}\label{precequationSSigma}
\begin{align}
\frac{\dd \bm{S}}{c\,\dd t} &= \frac{\partial H}{\partial \bm{S}} \times \bm{S} + \frac{\partial H}{\partial \bm{\Sigma}} \times \bm{\Sigma}\,,\\\
\frac{\dd \bm{\Sigma}}{c\,\dd t} &=  \frac{1}{\nu} \,\frac{\partial H}{\partial \bm{\Sigma}} \times \bm{S} + \left( \frac{\partial H}{\partial \bm{S}} + \frac{\delta}{\nu} \frac{\partial H}{\partial \bm{\Sigma}} \right) \times \bm{\Sigma}\,.
\end{align}
\end{subequations}
Projecting again on the $(\n, \la, \el)$ basis, with $\bm{p} \equiv p_r \,\n +p_\varphi\, \la $, we get a first-order differential system for 10 variables: $r,p_r,\varphi,p_\varphi$ and the six components of $\bm{S}$ and $\bm{\Sigma}$.

In the following we consider the perturbation of the EFT Hamiltonian and corresponding EFT EoM, and also precession equations, up to 4PN order, \textit{i.e.} including the non-local 4PN tail term (already dealt with in our previous paper~\cite{BLL25}), and taking into account spin terms up to the NNLO level for spin-orbit and spin-spin interactions, and LO level for cubic and quartic spin terms. All the details on the spin dynamics are provided in Appendices~\ref{app:hamiltonianNS},~\ref{app:spinEFT} and~\ref{app:EoM formulae}.

\subsection{Linearization  of the equations of motion}

The decomposition on the basis $(\n, \la, \el)$ of the equations of motion~\eqref{acceleration} and precession~\eqref{precession} leads to a system of nine equations, of the form ($\alpha = 1, \cdots, 9$)
\begin{equation}
\label{Xdot}
    \frac{\dd X^\alpha}{\dd t} = f^\alpha(X^\beta)\,,
\end{equation}
where the $X^\alpha$ denote the  dynamical variables given in Eq.~\eqref{defXalpha}. Given some specific background configuration characterised by a solution $X_0^\alpha(t)$ of~\eqref{Xdot}, the evolution of the  linear perturbation $\delta X^\alpha\equiv X^\alpha - X_0^\alpha$  will be governed by the linear system 
\begin{equation}
\label{deltaXdot}
     \frac{\dd \delta \bm{X}}{\dd t} = \bm{M}\cdot\delta \bm{X}\,,
\end{equation}
where $\delta \bm{X}$ stands for the column vector whose components are $\delta X^\alpha$ and $\bm{M}$ is the $9\times 9$ matrix obtained by linearising the system \eqref{Xdot} around the background solution,
namely
\begin{equation}
    M_{\alpha\beta}\equiv\frac{\partial f^\alpha}{\partial X^\beta}(X_0)\,.
\end{equation}
If one assumes a harmonic time dependence of the linear perturbations, \textit{i.e.} 
$\delta \bm{X}(t) = \de^{\di \sigma t}\, \bm{\Xi}$, then the time-independent vector $\mathbf{\Xi}$ must be an eigenvector, and $\di\sigma_p$ an eigenvalue, of the matrix $\bm{M}$. The proper frequencies $\sigma_p$ of the linearized system are thus characterized by the relation
\begin{equation}
\label{det}
    \det\bigl(\bm{M} - \di \sigma_p 1\!\!1\bigr) = 0\,.
\end{equation}
The system is stable only if all the $\sigma_p$'s are purely real, \textit{i.e.}, $\sigma_p^2$ is positive. Note that the same procedure applies to the Hamiltonian equations, with the only difference that the first-order system \eqref{Xdot} contains 10 equations for $\{r, p_r, \varphi, p_\varphi, \bm{S}, \bm{\Sigma}\}$ instead of 9 for $\{r, u, \Omega, \bm{S}, \bm{\Sigma} \}$ in the EoM case.

\subsection{Circular orbit with aligned spins}

We now focus on an exact circular orbit  of radius $r_0$ and orbital frequency $\Omega_0$ (neglecting radiation reaction in the EoM and Hamiltonian), with  spins {\it aligned} with the orbital angular momentum, \textit{i.e.} $\bm{S}_a\propto\bm{\ell}$. This configuration is described by the background variables
\begin{equation}
    X_0^\alpha = \bigl\{r_0, 0, \Omega_0, 0, 0, S_\ell, 0, 0, \Sigma_\ell  \bigr\}\,,
\end{equation}
noting that the background variables\footnote{To avoid clutter, we omit the subscript $0$ on $S_\ell$ and $\Sigma_\ell$. This introduces no ambiguity, as these two variables need not be perturbed to derive the criterion in the aligned case, as explained below.} $S_\ell$ and $\Sigma_\ell$ are necessarily time-independent since $S_{1\ell}$ and $S_{2\ell}$ are constant in this aligned case. Substituting these background values into the matrix $\bm{M}$, we obtain the matrix $\bm{M}_0$ valid for a circular orbit with aligned spins. 

Considering first  the components of $\bm{M}_0$ only up to 1PN order, we can easily compute the corresponding eigenvalues. We find the eigenvalue $0$ with a degeneracy of order 3, and three other distinct eigenvalues for which:
\begin{equation}
\frac{\sigma_p^2}{\Omega_0^2} \in \bigg\{1-6x\,,\  1-\biggl(\nu +\frac{3}{2}-\frac{3}{2} \sqrt{1-4 \nu }\biggr)  x\,,\   1-\biggl(\nu +\frac{3}{2}+\frac{3}{2} \sqrt{1-4 \nu }\biggr)  x\bigg\} +{\cal O}(x^{3/2})\,.
\end{equation}
Stability requires all values of $\sigma_p^2$ to be positive, but one observes that since $0\leqslant \nu \leqslant 1/4$ the condition $1-6x>0$  automatically implies that the two other values of $\sigma_p^2$ are positive as well. Therefore, in the specific case of a circular orbit with aligned spins (it could be different for another background configuration), the stability criterion is entirely determined by the $3\times 3$ upper left submatrix of $\bm{M}_0$, which we denote $\hat{\bm{M}}_0$. The discussion above justifies \emph{a posteriori} the analysis of Favata~\cite{F11b}, who did not consider the spin perturbations in the derivation of the stability criterion for circular orbits with aligned spins. 

We have now reduced the problem to the computation of the eigenvalues of a $3\times 3$ matrix of the form
\begin{equation}
\hat{\bm{M}}_0=\left(
    \begin{matrix}
0 & 1 & 0\cr
\alpha_0 & 0 & \beta_0\cr
0 & \gamma_0 & 0
    \end{matrix}
    \right)\,.
\end{equation}
The resulting stability criterion is then
\begin{equation}
\label{criterion_EOM}
    \hat C \equiv -\alpha_0-\beta_0\gamma_0>0\,.
\end{equation}
The criterion retains the same form as in the spinless case~\cite{BLL25}, but the explicit expressions for the coefficients $\alpha_0$, $\beta_0$ and $\gamma_0$ are now much more involved since they also depend on the spins $S_\ell$ and $\Sigma_\ell$.

We can also derive the criterion directly from the total Hamiltonian by following the procedure in~\cite{BLL25}. Let us summarise the main steps. The Hamiltonian can be expressed in terms of the variables $r$, $\varphi$ (where $\varphi$ is the orbital phase) and their conjugate momenta $p_r$ and $p_\varphi$. According to our discussion above, when perturbing a circular orbit with aligned spins we can ignore the spins as dynamical variables.  
A  circular orbit is characterized by a radius $r=r_0$ and the condition $p_r=0$. The latter condition allows us to express the (non-conserved\footnote{The angular momentum $p_\varphi$ is not conserved due to the nonlocal tail term at 4PN order.}) angular momentum $p_\varphi \equiv p_\varphi^0$ as a function of $r_0$ and $\varphi_{0} \equiv \Omega_0 t$ by solving, iteratively up to 4PN order,
\begin{equation}
  p_r=  \frac{\delta H}{\delta r}[r_0,0,\varphi_{0},p_{\varphi}^{0}; S_\ell, \Sigma_\ell]=0\,,
\end{equation}
where, as we said, the spins $S_\ell$, $\Sigma_\ell$ are just non-dynamical ``spectator'' variables. Here and below, we use functional derivatives to account for the tail part of the Hamiltonian. For the instantaneous part, these derivatives reduce to ordinary partial derivatives. Knowing $r_0$ and $p_\varphi^{0}(r_0)$, we can use another of Hamilton's equations to determine the orbital frequency as a function of $r_0$: 
\begin{equation}
\label{eq:orbital frequency hamiltonian formalism}
   \Omega_0= \frac{\delta H}{\delta p_\varphi}[r_0,0,\varphi_{0},p_{\varphi}^{0}; S_\ell, \Sigma_\ell]\,.
\end{equation}
The linearised equations of motion then read
\begin{subequations}
	\label{lin_Hamiltonian_eqs}
	\begin{align}
		&\label{eq: delta R}\delta \dot{r} =\sigma_0\, \delta p_r\,,  \\
		&\label{eq:deltaPR}\delta\dot{p}_r = -\pi_0\,\delta r-\rho_0\,\delta p_\varphi \,,  \\
		&\delta \dot{\varphi} = \zeta_0\, \delta r +\tau_0\, \delta p_\varphi \,,\\
		&\label{eq: delta P_Psi}\delta \dot{p}_\varphi = -\theta_0\, \delta p_r \,,
	\end{align}
\end{subequations}
where we have introduced the coefficients
\begin{subequations}
   \begin{align}
     &\sigma_0 = \frac{\delta^2 H}{\delta p_r^2}[r_0,0,\varphi_{0},p_{\varphi}^{0}; S_\ell, \Sigma_\ell] \,,\qquad
        \pi_0 =\frac{\delta^2 H}{\delta r^2}[r_0,0,\varphi_{0},p_{\varphi}^{0}; S_\ell, \Sigma_\ell] \,, \\
    & \label{eq: def rho0} \rho_0 = \frac{\delta^2 H}{\delta p_{\varphi} \delta r }[r_0,0,\varphi_{0},p_{\varphi}^{0}; S_\ell, \Sigma_\ell]   \,,\qquad
     \zeta_0 = \frac{\delta^2 H}{\delta r\, \delta p_{\varphi}}[r_0,0,\varphi_{0},p_{\varphi}^{0}; S_\ell, \Sigma_\ell]  \,,\\
   & \tau_0 = \frac{\delta^2 H}{\delta p_\varphi^2}[r_0,0,\varphi_{0},p_{\varphi}^{0}; S_\ell, \Sigma_\ell] \,,\qquad
    \theta_0 =\frac{\delta^2 H_{\rm tail}}{\delta p_{r}\, \delta \varphi}[r_0,0,\varphi_{0},p_{\varphi}^{0}; S_\ell, \Sigma_\ell]\,.
   \end{align}
\end{subequations}
The proper frequencies $\sigma_p$ of the linear system~\eqref{lin_Hamiltonian_eqs} are determined by the condition~\eqref{det}, which yields $\sigma_p^2=\pi_0\,\sigma_0 -\rho_0 \, \theta_0$, where  $\zeta_0$ and $\tau_0$ do not appear.\footnote{As a shortcut, one can  take the time derivative of \eqref{eq:deltaPR}, then substitute \eqref{eq: delta R} and \eqref{eq: delta P_Psi}. This gives  a second-order equation for $\delta p_r$ only, with frequency $\sigma_p^2=\pi_0\,\sigma_0 -\rho_0 \, \theta_0$.} The stability criterion is thus
\begin{equation}
\label{criterion_Hamiltonian}
   \hat{C} \equiv \pi_0\,\sigma_0 -\rho_0 \, \theta_0>0\,.
\end{equation}
As we have checked explicitly (in the EFT coordinate system we use here), this expression  yields the same result as the EoM criterion \eqref{criterion_EOM}.

\section{Discussion of the stability criterion}
\label{sec:criterion}

\subsection{The criterion for black holes}

Upon using the expression~\eqref{criterion_EOM}, based on the EoM, or~\eqref{criterion_Hamiltonian}, based on the Hamiltonian, we have obtained the stability criterion for circular orbits with spinning bodies up to 4PN order. The criterion depends on the EFT variables but we can obtain a gauge-invariant expression by expressing it in terms of the orbital frequency $\Omega$ measured at a large distance from the system. Defining, as usual,
\begin{equation}\label{xdef}
x \equiv \left(\frac{G m \Omega}{c^3}\right)^{2/3}\,,
\end{equation}
and introducing the dimensionless spin variables
\begin{equation}
\label{notspins}
    \chi_S\equiv \frac{S_\ell}{G m^2}\,,\qquad  \chi_\Sigma\equiv \frac{\Sigma_\ell}{G m^2}\,,
\end{equation}
we obtain the full 4PN expression of the invariant criterion $C$ (a renormalised version of $\hat C$, defined so that $C=1$ for $x=0$), valid for \emph{any} compact object, as reported in Appendix~\ref{app:ISCOcriterion}. Due to square ($SS$), cubic ($SSS$) and quartic ($SSSS$) spin terms, the result depends on dimensionless coefficients reflecting the internal structure of the bodies (denoted $\kappa_a$, $\lambda_a$ and $\iota_a$, respectively). 

Here we focus on the simplest version of the criterion, where the two objects are black holes, for which $\kappa_a=\lambda_a=\iota_a=1$. We find (recall that $\delta\equiv(m_1-m_2)/m$) 
\begin{align}
\label{critereTN}
C_{\rm BH}^{\rm 4PN} &= 1 - 6 x + x^{3/2} \left( 14 \chi_S + 6 \delta \chi_\Sigma \right) + x^2 \Bigl[ 14 \nu - 12 \chi_S^2 - 12 \delta \chi_S \chi_\Sigma+  \left( -3 + 12 \nu \right)\chi_\Sigma^2  \Bigr]  \\
&+ x^{5/2} \Bigl[  \left( -22 - 32 \nu \right)\chi_S -  \left( 18 + 15 \nu \right)\delta \chi_\Sigma \Bigr] \nn \\
&+ x^3 \left\{ \left( \frac{397}{2} - \frac{123 \pi^2}{16} \right) \nu - 14 \nu^2 +  \left( 74 + 20 \nu \right) \chi_S^2 + \left( 30 - 109 \nu - 20 \nu^2 \right)\chi_\Sigma^2  + \left( 96 + 20 \nu \right) \delta \chi_S \chi_\Sigma \right\} \nn \\
&+ x^{7/2} \left\{  \left( -\frac{568}{3}\nu + \frac{86}{3} \nu^2 \right)\chi_S + \left( -\frac{269}{4}\nu + 14 \nu^2 \right) \delta \chi_\Sigma - 64 \chi_S^3  - 112 \delta \chi_S^2 \chi_\Sigma -12\left(1 - 4 \nu \right)\delta \chi_\Sigma^3   -64 \left(1 -4\nu \right) \chi_S \chi_\Sigma^2 \right\} \nn \\
&+ x^4 \Bigg\{ \nu \left( -\frac{215729}{180} + \frac{58265 \pi^2}{1536} + \frac{5024}{15} \gamma_{\rm E} + \frac{2512}{15} \ln x + \frac{1184}{15} \ln 2 + \frac{2916}{5} \ln 3 \right) + \nu^2 \left( -\frac{4223}{6} + \frac{451 \pi^2}{16} \right) + \frac{196}{27} \nu^3 \nn \\
& + \left( -\frac{289}{3} + \frac{251}{3} \nu - \frac{40}{3} \nu^2 \right)\chi_S^2    + \left( -63 + \frac{843}{4} \nu + \frac{185}{3} \nu^2 + \frac{40}{3} \nu^3 \right) \chi_\Sigma^2 +  \left( -156 + 5 \nu - \frac{40}{3} \nu^2 \right)\delta \chi_S \chi_\Sigma \Bigg\} + \mathcal{O}(x^{9/2})\,.\nn
\end{align}
The first terms agree with the result of Favata~\cite{F11b}, who computed the spin contributions up to order 2.5PN. Recall that the two methods we have used to obtain~\eqref{critereTN} and the more general version~\eqref{criteregeneral} in App.~\ref{app:ISCOcriterion} -- namely perturbation of the Hamiltonian and perturbation of the EoM -- both rely on the EFT coordinate system. As an additional check, we have also computed the criterion from the EoM in harmonic coordinates, up to order 3.5PN (corresponding to the LO, NLO and NNLO spin-orbit, the LO and NLO spin-spin and the LO cubic spin contributions), and verified that it coincides with the above expression truncated at the same order.

When the two spins are zero, we recover the spinless criterion at 4PN found in our previous work~\cite{BLL25}, 
\begin{align}\label{criterion_nospin}
 C_{\rm spinless}^{\rm 4PN} &= 1 -6 x+14\nu\, x^2 + \nu \left[ \frac{397}{2}-\frac{123 \pi^2}{16} -14\nu  \right] x^3
+ \nu \Bigg[ -\frac{215729}{180} + \frac{58265 \pi^2}{1536} +\frac{5024}{15}\,\gamma_{\rm E} 
 \cr
&  
+\frac{2512}{15}\,\ln x  +\frac{1184}{15}\,\ln 2+\frac{2916}{5}\,\ln 3 
 + \left( -\frac{4223}{6}+\frac{451\pi^2}{16}\right) \nu  +\frac{196}{27}\,\nu^2 \Bigg]x^4 + \calO(x^5)\,.
\end{align}
While the spinless criterion was derived in~\cite{BLL25} based on harmonic coordinates and ADM coordinates, the derivation here is based on EFT coordinates.

As shown by first-order GSF ISCO-shift calculations~\cite{BarackS09,BDS10} (see also the recent work~\cite{TNP26}), the ISCO frequency is given by the solution of $(n/\Omega)^2=0$, where $n=2\pi/P$ is the radial frequency of the circular orbit ($P$ is the orbital period) and $\Omega$ is the orbital frequency, both being expressed in terms of $x$ for circular orbits. Since for circular orbits $\Omega=K n$ where $K$ is the orbital precession (or periastron advance), with $k=K-1$ being the relativistic precession, the condition $C(x)=0$ is equivalent to $K^{-2}(x)=0$. Quite interestingly, this correspondence seems to hold for any mass ratio. We observe that Eq.~\eqref{criterion_nospin} allows us, with the identification $C=K^{-2}$, to recover the precession for circular orbits (see \textit{e.g.} Eq.~(5.11) in~\cite{BBBFMb} for comparison). Moreover, it was shown very recently in~\cite{TrestiniSteinhoffInPreparation} that this link between the ISCO frequency and the orbital precession can be extended — again, for any mass ratio — to spin contributions (see therein Eq. (7.1), which perfectly agrees with our general result~\eqref{criteregeneral}).

\subsection{The Kerr ISCO}
\label{subsection:Kerr}

It is instructive to first compare our PN criterion with the stability criterion for a test (spinless) particle orbiting a Kerr black hole, with mass $M$ and dimensionless spin parameter $\chi=a/M$ where $a$ is the Kerr parameter.\footnote{Here and below, depending on the context, we often pose $c=G=1$.} The stability criterion in this case is known exactly and given by the expression~\cite{BPT72}
\begin{align}\label{exactKerrcriterion}
C_\text{Kerr}(x, \chi) = 1 - \frac{x}{\beta^{2/3}}\left[ 6 - \chi \,\frac{x^{1/2}}{\beta^{1/3}} \left( 8 - 3 \chi \, \frac{x^{1/2}}{\beta^{1/3}} \right)\right]\,, \qquad \beta \equiv 1-\chi \,x^{3/2}\,,
\end{align}
where  $x\equiv\left({G M \Omega}/{c^3}\right)^{2/3}$, which coincides with \eqref{xdef} in the test-mass limit. The ISCO is then given by the solution of the equation $C_\text{Kerr}(x,\chi)=0$, 
\begin{align}\label{exactKerrISCO}
x_\text{ISCO}^\text{Kerr}(\chi) = \left[ \chi + \left(\rho^{\rm Kerr}_{\rm ISCO}(\chi)\right)^{3/2}\right]^{-2/3}\,,
\end{align}
where\footnote{The quantity $\rho$ (denoted $w^{-1}$ in~\cite{F11b}) corresponds to $\rho=r_{\rm BL}/M$, where $r_{\rm BL}$  is the usual Boyer-Lindquist radial coordinate.}~\cite{Bardeen:1972fi}
\begin{equation}
\label{r_ISCO_Kerr}
    \rho^{\rm Kerr}_{\rm ISCO}(\chi)\equiv  3+Z_2-\text{sign}(\chi)\sqrt{(3-Z_1)(3+Z_1+2Z_2)}
\end{equation}
with 
\begin{equation} 
	Z_1 \equiv 1+ (1-\chi^2)^{1/3}\left[(1+\chi)^{1/3}+(1-\chi)^{1/3}\right]\,,\qquad 
	Z_2 \equiv \sqrt{3\chi^2+Z_1^2}\,.
\end{equation}
When $\chi=0$ we recover the Schwarzschild ISCO, $x_\text{ISCO}^\text{Schw}=1/6$ or $r_{\rm BL}=6 M$. As is well known, for maximum and \textit{prograde} spin $\chi=1$ the ISCO reaches the horizon of the black hole, $x_\text{ISCO}^\text{Kerr}(1)=2^{-2/3}$ or $r_{\rm BL}=M$, and is therefore located in the strong field region. By contrast, for maximum but \textit{retrograde} spin $\chi=-1$ the ISCO is located at $x_\text{ISCO}^\text{Kerr}(-1)=26^{-2/3}$ or $r_{\rm BL}=9M$ and is therefore in a relatively weak field region. As a consequence, one expects  the post-Newtonian method to accurately predict the ISCO only in the latter case.

If we assume our body 1 to be the test particle, for which one ignores the spin and mass ($m_1=0$, $S_1=0$), and our body 2 to be the Kerr BH ($m_2=M$, $S_2=M a$), we have, from \eqref{definition spin variables S and Sigma} and \eqref{notspins},  $\chi_S=\chi_\Sigma =\chi$. Our criterion~\eqref{critereTN} in this case thus reduces to 
\begin{align}\label{criterion_Kerr_4PN}
C^\text{4PN}_{\rm Kerr}(x,\chi) &= 1 - 6 x + 8 \chi x^{3/2} - 3\chi^2 x^2- 4 \chi x^{5/2} 
+ 8\chi^2  x^3- 4\chi^3 x^{7/2} - \frac{10}{3}\chi^2 x^4    + \mathcal{O}(x^{9/2})\,.
\end{align}
It is satisfying to confirm that the expansion of \eqref{exactKerrcriterion} up to 4PN order perfectly agrees with the above expression. Note that, at this order, the terms quartic in spins cancel out.

\subsection{Shift of the Kerr ISCO in the small-mass limit}

We now go one step further and consider a particle with small (but non zero)  mass  $m_1$ and spin $\chi_1$ orbiting a large BH with mass $m_2\gg m_1$ and spin $\chi_2$. In this case, we need to introduce the orbital frequency parameter $y$ that involves solely the large mass $m_2$ instead of the total mass $m$ as in \eqref{xdef}, defined by
%
\begin{align}\label{yvar}
	y\equiv \left(\frac{G m_2 \Omega}{c^3}\right)^{2/3} = x \left(1+\frac{m_1}{m_2} \right)^{-2/3}\,.
\end{align}
The spin parameters~\eqref{notspins} are given in terms of the individual spins $\chi_a$ by
\begin{equation}\label{Svar}
\chi_S = \left(\frac{m_1}{m}\right)^2 \chi_1 + \left(\frac{m_2}{m}\right)^2 \chi_2\,,\qquad\quad
\chi_{\Sigma}= \left(\frac{m_2}{m}\right) \chi_2 -\left(\frac{m_1}{m}\right) \chi_1\,. 
\end{equation}

We wish to know by how much the Kerr ISCO is shifted  due to the particle's finite mass $m_1$, at first order in the mass ratio $m_1/m_2$ or,  equivalently, at first order in the symmetric mass ratio $\nu$. We first reexpress the 4PN criterion~\eqref{critereTN} in terms of the variables~\eqref{yvar} and~\eqref{Svar} and compute the limit $\nu\to 0$. In this extreme mass ratio (EMR) limit, the spin $\chi_1$ of the particle disappears and we get 
\begin{align}\label{critere4PNKerr}
C^\text{4PN}_{\rm EMR}(y,\chi_1,\chi_2) &= 1 - 6 y + \left(8 y^{3/2} - 4 y^{5/2}\right)\chi_2 
+ \left(- 3 y^2 + 8  y^3 - \frac{10}{3} y^4\right) \chi_2^2 - 4 y^{7/2} \chi_2^3 + \mathcal{O}\left(\nu\right)\,,
\end{align}
which coincides with \eqref{criterion_Kerr_4PN}.
%
%
%
For the finite-mass correction $\mathcal{O}(\nu)$ in~\eqref{critere4PNKerr}, we find 
\begin{align}
\label{delta_C4PN}
\delta C^\text{4PN}_\text{EMR}(y,\chi_1,\chi_2) =& - 4 y + y^{3/2} \Bigl[6\chi_1-2\chi_2\Bigr] + y^2 \Bigl[14-6\chi_1\chi_2+2\chi_2^2\Bigr]
+ y^{5/2} \biggl[-18\chi_1-\frac{101}{3}\chi_2\biggr]
\nn\\
&
+ y^3 \left[\frac{397}{2}-\frac{123}{16}\pi^2+36\chi_1\chi_2+31 \chi_2^2\right] 
+ y^{7/2} \left[-\frac{1465}{12}\chi_2-20\chi_1 \chi_2^2-\frac{16}{3}\chi_2^3\right] 
\nn\\
&
+ y^4 \left[-\frac{215729}{180} + \frac{5024}{15}\gamma_\text{E}
+\frac{58265}{1536}\pi^2 
+ \frac{1184}{15}\ln 2 + \frac{2916}{5}\ln 3  
\right.\nn\\&
\qquad\left.
\quad
+\frac{2512}{15}\ln y 
- 30\chi_1\chi_2+\frac{427}{36}\chi_2^2\right]\,.
\end{align}
Since we know the exact criterion in the limit $\nu=0$, we can replace the $\nu=0$ part of the approximate  criterion (given by \eqref{critere4PNKerr} at 4PN order) by the exact Kerr criterion~\eqref{exactKerrcriterion}, and thus introduce the \emph{refined} criterion 
\begin{align}\label{shiftcritere}
    C(y,\chi_1,\chi_2) = C_\text{Kerr}(y,\chi_2) + \nu \,\delta C_\text{EMR}(y,\chi_1,\chi_2) + \mathcal{O}\left(\nu^2\right)\,.
\end{align}
At 4PN, this expression  coincides with our  criterion when using \eqref{delta_C4PN} for $\delta C_\text{EMR}$, but  a similar decomposition applies to any approximate criterion, \textit{e.g.}, at different PN orders. This enables a direct comparison of the finite-mass correction with numerical conservative GSF computations, where the background solution is \textit{exactly} given by the Kerr criterion~\eqref{exactKerrcriterion}
and the Kerr ISCO~\eqref{exactKerrISCO}. 
 
We then define the shift in the ISCO position $\delta y_\text{ISCO}^\text{EMR}(\chi_1,\chi_2)$ such that
\begin{align}\label{defshifty}
 y_\text{ISCO}(\chi_1,\chi_2) = y_\text{ISCO}^\text{Kerr}(\chi_2) + \nu \,\delta y_\text{ISCO}^\text{EMR}(\chi_1,\chi_2) + \mathcal{O}\left(\nu^2\right)\,,
\end{align}
where $y_\text{ISCO}(\chi_1,\chi_2)$ is the solution of 
$C[y_\text{ISCO}(\chi_1,\chi_2),\chi_1,\chi_2] = 0$. Inserting~\eqref{defshifty} into this equation, expanding at first order in $\nu$, and taking into account the fact that we have identically $C_\text{Kerr}[y_\text{ISCO}^\text{Kerr}(\chi_2),\chi_2] \equiv 0$, we obtain 
\begin{align}\label{resshifty}
 \delta y_\text{ISCO}^\text{EMR}(\chi_1,\chi_2) = - \frac{\delta C_\text{EMR}[y_\text{ISCO}^\text{Kerr}(\chi_2),\chi_1,\chi_2]}{C^{\,\prime}_{\!\text{Kerr}}[y_\text{ISCO}^\text{Kerr}(\chi_2),\chi_2]}\,,
\end{align}
with the notation $C^{\,\prime}_{\!\text{Kerr}}(y,\chi_2) \equiv (\partial C_\text{Kerr}/\partial y)(y,\chi_2)$. 

We wish to compare this result to the numerical first-order GSF calculations~\cite{IBDLNSTW14}. Following the notation and conventions of~\cite{IBDLNSTW14}, we introduce the ISCO shift parameter $c_\Omega(\chi_1,\chi_2)$ defined for the orbital frequency $\Omega$ such that
\begin{align}\label{OmegaISCO}
 (m_1+m_2) \Omega_\text{ISCO} = m_2 \Omega_\text{ISCO}^\text{Kerr} \Bigl[ 1 + \nu \,c_\Omega(\chi_1,\chi_2) + \mathcal{O}(\nu^2)\Bigr] \,,
\end{align}
with $m_2 \Omega_\text{ISCO}^\text{Kerr} = (y_\text{ISCO}^\text{Kerr})^{3/2}$. Note that we extend the case investigated in~\cite{IBDLNSTW14} by allowing $c_\Omega$ to depend not only on the  large BH spin $\chi_2$ but also on the spin $\chi_1$ of the particle. Thus we have
\begin{align}
\label{cOmega}
 c_\Omega(\chi_1,\chi_2) = 1 - \frac{3}{2}\frac{\delta C_\text{EMR}[y_\text{ISCO}^\text{Kerr}(\chi_2),\chi_1,\chi_2]}{y_\text{ISCO}^\text{Kerr}(\chi_2)C'_{\!\text{Kerr}}[y_\text{ISCO}^\text{Kerr}(\chi_2),\chi_2]}\,.
\end{align}
Inserting \eqref{exactKerrcriterion}, \eqref{exactKerrISCO} and \eqref{delta_C4PN} in this relation, we obtain an explicit expression for the 4PN estimate $c^\text{4PN}_\Omega(\chi_1,\chi_2)$.

\begin{figure}[htbp]
    \centering
\includegraphics[width=0.7\linewidth]{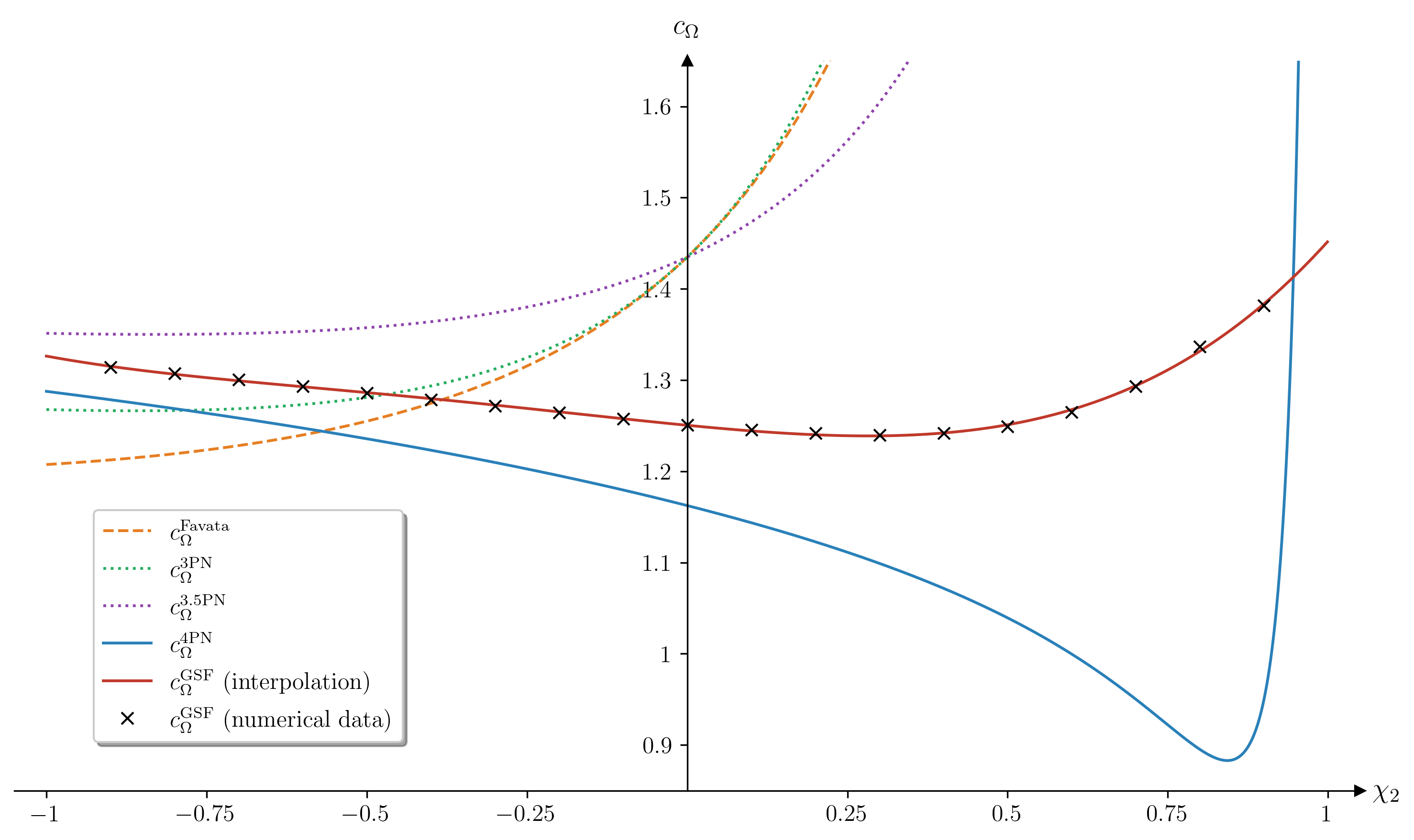}
    \caption{\justifying Comparison of the ISCO shift coefficient $c_\Omega(0,\chi_2)$ (defined in~\eqref{cOmega}) based on our 4PN criterion (solid blue curve) with GSF numerical results from~\cite{IBDLNSTW14} (black crosses, interpolated by the red curve). We also plot the coefficients derived from lower-order criteria:  Favata's hybrid (spinless contributions up to 3PN + spin contributions up to 2.5PN) criterion~\cite{F11b} (orange dashed curve) and the 3PN and 3.5PN truncations of our criterion (green and purple dotted curves, respectively).}
\label{fig:GSF4PNcomparison}
\end{figure}

When $\chi_1=0$ we can compare this 4PN-based prediction with the high-precision GSF numerical calculations in~\cite{IBDLNSTW14}. The results are shown in Fig.~\ref{fig:GSF4PNcomparison}. For comparison, we display the coefficients calculated by Favata \cite{F11b}, based on a criterion that includes spin terms up to 2.5PN and non-spin terms up to 3PN. We also show the coefficients obtained by truncating  our criterion at 3PN and 3.5PN. We observe a significant improvement at 4PN order: the 4PN result matches the GSF numerical calculations better than lower-order PN estimates (particularly for negative and moderately positive $\chi_2$), and the resulting curve is overall closer to the GSF curve. As shown in Table~\ref{tab:rel_error}, the relative difference with the GSF result remains below 10\% for negative or small positive $\chi_2$ but sharply increases for higher positive spins. For the highest values, close to maximality, the curve exhibits a sharp turn and cannot be trusted: this corresponds to a regime where the PN expansion is not expected to be reliable, as the ISCO is close to the BH horizon. We also note that the 4PN coefficient diverges when $\chi_2\rightarrow 1$.
\begin{table}[htbp]
\centering
\setlength{\tabcolsep}{12pt}
\renewcommand{\arraystretch}{1.3}
\resizebox{0.8\textwidth}{!}{
\begin{tabular}{ccccccc}
\toprule
$\chi_2$ & $c_\Omega^{\mathrm{GSF}}$ & $c_\Omega^{\mathrm{4PN}}$ & $(c_\Omega^{\mathrm{4PN}} - c_\Omega^{\mathrm{GSF}})/c_\Omega^{\mathrm{GSF}}$ & $c_{\Omega, \chi_1}^{\mathrm{MPD}}$ & $c_{\Omega, \chi_1}^{\mathrm{4PN}}$ & $(c_{\Omega, \chi_1}^{\mathrm{4PN}} - c_{\Omega, \chi_1}^{\mathrm{MPD}})/c_{\Omega, \chi_1}^{\mathrm{MPD}}$ \\
\midrule
$-0.9$ & $\phantom{-}1.3140$ & $\phantom{-}1.2785$ & \makebox[4em][l]{$-3\%$} & $\phantom{-}0.2364$ & $\phantom{-}0.2316$ & \makebox[4em][l]{$-2\%$} \\
$-0.8$ & $\phantom{-}1.3073$ & $\phantom{-}1.2687$ & \makebox[4em][l]{$-3\%$} & $\phantom{-}0.2424$ & $\phantom{-}0.2380$ & \makebox[4em][l]{$-2\%$} \\
$-0.7$ & $\phantom{-}1.3004$ & $\phantom{-}1.2584$ & \makebox[4em][l]{$-3\%$} & $\phantom{-}0.2487$ & $\phantom{-}0.2448$ & \makebox[4em][l]{$-2\%$} \\
$-0.6$ & $\phantom{-}1.2933$ & $\phantom{-}1.2474$ & \makebox[4em][l]{$-4\%$} & $\phantom{-}0.2553$ & $\phantom{-}0.2520$ & \makebox[4em][l]{$-1\%$} \\
$-0.5$ & $\phantom{-}1.2861$ & $\phantom{-}1.2357$ & \makebox[4em][l]{$-4\%$} & $\phantom{-}0.2625$ & $\phantom{-}0.2598$ & \makebox[4em][l]{$-1\%$} \\
$-0.4$ & $\phantom{-}1.2788$ & $\phantom{-}1.2231$ & \makebox[4em][l]{$-4\%$} & $\phantom{-}0.2700$ & $\phantom{-}0.2680$ & \makebox[4em][l]{$-1\%$} \\
$-0.3$ & $\phantom{-}1.2715$ & $\phantom{-}1.2097$ & \makebox[4em][l]{$-5\%$} & $\phantom{-}0.2782$ & $\phantom{-}0.2768$ & \makebox[4em][l]{$-0.5\%$} \\
$-0.2$ & $\phantom{-}1.2643$ & $\phantom{-}1.1952$ & \makebox[4em][l]{$-5\%$} & $\phantom{-}0.2868$ & $\phantom{-}0.2861$ & \makebox[4em][l]{$-0.3\%$} \\
$-0.1$ & $\phantom{-}1.2574$ & $\phantom{-}1.1795$ & \makebox[4em][l]{$-6\%$} & $\phantom{-}0.2962$ & $\phantom{-}0.2959$ & \makebox[4em][l]{$-0.1\%$} \\
$\phantom{-}0.0$ & $\phantom{-}1.2510$ & $\phantom{-}1.1624$ & \makebox[4em][l]{$-7\%$} & $\phantom{-}0.3062$ & $\phantom{-}0.3062$ & \makebox[4em][l]{$\phantom{-}0.0\%$} \\
$\phantom{-}0.1$ & $\phantom{-}1.2456$ & $\phantom{-}1.1436$ & \makebox[4em][l]{$-8\%$} & $\phantom{-}0.3170$ & $\phantom{-}0.3167$ & \makebox[4em][l]{$-0.1\%$} \\
$\phantom{-}0.2$ & $\phantom{-}1.2416$ & $\phantom{-}1.1227$ & \makebox[4em][l]{$-10\%$} & $\phantom{-}0.3287$ & $\phantom{-}0.3269$ & \makebox[4em][l]{$-1\%$} \\
$\phantom{-}0.3$ & $\phantom{-}1.2399$ & $\phantom{-}1.0990$ & \makebox[4em][l]{$-11\%$} & $\phantom{-}0.3414$ & $\phantom{-}0.3360$ & \makebox[4em][l]{$-2\%$} \\
$\phantom{-}0.4$ & $\phantom{-}1.2418$ & $\phantom{-}1.0717$ & \makebox[4em][l]{$-14\%$} & $\phantom{-}0.3551$ & $\phantom{-}0.3420$ & \makebox[4em][l]{$-4\%$} \\
$\phantom{-}0.5$ & $\phantom{-}1.2492$ & $\phantom{-}1.0393$ & \makebox[4em][l]{$-17\%$} & $\phantom{-}0.3699$ & $\phantom{-}0.3407$ & \makebox[4em][l]{$-8\%$} \\
$\phantom{-}0.6$ & $\phantom{-}1.2650$ & $\phantom{-}0.9996$ & \makebox[4em][l]{$-21\%$} & $\phantom{-}0.3856$ & $\phantom{-}0.3224$ & \makebox[4em][l]{$-16\%$} \\
$\phantom{-}0.7$ & $\phantom{-}1.2932$ & $\phantom{-}0.9501$ & \makebox[4em][l]{$-27\%$} & $\phantom{-}0.4014$ & $\phantom{-}0.2610$ & \makebox[4em][l]{$-35\%$} \\
$\phantom{-}0.8$ & $\phantom{-}1.3368$ & $\phantom{-}0.8951$ & \makebox[4em][l]{$-33\%$} & $\phantom{-}0.4150$ & $\phantom{-}0.0700$ & \makebox[4em][l]{$-83\%$} \\
$\phantom{-}0.9$ & $\phantom{-}1.3816$ & $\phantom{-}0.9498$ & \makebox[4em][l]{$-31\%$} & $\phantom{-}0.4152$ & $-0.7018$ & \makebox[4em][l]{$-269\%$} \\
\bottomrule
\end{tabular}}
\caption{\justifying 
	Values of the ISCO shift coefficients for $\chi_2 \in [-0.9,0.9]$. The second and third columns respectively display $c_{\Omega}(0,\chi_2)$ computed in the GSF framework and with our 4PN criterion (corresponding to the red and blue curves in Fig. \ref{fig:GSF4PNcomparison}); the fourth column indicates the relative difference between the two; the fifth column displays $c_{\Omega, \chi_1}(\chi_2)$ computed using the MPD equations of motion, see the next Sec.~\ref{sec:MPD} and the red curve in Fig.~\ref{fig:c_{Omega,chi1}}; the sixth column shows $c_{\Omega, \chi_1}(\chi_2)$ computed using our 4PN criterion, also shown as the blue curve in Fig.~\ref{fig:c_{Omega,chi1}}; the last column indicates again the relative difference between the two.}
\label{tab:rel_error}
\end{table}

\begin{figure}[h!]
    \centering
\includegraphics[width=0.65\linewidth]{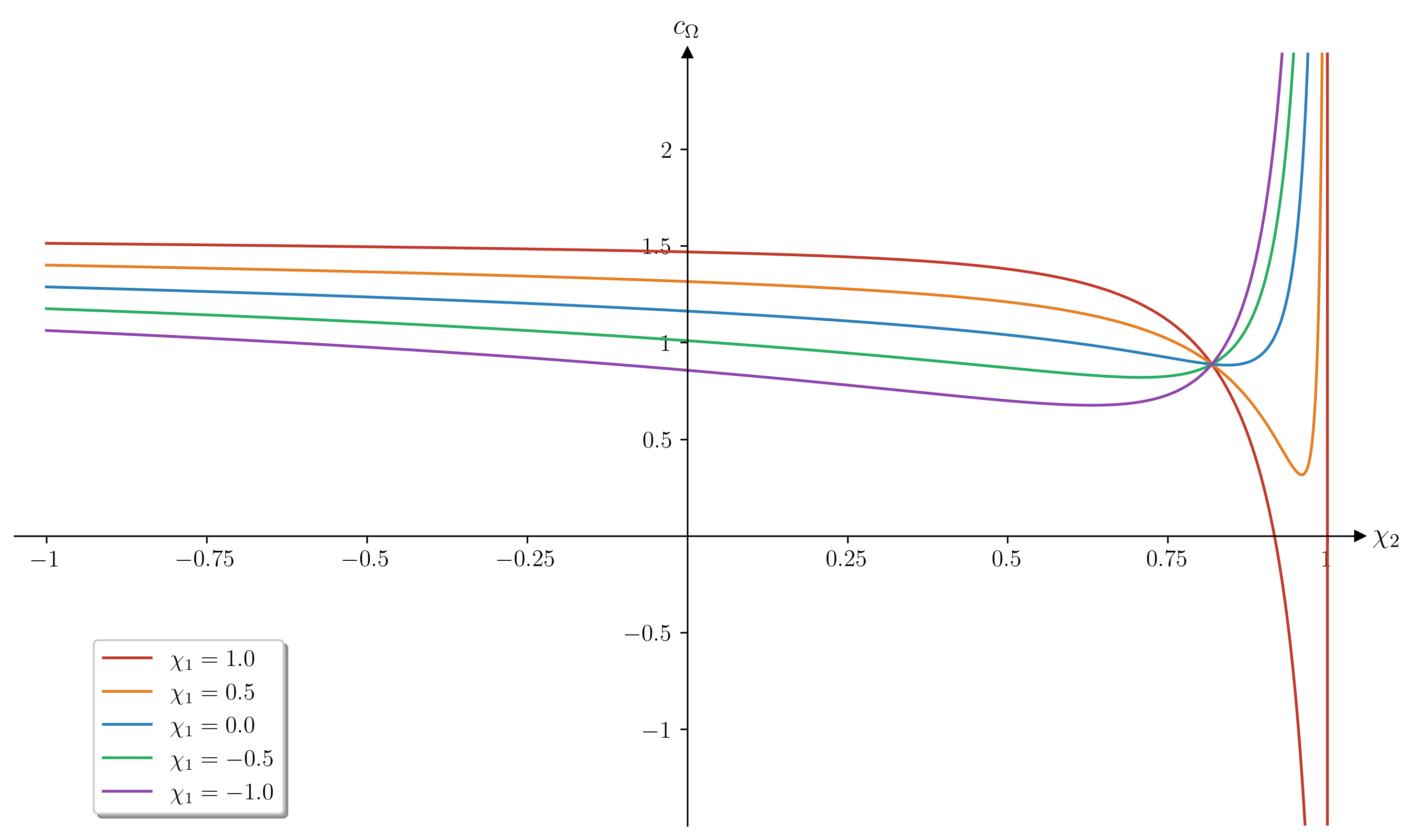}
    \caption{\justifying ISCO shift coefficient $c^{\rm 4PN}_\Omega(\chi_1,\chi_2)$ (defined in~\eqref{cOmega}) as a function of the spin $\chi_2$ of the large BH, for several values of the spin $\chi_1$ of the small BH.}
    \label{fig:shiftISCO}
\end{figure}

For a spinning small body, \textit{i.e.} $\chi_1\neq 0$, the coefficient $c_\Omega$ is modified. In fact, one can see from the expressions~\eqref{delta_C4PN} and~\eqref{cOmega} that $c_\Omega^\text{4PN}$ is a linear function of $\chi_1$, namely
\begin{equation}
\label{c_Omega,chi1}
    c_\Omega^\text{4PN}(\chi_1,\chi_2) = c_\Omega^\text{4PN}(0,\chi_2) +c^\text{4PN}_{\Omega,\chi_1}(\chi_2)\, \chi_1\,. 
\end{equation}
We have plotted on Fig.~\ref{fig:shiftISCO} this coefficient as a function of $\chi_2$ for several values of $\chi_1$. It turns out that $c^\text{4PN}_{\Omega,\chi_1}$ vanishes for $\chi_2\simeq 0.819$, which explains why all curves intersect at the same point. In the interval where the 4PN estimate is reliable, $c^\text{4PN}_{\Omega,\chi_1}$ is always positive, so that a positive $\chi_1$ increases $\Omega_{\rm ISCO}$ (making the ISCO more ``inward'') while a negative $\chi_1$ has the opposite effect.
When the large BH is not rotating, i.e. $\chi_2=0$, one finds
\begin{equation}
\label{cOmega4PN0}
c_\Omega^{\rm 4PN}(\chi_1,0) = \frac{89371}{155520} + \frac{157}{405}\gamma_\text{E} - \frac{12583}{1327104}\pi^2 - \frac{83}{810}\ln(2) + \frac{1559}{3240}\ln(3) + \frac{\sqrt{6}}{8}\chi_1\,,
\end{equation}
where the coefficient $c^\text{4PN}_\Omega(0,0)$ was already given in~\cite{BLL25} and the coefficient $c^\text{4PN}_{\Omega,\chi_1}(0)$ in~\cite{F11b}.

\subsection{Spinning test-particle in Kerr}
\label{sec:MPD}

In this subsection, we second~\cite{F11b} and make the relevant comparison between our result $c^{\rm 4PN}_{\Omega,\chi_1}(\chi_2)$ and the coefficient derived analytically from the Mathisson-Papapetrou-Dixon (MPD) equations of motion for a spinning test-particle moving in the equatorial plane of a Kerr black hole. Thus, we consider a spinning particle around a BH of mass $m_2=M$ and spin  $M a=M^2 \chi$. The particle's spin  $m_1 s=m_1^2 \chi_1$ is aligned with the BH spin; the motion of the particle is equatorial. Following~\cite{Saijo:1998mn} (see also the Appendix B in~\cite{F11b}), the equations of motion for the spinning particle, written in terms of the usual Boyer-Lindquist coordinates ($r\equiv r_\text{BL}$), are of the form
\begin{subequations}
\begin{align}
\label{tdot}
  \Sigma_s \Lambda_s \frac{\dd t}{\dd\tau} &= \! a \left( 1 + \frac{3M s^2}{r \Sigma_s} \right) \! \left[ \tilde{L}_z - (a + s) \tilde{E} \right] \! + \! \frac{r^2+a^2}{\Delta}P_s\,, \\
\label{phidot}
  \Sigma_s \Lambda_s \frac{\dd\varphi}{\dd\tau} &= \left( 1 + \frac{3M s^2}{r \Sigma_s} \right) \left[ \tilde{L}_z - (a + s) \tilde{E} \right] + \frac{a}{\Delta}P_s\,, \\
\label{rdot}
  \Sigma_s \Lambda_s \frac{\dd r}{\dd\tau} &= \pm \sqrt{R_s}\,,
\end{align}
\end{subequations}
with
\begin{subequations}
\label{Sigma_s_etal}
    \begin{align}
        \Sigma_s &\equiv r^2 \left( 1 - \frac{M s^2}{r^3} \right), \qquad\qquad
\Lambda_s \equiv 1- \frac{3 M s^2 r [\tilde{L}_z - (a+s)\tilde{E} ]^2}{\Sigma_s^3}\,, \\
P_s &\equiv \left[ (r^2+a^2) + a s \left( 1 + \frac{M}{r} \right) \right] \tilde{E} - \left( a + s \frac{M}{r} \right) \tilde{L}_z, \qquad\quad
\Delta \equiv r^2-2M r + a^2\,,
 \\
R_s &\equiv P_s^2 - \Delta \left\{ \frac{\Sigma_s^2}{r^2} + \bigl[ \tilde{L}_z - (a+s)\tilde{E} \bigr]^2 \right\}\,,
    \end{align}
\end{subequations}
where $\E$ is the particle's conserved energy (per unit mass)  and $\Lp_z$ its conserved angular momentum (per unit mass). The quantity $R_s$, which appears on the right-hand side of the radial EoM~\eqref{rdot}, is quadratic in $\E$, of the form 
\begin{equation}
    R_s =\alpha \tilde{E}^2 - 2 \beta \tilde{E} + \gamma = \alpha(r) \bigl[\tilde{E}-\tilde{E}_+(r,\tilde{L}_z)\bigr]\bigl[\tilde{E}-\tilde{E}_-(r,\tilde{L}_z)\bigr]\,,
\end{equation}
where the coefficients $\alpha$, $\beta$ and $\gamma$ can be read immediately from the expressions in~\eqref{Sigma_s_etal}. As a consequence, the radial EoM~\eqref{rdot} can be expressed as 
\begin{equation}
\label{radial_eq}
    \dot{r}^2 = \Xi(r,\tilde{E}, \tilde{L}_z) \bigl[\tilde{E}-\Ve\bigr]\,, \qquad \Ve\equiv \tilde{E}_+(r,\tilde{L}_z)  = \frac{\beta + \sqrt{\beta^2 - \alpha \gamma}}{\alpha}\,,
\end{equation}
where the dimensionless quantity $\Ve$ plays the role of an effective potential for the particle's motion. It is convenient to express it in terms of the dimensionless quantities
\begin{align}\label{rescale}
    \rho\equiv \frac{r}{M}\,,\qquad  \ell\equiv \frac{\tilde{L}_z}{M}\,,\qquad \chi\equiv \frac{a}{M}\,, \qquad \sigma\equiv \frac{s}{M}\,, \qquad \omega\equiv M\,\Omega\,.
\end{align}

On a circular orbit of radius $r$, characterized by $\dot r=0$ and $\ddot r=0$, we have
\begin{equation}
\label{circular}
    \E=\Ve(\rho,\ell)\,, \qquad \frac{\partial\Ve}{\partial \rho}(\rho,\ell)=0\,,
\end{equation}
while the ISCO is characterised by the additional condition
\begin{equation}
\label{isco}
    \frac{\partial^2\Ve}{\partial \rho^2}(\rho,\ell)=0\,.
\end{equation}
When the particle has no spin, \textit{i.e.} $\sigma=0$, the solution to the equations~\eqref{circular}--\eqref{isco} is given by~\cite{Bardeen:1972fi}
\begin{equation}\label{sols=0}
  \E_0=\frac{\chi +\sqrt{\rho_0} (\rho_0-2 )}{\rho_0^{3/4} \sqrt{2 \chi+\sqrt{\rho_0} (\rho_0-3)}}\,,\qquad \ell_0=\frac{\chi^2-2 \chi 
   \sqrt{\rho_0}+\rho_0^2}{\rho_0^{3/4} \sqrt{2 \chi+\sqrt{\rho_0} (\rho_0-3 )}}\,,\qquad   \rho_0=\rho^{\rm Kerr}_{\rm ISCO}(\chi)\,,
\end{equation}
where $\rho^{\rm Kerr}_{\rm ISCO}(\chi)$ corresponds to the ISCO of a \emph{spinless} particle, given in \eqref{r_ISCO_Kerr}. Note that the first two relations apply to any circular orbit, whereas the last one defines uniquely the ISCO.

There is no analytical solution for a generic value of the particle's spin $\sigma$ but one can solve Eqs.~\eqref{circular} and~\eqref{isco} perturbatively, in an expansion with respect to $\sigma \to 0$. By substituting into these equations
\begin{equation}
\label{rho1_etal}
    \rho=\rho_0+\sigma\, \rho_1+{\cal O}(\sigma^2)\,,\qquad  \ell=\ell_0+\sigma\, \ell_1+{\cal O}(\sigma^2)\,,\qquad  \E=\E_0+\sigma\,\E_1+{\cal O}(\sigma^2)\,,
\end{equation}
and expanding at first order in $\sigma$, one  obtains the expressions of $\rho_1$ and $\ell_1$ from the second equation in~\eqref{circular} and from~\eqref{isco}, while  the first equation in~\eqref{circular} provides $\E_1$. The  analytical expressions of the linear-perturbation coefficients $\rho_1$, $\ell_1$ and $\tilde{E}_1$,  as functions of $\chi$, are given in Appendix~\ref{app:MPD}, where they are also plotted (in Fig.~\ref{fig:rho1_E1_L1}).
 
The angular frequency $\omega$ is obtained from the ratio of~\eqref{phidot} and~\eqref{tdot},
\begin{equation}
	\omega = M \frac{\dd\varphi/\dd\tau}{\dd t/\dd\tau} = \omega_0+ \sigma\, \omega_1+{\cal O}(\sigma^2)\,.
\end{equation}
%
%
%
At zeroth order in $\sigma$, one recovers the Kerr ISCO frequency, $\omega_0=(\chi +\rho_0{}^{\!\!3/2})^{-1}$, corresponding to~\eqref{exactKerrISCO}. The linear term is directly related to the coefficient $c_{\Omega,\chi_1}$ defined by~\eqref{OmegaISCO} and~\eqref{c_Omega,chi1}, namely $c_{\Omega,\chi_1}(\chi)=(\partial c_\Omega/\partial\chi_1)(0,\chi)$,  therefore obtaining the MPD prediction for this coefficient. Substituting the explicit expressions for $\rho_1$, $\ell_1$ and $\tilde{E}_1$ given by~\eqref{solution1} in App.~\ref{app:MPD}, we find
\begin{equation}
\label{eq: c_Omega,chi1 for spinning test particle}
    c^\text{MPD}_{\Omega,\chi_1}(\chi)=\frac{\omega_1}{\omega^{\rm Kerr}_{\rm ISCO}} = \frac{A[\rho^{\rm Kerr}_{\rm ISCO}(\chi),\chi]}{B[\rho^{\rm Kerr}_{\rm ISCO}(\chi),\chi]}\,,
\end{equation}
with 
\begin{subequations}
\begin{align}\label{A}
    A[\rho,\chi] &= 3 \Big[\sqrt{\rho } (\rho -3)+2 \chi \Big] \Big[
    -\rho ^{5/2} (5 (\rho -4) \rho +12)+2 \rho ^2 (\rho  (3 \rho
   -16)+8) \chi   +2 \rho ^{3/2} (3 \rho +2) \chi^2
   \nn
   \\
  &\qquad  \qquad \qquad \qquad \qquad \quad
   +4 \rho ^2 \chi ^3 -13 \sqrt{\rho } \chi ^4+6 \chi ^5\Big]\,,
\\
\label{B}
    B[\rho,\chi] &= \rho ^{9/2} \Bigl[\rho  (\rho  (6 \rho -71)+216)-180\Bigr]
    +2 \rho ^3 (2 \rho
   -3) \Bigl[\rho  (21 \rho -76)+12\Bigr] \chi
    +\rho ^{5/2} \Bigl[\rho  (7 (39-4 \rho ) \rho
   -472)+180\Bigr] \chi ^2
   \nn
   \\
   &
    -4 \rho ^2 \Bigl[\rho  (8 \rho -27)+38\Bigr] \chi ^3
   +\rho ^{3/2} \Bigl[(79-18 \rho ) \rho +24\Bigr] \chi ^4
   +2 (13-18 \rho ) \rho  \chi ^5 -9
   \sqrt{\rho}\, \chi ^6 \,.
\end{align}
\end{subequations}
\begin{figure}[htbp]
\centering
\includegraphics[width=0.6\linewidth]{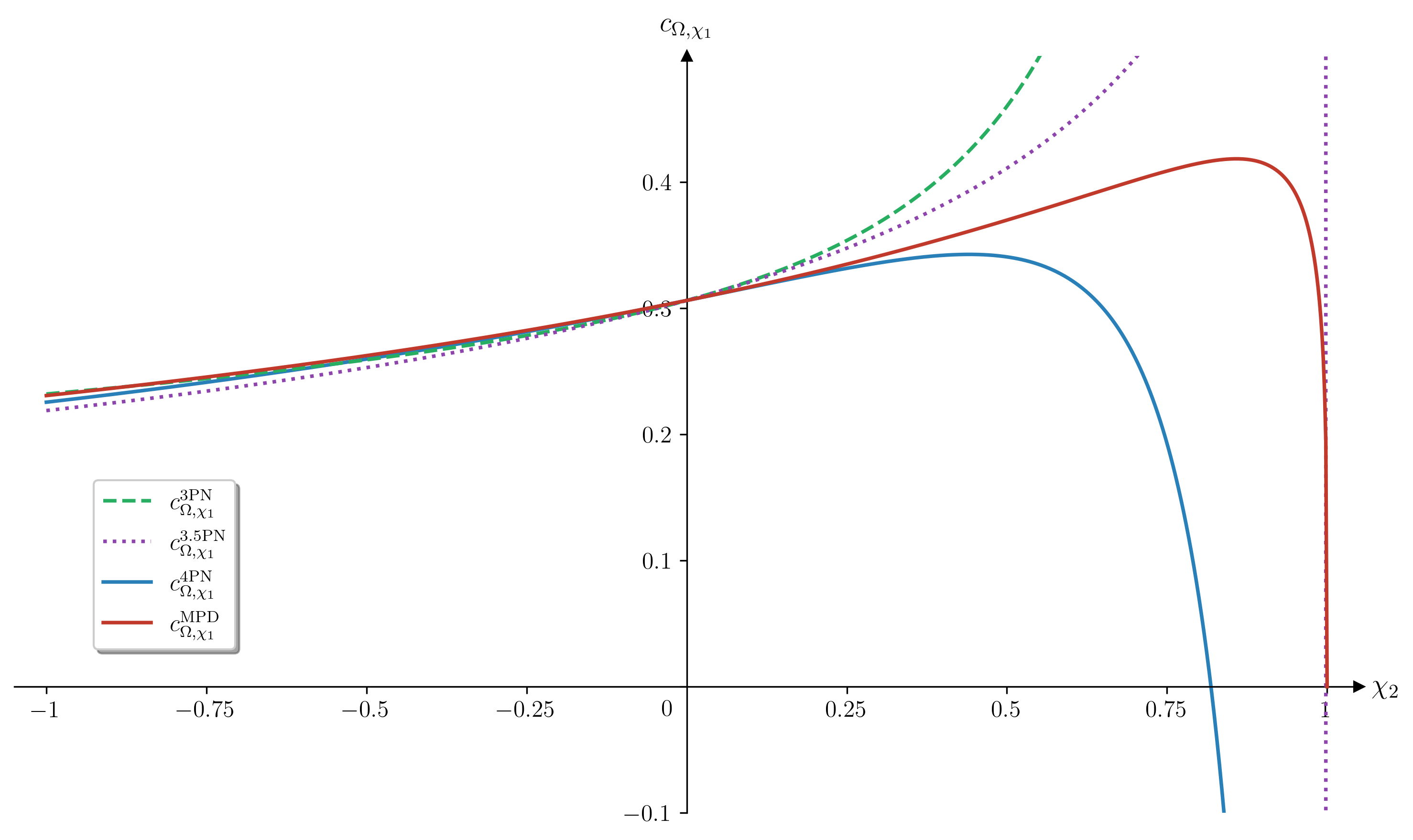}
\caption{\justifying  Comparison between the ISCO shift coefficients $c^\text{4PN}_{\Omega,\chi_1}$ (blue curve) and $c^\text{MPD}_{\Omega,\chi_1}$   (red curve). We also plot the 3PN and 3.5PN coefficients (dashed green and dotted purple curves, respectively) for comparison.}
\label{fig:c_{Omega,chi1}}
\end{figure}

The Fig.~\ref{fig:c_{Omega,chi1}} shows the coefficient $c^\text{MPD}_{\Omega,\chi_1}$ along  with the corresponding  $c^\text{4PN}_{\Omega,\chi_1}$ from our 4PN criterion. The agreement is excellent for negative $\chi$ and moderate positive $\chi$, with the relative difference remaining within $2$\% for $-1\leqslant \chi\lesssim 0.3$ (see Table~\ref{tab:rel_error}). For $\chi \gtrsim 0.5$, the 4PN prediction significantly underestimates the correct value. For comparison, we  also plot the estimates based on the 3PN and 3.5PN criteria, which are clearly less accurate than the 4PN result. 

In the simpler case of a spinning particle orbiting a Schwarzschild BH, for which we have $\rho_0=6$ and $\chi=0$, all previous equations, as well as those given in Appendix~\ref{app:MPD}, simplify  drastically, yielding
\begin{equation}
    \rho=6-\frac{2\sqrt{6}}{3}\sigma+{\cal O}(\sigma^2)\,,\qquad  \E=\frac{2\sqrt{2}}{3}-\frac{\sqrt{3}}{108}\sigma +{\cal O}(\sigma^2)\,,\qquad \ell=2\sqrt{3}+\frac{\sqrt{2}}{3}\sigma+{\cal O}(\sigma^2)\,.
\end{equation}
This entails
\begin{equation}
    \omega=6^{-3/2}+ \frac{1}{48}\sigma\, +{\cal O}(\sigma^2)\,, \qquad     c^\text{MPD}_{\Omega,\chi_1}(0)=\frac{\sqrt{6}}{8}\,,
\end{equation}
which agrees with the coefficient of the last term in \eqref{cOmega4PN0}. Note that, in the Schwarzschild limit, our approximate 4PN criterion provides the exact value for $c_{\Omega,\chi_1}$. Indeed, recall that in this limit, the $\nu=0$ part of the criterion is just $1-6y$, which coincides with the exact criterion for a spinless  particle around Schwarzschild while, in~\eqref{delta_C4PN} with $\chi_2=0$, the terms linear in $\chi_1$ come only from the 1.5PN and 3.5PN contributions. As a consequence, all criteria at and above 2.5PN order give the exact value $c_{\Omega,\chi_1}$ in the Schwarzschild case, which explains why all the curves in Fig.~\ref{fig:c_{Omega,chi1}} intersect on the vertical axis. 

\subsection{Corotating BH binary}

An interesting particular case is a circular black hole binary system with aligned spins, and where the two black holes are in corotation. This corotating case is also interesting mathematically, because the system admits a global helical Killing symmetry, and is amenable to numerical calculations~\cite{GGB1,GGB2,CPf04,CCGPf06}.
Each component, considered as a Kerr BH, is characterised by its mass $m_a$ and its spin $S_a$. The mass can be expressed in terms of the irreducible mass $\mu_a$ and the spin, according to the Christodoulou mass formula (with $c=G=1$),
\begin{equation}
\label{christodoulou}
    m_a^2=\mu_a^2+\frac{S_a^2}{4\mu^2_a}\,,
\end{equation}
whereas the spin can be related to the proper rotation frequency or angular velocity of the black hole $\omega_a$, deﬁned as the angular velocity of the outgoing photons located at the BH horizon, 
\begin{equation}
\label{spin_omega}
    S_a = 4m_a \mu_a^2\omega_a\,.
\end{equation}
By combining these two relations, one can easily derive the spin and the mass in terms of the irreducible mass $\mu_a$ and angular rotation $\omega_a$:\footnote{In the case of slow rotation we have (with $I_a=4\mu_a^3$ being the moment of inertia of the black hole)
\begin{align*}
S_a = I_a\,\omega_a + \mathcal{O}\left(\omega_a^3\right)\,,\qquad m_a=\mu_a+\frac{1}{2}I_a\,\omega_a^2+ \mathcal{O}\left(\omega_a^4\right)\,.
\end{align*}
}
\begin{equation}
\label{spin_mass}
S_a = \frac{4 \mu_a^3 \omega_a}{\sqrt{1-4 \mu_a^2 \omega_a^2}}\,,\qquad m_a=\frac{\mu_a}{\sqrt{1-4 \mu_a^2 \omega_a^2}}\,.
\end{equation}
In particular the dimensionless spin variable $\chi_a = S_a/m_a^2$ can be expressed in terms of the proper frequency $\omega_a$ or equivalently $y_a\equiv(m_a\omega_a)^{2/3}$ as 
\begin{equation}
\label{chiaxa}
\chi_a = \frac{4 y_a^{3/2}}{ 1 + 4 y_a^3 }\,.
\end{equation}

In the case of corotation, the spin, or alternatively the proper rotation frequency, is no longer a free parameter but is now fixed by the circular orbital frequency $\Omega$. We expect that in first approximation $\omega_a$ should be equal to $\Omega$, as was assumed in an initial computation of corotating black holes in~\cite{B02ico}. But actually the corotation condition involves post-Newtonian corrections and reads in the general case~\cite{BBL13}
\begin{equation}\label{corotcond}
    z_a \, \omega_a=\Omega-\Omega_a\,,
\end{equation}
where $z_a$ is the redshift variable associated to the particle moving on the exact circular orbit~\cite{Det08}, and where $\Omega_a$ is the precession frequency of the spin of the corotating BH. Since the spins appear linearly in the criterion at order 1.5PN, at 4PN one needs to determine them only up to order 2.5PN. Furthermore the leading-order corotating spin is already 1.5PN according to~\eqref{chiaxa} so we need the solution of Eq.~\eqref{corotcond} to 1PN relative order only. One finds (see Eq.~(6.12) in~\cite{BBL13})
\begin{equation}
\label{omega_A_x}
    \omega_a=\Omega\Bigl[1-\nu x+ {\cal O}(x^2)\Bigr] \quad\Leftrightarrow\quad y_a = \left(\frac{m_a}{m}\right)^{2/3} x \biggl[1-\frac{2}{3}\nu x+ {\cal O}(x^2)\biggr]\,.
\end{equation}
%
Substituting successively the relations~\eqref{chiaxa} and~\eqref{omega_A_x} into the general criterion~\eqref{critereTN}, one finally obtains the 4PN criterion in the corotating case:
\begin{align}
\label{criterion_corotation}
    C^\text{4PN}_{\rm corot} &= 1-6x +14\nu x^2 + \left[32+\left(\frac{253}{2}-\frac{123 \pi ^2}{16}\right) \nu -14 \nu ^2\right] x^3
    \nn\\
    &+\left[-16+
   \left(-\frac{238049}{180}+\frac{58265 \pi ^2}{1536}+\frac{5024 \gamma_{\rm E}
   }{15}+\frac{2512 \ln{x}}{15}+\frac{2916 \ln{3}}{5}+\frac{1184 \ln{2}}{15}\right)\nu 
   \right.
   \nn\\
   &\left. 
   \qquad +\left(\frac{451 \pi ^2}{16}-\frac{2927}{6}\right) \nu ^2+\frac{196 }{27}\nu ^3\right] x^4 + {\cal O}(x^5)\,.
\end{align}
It is interesting to compare the above criterion with the criterion associated with two black holes without spins and given by~\eqref{criterion_nospin}. The difference between the two criteria starts at 3PN and is given by
\begin{equation}\label{diff}
    C^\text{4PN}_{\rm corot}-C^\text{4PN}_{\rm non-spin}=8(4-9\nu)x^3-4 (4+31\nu-54\nu^2)x^4 +{\cal O}(x^5)\,.
\end{equation}
Note that when comparing the two criteria we are assuming that the two BH systems have the same $x$-parameter. In particular, with identical orbital frequency $\Omega$, the BHs must have the same mass in the two configurations. In the non-spin case, corresponding to the irrotational black hole binary, the mass is just the Schwarszchild mass of the non-rotating black hole. In the corotating case the mass is the total mass of the black hole and includes the contribution from the spin according to Eq.~\eqref{christodoulou}. Since $0\leqslant \nu\leqslant 1/4$, the coefficient of $x^3$ in~\eqref{diff} is always positive, while the coefficient of $x^4$, although negative, is sub-dominant. Thus we conclude that, for a given orbital frequency and identical BH masses, the corotating configuration is \textit{more stable} than the non-spin (irrotational) configuration.

Finally, let us note that  the test-mass limit ($\nu=0$) of~\eqref{criterion_corotation} can be compared to the corresponding exact criterion that follows from the Kerr criterion~\eqref{exactKerrcriterion}. Indeed, in the test-mass limit, the corotation condition simply reduces to $\omega_2=\Omega$ according to Eq.~\eqref{omega_A_x}. Using~\eqref{chiaxa} to replace $\chi$ by its expression in terms of $x=x_2$ in the Kerr criterion~\eqref{exactKerrcriterion}, we finally obtain the exact criterion 
\begin{equation}
    C_{\rm Kerr}^{\rm corot} = 1 + 32x^3 - \frac{6x}{(1+4x^3)^{1/3}}\left[ 1 + 4x^3 + \frac{8x^4}{(1+4x^3)^{1/3}}\right]\,.
\end{equation}
%
%
It is immediate to check that the above expression agrees, at 4PN order, with the limit $\nu=0$ of Eq.~\eqref{criterion_corotation}. Moreover, one finds that the exact ISCO is given by 
\begin{equation}\label{iscoKerrcorot}
   x^{\rm Kerr, corot}_{\rm ISCO}= \frac{1}{2}\left[
   \frac{\sqrt[3]{82 \left(3977+621
   \sqrt{41}\right)}+\sqrt[3]{82 \left(3977-621
   \sqrt{41}\right)}-82}{123}\right]^{1/3}\simeq 0.205\,.
\end{equation}
This corresponds to $r_{\rm BL}^{\rm ISCO}\simeq 4.76 M$, well outside the horizon of the BH in this case which is $r_{\rm BL}^{\rm H}\simeq 1.93 M$.\footnote{Besides~\eqref{iscoKerrcorot} another solution of the equation $C_{\rm Kerr}^{\rm corot}(x) = 0$ is simply the BH horizon which is of course not admissible as the ISCO.} Using the 4PN criterion, one finds the approximate value $x^{\rm 4PN, corot}_{\rm ISCO}\simeq 0.212$, in rather good agreement with~\eqref{iscoKerrcorot}.

\section{Discussion and conclusions}
\label{sec:discussion}

In this work, we have derived the ISCO stability criterion up to 4PN order for circular compact binaries with arbitrary masses and spins aligned with the orbital angular momentum. This result extends the 4PN criterion obtained in our previous work for the spinless case~\cite{BLL25}, which included the non-local tail term. It also improves upon Favata's criterion~\cite{F11b}, which included spin contributions only up to 2.5PN order. 

To derive our result, we performed a perturbation analysis based on the EFT   Hamiltonian formulation with all spin contributions up to 4PN order, which have been derived explicitly by Levi and Steinhoff~\cite{LS14,LS15a,LS15b,LS16a,LS16b,LS21}. At 4PN order, this requires spin-orbit and spin-spin terms up to NNLO (next-to-next-to-leading-order), as well as LO spin-cube and spin-four terms. In parallel, we also computed  the criterion from the 4PN EFT equations of motion derived from the EFT Hamiltonian, obtaining the same result. The spinless part of the criterion coincides with the criterion derived in~\cite{BLL25}, thus providing an additional check, as we previously used harmonic and ADM coordinates instead of the EFT coordinates. For the spin-dependent part of the criterion, we cross-checked our result up to 3.5PN order by using the contributions derived with harmonic coordinates. The 4PN spin contributions in harmonic coordinates are not directly available in the literature, which is why we resorted to the EFT Hamiltonian. Similarly, the spinless part of the 4PN EFT Hamiltonian is unavailable in the literature, so we provide it in Appendix~\ref{app:hamiltonianNS} in both a general frame and the CM frame. 

Our general criterion depends on generic spin-induced finite-size deformability coefficients, which may deviate from their BH values (fixed by convention to unity) for other compact objects, such as neutron stars. For simplicity, we have focused our discussion on the BH binary case, but the interested reader can find the full expression in Appendix~\ref{app:ISCOcriterion}. In the test-mass limit around a Kerr BH, we  verified that our criterion agrees, up to 4PN order, with the Kerr criterion, which is known exactly. 

We then computed the first order contribution in the symmetric mass ratio $\nu$ to  the ISCO shift with respect to the Kerr ISCO, allowing a direct comparison with conservative gravitational-self-force (GSF) numerical results~\cite{IBDLNSTW14}. For a Schwarzschild BH, we recover our previous spinless 4PN results~\cite{BLL25} which showed good agreement with numerical GSF calculations~\cite{BarackS09,LBB12}. For a Kerr BH, our 4PN estimate agrees with the numerical GSF calculations to within $10$\% for retrograde orbits ($-1\leqslant \chi \leqslant 0$), as well as for prograde orbits around a moderately spinning BH ($0\leqslant \chi \lesssim 0.2$). The agreement deteriorates for prograde orbits around a rapidly spinning BH. This is not surprising, as for extremal prograde orbital rotation ($\chi=+1$), the ISCO lies  on the horizon of the Kerr BH ($r_{\rm BL} = M$), which is well outside the domain of validity of the post-Newtonian approximation. By contrast, for extremal retrograde orbit ($\chi=-1$) the ISCO is at $r_{\rm BL} = 9M$, which explains why it is very well predicted by the PN expansion.

We also studied the influence of the small particle's spin on the ISCO using our 4PN criterion. The result  can be compared with an exact calculation based on the Mathisson-Papapetrou-Dixon (MPD) equations governing the motion of a spinning test particle orbiting a Kerr BH. In the small-spin limit, we derived an analytical expression for the ISCO shift. As before, the 4PN and MPD results show an excellent agreement for negative and moderately positive $\chi$, but the agreement worsens for higher values of $\chi$. It is highly significant that the post-Newtonian expansion (without any form of ``resummation'') is able to predict such a strong-field effect as the location of the ISCO.

Finally, we examined the stability criterion for corotating BH binaries, where the spin of each BH is fixed by the orbital frequency. We found that a corotating BH binary is more stable than its irrotational counterpart.

\acknowledgments

We thank David Trestini for an interesting discussion on the link between the ISCO frequency and the orbital precession for circular orbits.

\appendix

\section{Result for the complete ISCO criterion}
\label{app:ISCOcriterion}

We give here the complete result for the ISCO criterion for arbitrary-mass compact binaries with aligned spins. The result is presented in terms of arbitrary coefficients characterizing the internal structure of the bodies and called spin-induced deformability coefficients. Our notation is $\kappa_a$ for the $SS$ terms, $\lambda_a$ for the $SSS$ terms and $\iota_a$ for the $SSSS$ terms.\footnote{In Ref.~\cite{LS21} these parameters are denoted $C_{a(ES^2)}\equiv\kappa_a$, $C_{a(BS^3)}\equiv\lambda_a$ and $C_{a(ES^4)}\equiv\iota_a$.} Below we denote for convenience $\kappa_+ \equiv \kappa_1+\kappa_2$ and $\kappa_- \equiv \kappa_1-\kappa_2$ and \textit{idem} for $\lambda_\pm$ and $\iota_\pm$. For black holes we have $\kappa_a=\lambda_a=\iota_a=1$. For the spins our notation is defined in~\eqref{notspins}. We have
\begin{align}\label{criteregeneral}
	C^{\rm 4PN} &= 1 - 6 x 
	+ 2 x^{3/2} \big( 7 \chi_{S} + 3 \delta \chi_{\Sigma} \big) 
	+ x^2 \bigg\{ 14 \nu - 3 (2 + \kappa_{+}) \chi_S^2 
	+ 3 \Big[ \kappa_{-} -  (2 + \kappa_{+})\delta \Big] \chi_S \chi_\Sigma 
	\nn\\
	&
	+ \left[\frac{3}{2} \left( \delta \kappa_{-} - \kappa_+\right) + 3 (2 + \kappa_+) \nu \right] \chi_\Sigma^2 
	\bigg\}
	+ x^{5/2} \bigg\{ -2 \left(11 + 16 \nu\right) \chi_{S} -  3\left(6 + 5  \nu\right) \delta \chi_{\Sigma} \bigg\} \nonumber 
	\\
	&+ x^3 \bigg\{\Big( \frac{397}{2} - \frac{123 \pi^2}{16} \Big) \nu - 14 \nu^2  + \Big[ 86 - 6 (\delta \kappa_- + \kappa_+) + 5 (2 + \kappa_+) \nu \Big] \chi_S^2 \nn \\
	&  
	+ \Big[ 96 \delta + \left( 19 \kappa_- + 5 \delta (2 + \kappa_+) \right) \nu \Big] \chi_S \chi_\Sigma 
	+ \left[ 30 + \left( -102 + \frac{7}{2} (\delta \kappa_- - \kappa_+) \right) \nu - 5 (2 + \kappa_+) \nu^2 \right] \chi_\Sigma^2  
	\bigg\} \nn\\
	&+ x^{7/2} \bigg\{  \Big[ -\frac{568}{3}\nu + \frac{86}{3} \nu^2 \Big] \chi_S +  \Big[ -\frac{269}{4} \nu + 14 \nu^2 \Big] \delta \chi_\Sigma + 2 \left( -52 + \kappa_+ + 9 \lambda_+ \right) \chi_S^3 \nn \\
	& + \Big[  25 \kappa_- - \delta \kappa_+ - 27 \lambda_- + (27 \lambda_+-164) \delta \Big] \chi_S^2 \chi_\Sigma \nn \\
	&+ \Big[ -60 -29(\kappa_+- \delta \kappa_-)  + 27 (\lambda_+- \delta \lambda_- )+ 2  \big( 172 + 5 \kappa_+ - 27 \lambda_+ \big)\nu \Big] \chi_S \chi_\Sigma^2 \nn \\
	& + 3 \left[  - 5( \delta \kappa_+ - \kappa_-)  + 3 (\delta \lambda_+ - \lambda_-) + \nu \big( 20 \delta - 11 \kappa_- + \delta \kappa_+ + 9 \lambda_- - 3 \delta \lambda_+ \big) \right] \chi_\Sigma^3 \bigg\} \nn \\
	&+ x^4 \bigg\{ \nu \left( -\frac{215729}{180}+ \frac{58265 \pi^2}{1536} + \frac{2512}{15} \ln x+ \frac{5024}{15} \gamma_{\rm E}  + \frac{1184}{15} \ln 2 + \frac{2916}{5} \ln 3  \right)  
	+ \nu^2 \left( -\frac{4223}{6} + \frac{451 \pi^2}{16} \right)
	\nn
	\\
	& 
	+ \frac{196}{27} \nu^3 
	+ \bigg[ -\frac{3797}{42} - \frac{419}{28} \delta \kappa_- - \frac{83}{28} \kappa_+ 
	+ \nu \left( \frac{307}{42} + 12 \delta \kappa_- + \frac{1069}{28} \kappa_+ \right)
	- \frac{10}{3} (2 + \kappa_+) \nu^2 \bigg] \chi_S^2 
	\nn \\
	& 
	+ \bigg[ 12 (\delta \kappa_+ - \kappa_- - 15 \delta ) + \nu \left( -\frac{663}{14} \delta + \frac{943}{28} \kappa_- + \frac{733}{28} \delta \kappa_+ \right) 
	- \frac{2}{3} \nu^2 \Big( 67 \kappa_- + 5  (2 + \kappa_+) \delta\Big) \bigg] \chi_S \chi_\Sigma
	\nn \\
	& 
	+ \bigg[ -75 + 6 (\kappa_+-\delta \kappa_- ) + \nu \left( \frac{477}{2} + \frac{15}{8} \delta \kappa_- - \frac{111}{8} \kappa_+ \right) 
	+ \nu^2 \left( \frac{1307}{14} - \frac{31}{3} \delta \kappa_- - \frac{1331}{84} \kappa_+ \right)
	+ \frac{10}{3} (2 + \kappa_+) \nu^3 \bigg] \chi_\Sigma^2 
	\nn\\
	& 
	+ \frac{15}{4} \bigg[ 8(1+\kappa_+- \lambda_{+}) - 2 \iota_+ + 3 \kappa_-^2 - \kappa_+^2 \bigg] \chi_S^4 
	\nn \\
	&
	+ \frac{15}{2} \bigg[ 2 \big( \iota_- - \kappa_-\kappa_+ + 2\lambda_- - 2\kappa_- \big)+ \Big( 8(1+\kappa_+- \lambda_{+}) - 2 \iota_+ + 3 \kappa_-^2 - \kappa_+^2 \Big)\delta  \bigg] \chi_S^3 \chi_\Sigma 
	\nn \\
	&
	+ \frac{15}{4} \bigg[ 8+12(\kappa_+-\lambda_+)+ \kappa_+^2  + 5\kappa_-^2 - 6\iota_+  + 6  \big( \iota_- - \kappa_-\kappa_+ + 2\lambda_- - 2\kappa_- \big) \delta \nn \\
	& 
	\qquad
	- 6 \nu \big[ 8(1+\kappa_+ -\lambda_+) - 2\iota_+ + 3\kappa_-^2 - \kappa_+^2 \big] \bigg] \chi_S^2 \chi_\Sigma^2 \nn \\
	&+ \frac{15}{2} \bigg[   2\big( \iota_- - \kappa_-\kappa_+ + \lambda_- - \kappa_- \big)  + \big( \kappa_-^2 + \kappa_+^2 - 2\iota_+ + 2\kappa_+ - 2\lambda_+ \big)\delta\nn \\
	&
	\qquad
	- \nu \Big(6 \big( \iota_{-} - \kappa_{-}\kappa_{+} + 2\lambda_- - 2\kappa_- \big) +\big( 8(1+\kappa_+-\lambda_+) - 2\iota_+ + 3\kappa_-^2 - \kappa_+^2 \big)\delta \Big) \bigg] \chi_S \chi_\Sigma^3 \nn
	\\
	&
	+ \frac{15}{8} \bigg[  \kappa_+^2+\kappa_-^2   - 2\iota_+  + 2 \delta \big( \iota_- - \kappa_-\kappa_+ \big) 
	- 4 \nu \Big( \kappa_-^2 + \kappa_+^2 - 2\iota_+ + 2\kappa_+ - 2\lambda_+ +  \big( \iota_- - \kappa_-\kappa_+ + 2\lambda_- - 2\kappa_- \big)\delta \Big)
	\nn \\
	&
	\qquad
	+ 2 \nu^2 \Big( 8(1+\kappa_+-\lambda_+) - 2\iota_+ + 3\kappa_-^2 - \kappa_+^2 \Big) \bigg] \chi_\Sigma^4 
	\bigg\}\,.
\end{align}
This expression is also given in the ancillary file \cite{ancillaryfile}.
\section{Spinning particle orbiting a Kerr BH}
\label{app:MPD}

\begin{figure}[htbp]
	\centering
	\includegraphics[width=0.7\linewidth]{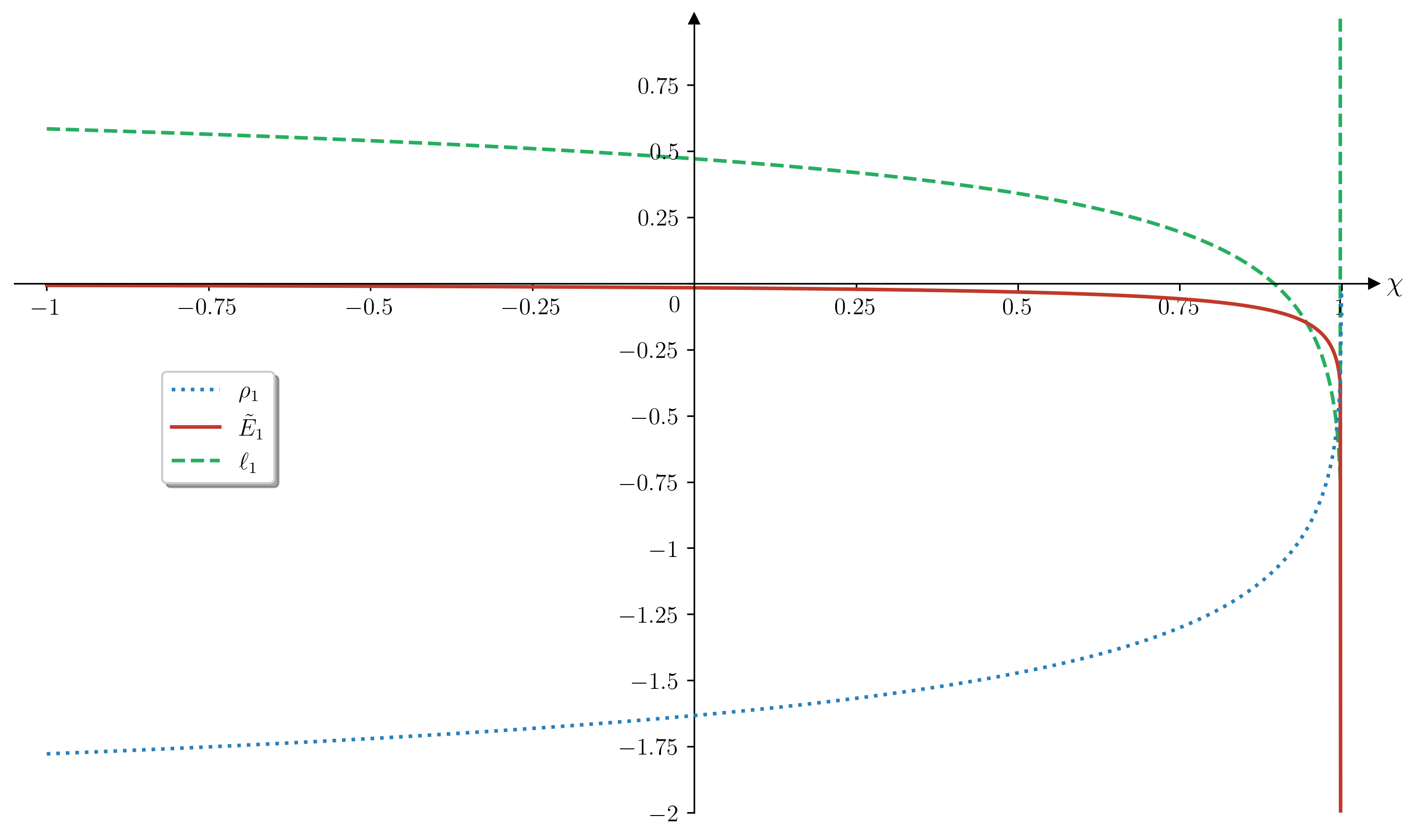}
	\caption{\justifying Linear-perturbation coefficients  $\rho_1$, $\tilde{E}_1$, $\ell_1$ introduced in~\eqref{rho1_etal}, characterising the deviation from the Kerr ISCO due to the spin $s=M\sigma$ of the particle, in terms of the BH spin $\chi$.}
	\label{fig:rho1_E1_L1}
\end{figure}

The effective potential defined in~\eqref{radial_eq} reads, following the notation~\eqref{rescale},
\begin{equation}
	\Ve \equiv V(\rho,\ell,\sigma)=\frac{A^V}{B^V}\,,
\end{equation}
with
\begin{subequations}
	\begin{align}
		A^V &\equiv \ell \Big[2\rho ^4 \chi - \rho ^2   \left(\rho ^3-3 \rho ^2-2 \chi ^2\right)\sigma+\rho  (\rho +1) \sigma ^2 \chi\Big]
		+(\rho ^3-\sigma ^2)(\rho ^2-2 \rho +\chi ^2)^{1/2}\Big[\rho ^3 \left(\ell^2 \rho +\rho ^3+\rho  \chi ^2+2 \chi ^2\right)
		\nn
		\\
		&
		\qquad\qquad\qquad\qquad
		+2 \rho    \chi  \left(3 \rho ^2+\chi ^2\right)\sigma- \left(\rho ^4-2 \rho ^3-2 \rho  \chi ^2-\chi ^2\right)\sigma ^2\Big]^{1/2}\,,
		\\
		B^V &\equiv \rho ^4 \left(\rho ^3+\rho  \chi ^2+2 \chi ^2\right)+2 \rho ^2  \chi  \left(3 \rho ^2+\chi ^2\right)\sigma +\rho   \left(-\rho ^4+2 \rho ^3+2 \rho  \chi ^2+\chi ^2\right)\sigma ^2\,.
	\end{align}
\end{subequations}
Below we denote the partial derivatives of $V$ evaluated on the zeroth order solution $(\rho_0,\ell_0,0)$ (corresponding to the spinless case of a circular and equatorial geodesic of Kerr) with a subscript, \textit{e.g.} 
\begin{equation}
	V_{\rho\sigma}\equiv \frac{\partial^2V}{\partial\rho\partial\sigma}(\rho_0,\ell_0,0)\,.
\end{equation}
After solving the circular orbit and ISCO equations at first order in $\sigma$, the linear coefficients  $\rho_1$, $\ell_1$ and $\tilde{E}_1$, introduced in~\eqref{rho1_etal}, are given explicitly by
\begin{align}\label{solution1}
	\rho_1 = \frac{V_{\rho\sigma}V_{\rho\rho\ell}-V_{\rho\ell}V_{\rho\rho\sigma}}{V_{\rho\ell}V_{\rho\rho\rho}} = \frac{A^\rho}{B^\rho}\,,\qquad 
	\ell_1 = - \frac{V_{\rho\sigma}}{V_{\rho\ell}} = \frac{A^\ell}{B^\ell}\,,\qquad 
	\tilde{E}_1 = \frac{V_{\sigma}V_{\rho\ell}-V_{\ell}V_{\rho\sigma}}{V_{\rho\ell}} = \frac{A^E}{B^E}\,,
\end{align}
%
%
where we take into account the facts that $V_{\rho}=V_{\rho\rho}=0$, see~\eqref{circular}--\eqref{isco}. We have the analytic solution
\begin{subequations}
	\begin{align}
		A^\rho\equiv & \left(\rho ^2-2 \rho +\chi ^2\right)\left[4 \rho ^{11/2}+9 \rho ^{9/2}-54 \rho
		^{7/2}+\left(-6 \rho ^5-37 \rho
		^4+198 \rho ^3\right) \chi+\left(60 \rho ^{7/2}-316 \rho
		^{5/2}+54 \rho ^{3/2}\right) \chi ^2
		\right.
		\nn
		\\
		&
		\left.
		\qquad\qquad\qquad\qquad
		+\left(-30 \rho ^3+246 \rho ^2-146 \rho \right) \chi ^3 +\left(123 \sqrt{\rho }-72 \rho ^{3/2}\right) \chi ^4-33 \chi ^5 \right]\,,
		\\
		B^\rho\equiv
		&\  
		\left(6 \rho ^3-71 \rho ^2+216 \rho -180\right)\rho ^5 
		+2 \left(42 \rho ^3-215 \rho ^2+252 \rho -36\right) \rho
		^{7/2} \chi 
		+\left(-28 \rho ^3+273 \rho ^2-472 \rho
		+180\right) \rho ^3 \chi ^2
		\nn
		\\
		& 
		-4 \left(8 \rho
		^2-27 \rho +38\right) \rho ^{5/2} \chi ^3
		+\left(-18 \rho ^2+79 \rho +24\right) \rho^2  \chi ^4
		+2 (13-18 \rho ) \rho^{3/2} \chi ^5-9 \rho \chi ^6\,,
		\\
		A^\ell\equiv & \  6 \rho ^{21/2}-90 \rho ^{19/2}+450 \rho ^{17/2}-912 \rho ^{15/2}+648 \rho
		^{13/2}
		+3 \left(32 \rho ^4-289 \rho ^3+825 \rho ^2-796 \rho +108\right) \rho ^5 \chi
		\nn\\
		&
		-2 \left(14 \rho ^4-281 \rho ^3+1220 \rho ^2-1650 \rho +552\right) \rho ^{9/2}
		\chi ^2
		+\left(-102 \rho ^4+903 \rho ^3-1883 \rho ^2+1476 \rho -60\right) \rho
		^3 \chi ^3
		\nn\\
		&
		-2 \left(9 \rho ^4-27 \rho ^3+19 \rho ^2+376 \rho -96\right) \rho ^{5/2}
		\chi ^4
		-\left(84 \rho ^3-545 \rho ^2+255 \rho +160\right)
		\rho ^2 \chi ^5
		\nn\\
		&
		-2 \left(105 \rho ^2-254
		\rho +34\right) \rho ^{3/2} \chi ^6
		+\left(18 \rho ^2-231 \rho +175\right) \rho  \chi ^7
		+12 (3 \rho -8) \sqrt{\rho } \chi ^8+18 \chi ^9\,,
		\\
		B^\ell\equiv &\  \rho ^{7/4}\left(\rho ^{3/2}-3 \sqrt{\rho }+2 \chi \right)^{1/2} B^\rho\,,
		\\
		A^E\equiv & -\left(\rho ^3-21 \rho ^2+84 \rho -108\right)\rho ^5 
		-6 \left(7 \rho ^2-38
		\rho +60\right) \rho ^{9/2} \chi +\left(37 \rho ^3-247 \rho ^2+468 \rho -60\right)
		\rho ^3 \chi ^2
		\nn
		\\
		&
		-2 \left(7 \rho ^3-58 \rho ^2+108 \rho -96\right) \rho ^{5/2} \chi ^3 -\left(\rho ^2+157 \rho +160\right) \rho ^2
		\chi ^4
		-2 \left(6 \rho ^2-127 \rho +34\right) \rho ^{3/2} \chi ^5
		\nn
		\\
		&
		+(175-117 \rho ) \rho  \chi ^6
		+6
		(3 \rho -16) \sqrt{\rho } \chi ^7
		+18 \chi ^8\,,
		\\
		B^E\equiv & \ B^\ell\,,
	\end{align}
\end{subequations}
where all the coefficients are evaluated with $\rho=\rho_{\rm ISCO}^{\rm Kerr}(\chi)$ given in~\eqref{exactKerrISCO}. The dependence of $\rho_1$, $\ell_1$ and $\tilde{E}_1$ over the Kerr BH spin is shown in Fig.~\ref{fig:rho1_E1_L1}. 

\section{Non-spin terms in the EFT Hamiltonian up to 4PN order}
\label{app:hamiltonianNS}

The non-spin (NS) part of the Hamiltonian in EFT coordinates is given explicitly only up to 2PN in~\cite{LS16a}. In order to derive its expression up to 4PN, which is of the form (we drop the NS subscript in the rest of this appendix)
\begin{align}
    H_\text{NS}=H_\text{NS}^{\rm N}+\frac{1}{c^2}\,H_\text{NS}^{\rm 1PN}+\frac{1}{c^4}\,H_\text{NS}^{\rm 2PN}+\frac{1}{c^6}\,H_\text{NS}^{\rm 3PN}+\frac{1}{c^8}\,\left(H_\text{NS}^{\rm 4PN}+H_{\rm tail}^{\rm 4PN}\right)\,,
\end{align}
we start from the Hamiltonian in ADM coordinates up to 4PN  obtained in~\cite{DJS14,JaraS15} (equivalent to the 4PN Fokker Lagrangian in harmonic coordinates~\cite{BBBFMc,BBFM17,FS19}) and reviewed in~\cite{BlanchetLR_2024}, and apply to it an infinitesimal canonical transformation. This transformation is discussed in~\cite{LS14} as a mean to cross-check the results obtained in the EFT formalism with the earlier results in the ADM formalism. Up to 4PN, the difference $\Delta H^{\rm 4PN}$ between these Hamiltonians can be determined by the infinitesimal generator $g$ of the canonical transformation between the EFT and ADM Hamiltonians at 2PN order only, say $g = c^{-4}\g +\mathcal{O}(c^{-6})$:
\begin{equation}
\label{eq: formula Delta H 4PN full}
    \Delta H^{\rm 4PN} \equiv H^{\rm 4PN}_{\rm EFT}-H^{\rm 4PN}_{\rm ADM} = \bigl\{H^{\rm 2PN}_{\rm ADM}, \g\bigr\}+\frac12\bigl\{\bigl\{H^{\rm N}_{\rm ADM},\g\bigr\},\g\bigr\} \,.
\end{equation}
%
%
%
In a general frame, $\g$ is given explicitly by~\cite{LS14}
\begin{equation}
\g = \frac{G}{2} \left[ \frac{(n_{12} p_2) p_1^2}{m_1} - \frac{(n_{12} p_1) p_2^2}{m_2} \right] + \frac{G^2}{4 r_{12}} \bigl[m_1^2 (n_{12} p_2) -m_2^2 (n_{12} p_1) \bigr]\,,
\end{equation}
where we have introduced the notations $r_{12}\equiv \vert \bm{x}_{1}-\bm{x}_2 \rvert $,  $\bm{n}_{12} \equiv (\bm{x}_{1}-\bm{x}_2)/r_{12}$,  $(n_{12} p_1) \equiv \bm{n}_{12} \cdot \bm{p}_1$, $(n_{12} p_2) \equiv \bm{n}_{12} \cdot \bm{p}_2$ and $(p_{1} p_2) \equiv \bm{p}_{1} \cdot \bm{p}_2$. Decomposing \eqref{eq: formula Delta H 4PN full} order by order, taking into account the fact that the Newtonian and 1PN contributions to the Hamiltonian remain unchanged under the coordinate transformation, we have
\begin{subequations}
\begin{align}
    \Delta H^{\rm 2PN} &= \bigl\{H^{\rm N},\g\bigr\}\,,
    \\
    \Delta H^{\rm 3PN} &= \bigl\{H^{\rm 1PN},\g\bigr\}\,,
    \\
    \Delta H^{\rm 4PN} &= \bigl\{H^{\rm 2PN}_{\rm ADM},\g\bigr\} + \frac12\bigl\{\Delta H^{\rm 2PN},\g\bigr\}\,.
\end{align}
\end{subequations}
Adding the $\Delta H$ contributions to the ADM Hamiltonian, one finally gets the full EFT Hamiltonian at 4PN order in a general frame: 
\begin{subequations}
\begin{align}
H^{\rm N}_{\rm EFT} &= \frac{p_1^2}{2m_1}-\frac{G m_1m_2}{2r_{12}}  + 1 \leftrightarrow 2\,,\\
H^{\rm 1PN}_{\rm EFT} &= -\frac{p_1^4}{8 m_1^3} 
+ \frac{G m_1 m_2}{r_{12}} \left[ -\frac{3 p_1^2}{2 m_1^2}  + \frac{(n_{12} p_1)(n_{12} p_2)}{4 m_1 m_2} + \frac{7 (p_1 p_2)}{4 m_1 m_2} \right] + \frac{G^2 m_1^2 m_2}{2 r_{12}^2}  + 1 \leftrightarrow 2\,,\\
H^{\rm 2PN}_{\rm EFT}&= \frac{p_1^6}{16 m_1^5} 
+ \frac{G m_1 m_2}{r_{12}} \left[ -\frac{3 (n_{12} p_1)^2 (n_{12} p_2)^2}{16m_1^2 m_2^2} - \frac{3 p_1^2 p_2^2}{16 m_1^2 m_2^2} + \frac{(n_{12} p_1)(n_{12} p_2) p_1^2}{2 m_1^3 m_2}  + \frac{(n_{12} p_2)^2 p_1^2}{8 m_1^2 m_2^2}  \right.
\nn\\
&\qquad\qquad\qquad\qquad\quad 
+ \left. \frac{5  p_1^4}{8 m_1^4} - \frac{3 (n_{12} p_1)(n_{12} p_2) (p_1 p_2)}{4 m_1^2 m_2^2} - \frac{p_1^2 (p_1 p_2)}{2 m_1^3m_2}  - \frac{(p_1 p_2)^2}{8 m_1^2 m_2^2} \right]
\nn\\
&\qquad\qquad\quad
+ \frac{G^2 m_1 m_2}{r_{12}^2} \left[ -\frac{m_2 (n_{12} p_1)^2}{2 m_1^2}  + \frac{17  p_1^2}{4 m_1} + \frac{11 m_2 p_1^2}{4 m_1^2} - \frac{7  (p_1 p_2)}{m_1}   \right]
\nn\\
&\qquad\qquad\quad
- \frac{G^3 m_1 m_2}{r_{12}^3} \left( \frac{m_1^2}{2} + \frac{5 m_1 m_2}{8}  \right)
+ 1 \leftrightarrow 2\,, 
\\
H^{\rm 3PN}_{\rm EFT} &= -\frac{5}{128}\frac{p_1^{8}}{m_1^7}
+ \frac{G m_1 m_2}{r_{12}}\left[
\frac{5}{32} \frac{(n_{12} p_1)^3 (n_{12} p_2)^3}{m_1^3 m_2^3}
+ \frac{3}{16} \frac{(n_{12} p_1)^2 (n_{12} p_2)^2 p_1^2}{m_1^4 m_2^2}
\right.
\nn\\
& \left. 
 \qquad
 - \frac{9}{16} \frac{(n_{12} p_1)(n_{12} p_2)^3 p_1^2}{m_1^3 m_2^3} 
- \frac{1}{4} \frac{(n_{12} p_1)(n_{12} p_2) p_1^4}{m_1^5 m_2} 
 + \frac{1}{4} \frac{p_1^4 p_2^2}{m_1^4 m_2^2}
 + \frac{1}{4} \frac{p_1^4 (p_1 p_2) }{m_1^5 m_2} 
 \right.
\nn\\
& \left. 
 \qquad
- \frac{5}{16} \frac{(n_{12} p_2)^2 p_1^4}{m_1^4 m_2^2} 
- \frac{7}{16} \frac{p_1^6}{m_1^6} 
+ \frac{15}{32} \frac{(n_{12} p_1)^2 (n_{12} p_2)^2 (p_1 p_2)}{m_1^3 m_2^3}
+ \frac{3}{4} \frac{(n_{12} p_1)(n_{12} p_2) (p_1p_2) p_1^2}{m_1^4 m_2^2} 
\right.
\nn\\
& \left. 
 \qquad
+ \frac{1}{16} \frac{(n_{12} p_1)^2 (p_1p_2)p_2^2}{m_1^3 m_2^3} 
+ \frac{1}{32} \frac{p_1^2 (p_1 p_2) p_2^2}{m_1^3 m_2^3}
- \frac{5}{16} \frac{(n_{12} p_1)(n_{12} p_2) (p_1 p_2)^2}{m_1^3 m_2^3}
+ \frac{1}{8} \frac{p_1^2 (p_1 p_2)^2}{m_1^4 m_2^2} 
\right.
\nn\\
& \left. 
 \qquad
 - \frac{1}{16} \frac{(p_1 p_2)^3}{m_1^3 m_2^3} 
- \frac{1}{16} \frac{(n_{12} p_1)^2 p_1^2 p_2^2}{m_1^4 m_2^2} 
+ \frac{7}{32} \frac{(n_{12} p_1)(n_{12} p_2) p_1^2 p_2^2}{m_1^3 m_2^3}
\right]
\nn\\
\qquad
& +\frac{G^2 m_1 m_2}{r_{12}^2}\left[
\frac{5}{12} \frac{(n_{12} p_1)^4}{m_1^3} 
- \frac{3}{2} \frac{(n_{12} p_1)^3 (n_{12} p_2)}{m_1^2 m_2} 
+ \frac{11}{6} \frac{(n_{12} p_1)^2 (n_{12} p_2)^2}{m_1 m_2^2} 
+ \frac{17}{16} \frac{(n_{12} p_1)^2 p_1^2}{m_1^3} 
\right.
\nn\\
& \left. 
 \qquad
+ \frac{1}{4} \frac{m_2 (n_{12} p_1)^2 p_1^2}{m_1^4}
- \frac{17}{8} \frac{(n_{12} p_1)(n_{12} p_2) p_1^2}{m_1^2 m_2} - \frac{37}{12} \frac{(n_{12} p_2)^2 p_1^2}{m_1 m_2^2} 
- \frac{85}{48} \frac{(n_{12} p_2)^2 p_1^2}{m_1^2 m_2} 
- \frac{11}{16} \frac{p_1^4}{m_1^3} 
\right.
\nn\\
& \left. 
 \qquad
 - \frac{29}{16} \frac{m_2 p_1^4}{m_1^4} 
- \frac{11}{8} \frac{(n_{12} p_1)^2 (p_1 p_2)}{m_1^2 m_2} 
+ \frac{89}{12} \frac{(n_{12} p_1)(n_{12} p_2) (p_1 p_2)}{m_1 m_2^2} 
+ \frac{3}{2} \frac{p_1^2 (p_1 p_2)}{m_1^3} 
\right.
\nn\\
& \left.
 \qquad
- \frac{57}{16} \frac{p_1^2 (p_1 p_2)}{m_1^2 m_2} 
 + \frac{25}{48} \frac{(p_1 p_2)^2}{m_1^2 m_2}
 + \frac{179}{48} \frac{p_1^2 p_2^2}{m_1^2 m_2}
\right]
\nn\\
&+\frac{G^3 m_1 m_2}{r_{12}^3}\left[
 \frac{5}{4} (n_{12} p_1)^2 
- \frac{13}{16} \frac{m_2 (n_{12} p_1)^2}{m_1} 
+ \frac{3}{2} \frac{m_2^2 (n_{12} p_1)^2}{m_1^2} 
+ \frac{79}{16} (n_{12} p_1)(n_{12} p_2)
\right.
\nn\\
& \left. 
 \qquad
 - \frac{1}{8} \frac{m_1 (n_{12} p_1)(n_{12} p_2)}{m_2} 
+ \frac{155}{16} (p_1 p_2)
+ \frac{91}{8} \frac{m_1 (p_1 p_2)}{m_2}
- \frac{419}{48} p_1^2 
- \frac{491}{48} \frac{m_2 p_1^2}{m_1} 
\right.
\nn\\
& \left. 
 \qquad
 - \frac{17}{4} \frac{m_2^2 p_1^2}{m_1^2} 
- \frac{3\pi^2}{64} \frac{m_2 (n_{12} p_1)^2}{m_1} 
+ \frac{3\pi^2}{64} (n_{12} p_1)(n_{12} p_2) 
+ \frac{\pi^2}{64} \frac{m_2 p_1^2 }{m_1} 
- \frac{\pi^2}{64} (p_1 p_2) 
\right]
\nn\\
&
+ \frac{G^4 m_1 m_2}{r_{12}^4}\left[
\frac{3}{8} m_1^3 + \frac{233}{24} m_1^2 m_2  - \frac{21\pi^2}{32} m_1^2 m_2  
\right]
+ 1 \leftrightarrow 2\,.
\end{align}   
\end{subequations}
We split the (local) 4PN contribution into successive powers of $G$: 
\begin{subequations}
\begin{align}
H^{\rm 4PN(0)}_{\rm EFT}&=   \frac{7}{256}\frac{p_1^{10}}{m_1^9}+ 1 \leftrightarrow 2\,,
\\
H^{\rm 4PN(1)}_{\rm EFT}&= 
\frac{G m_1 m_2}{r_{12}}\left[
- \frac{35}{256} \frac{(n_{12} p_1)^5 (n_{12} p_2)^3}{m_1^5 m_2^3} 
+ \frac{25}{128} \frac{(n_{12} p_1)^3 (n_{12} p_2)^3 p_1^2}{m_1^5 m_2^3} 
+ \frac{25}{64} \frac{(n_{12} p_1)^3 (n_{12} p_2) (p_1p_2)^2}{m_1^5 m_2^3} 
\right.
 \nn\\
& \left. 
 \qquad
- \frac{5}{32} \frac{(n_{12} p_1)^2 (n_{12} p_2)^4 p_1^2}{m_1^4 m_2^4} 
+ \frac{55}{256} \frac{(n_{12} p_1)(n_{12} p_2)^5 p_1^2}{m_1^3 m_2^5} 
- \frac{9}{64} \frac{(n_{12} p_1)^2 (n_{12} p_2)^2 p_1^4}{m_1^6 m_2^2} 
+ \frac{45}{128} \frac{p_1^8}{m_1^8} 
- \frac{9}{64} \frac{p_1^6 p_2^2}{m_1^6 m_2^2}
\right.
 \nn\\
& \left. 
 \qquad
- \frac{7}{128} \frac{p_1^4 p_2^4}{m_1^4 m_2^4} + \frac{33}{256} \frac{(n_{12} p_1)(n_{12} p_2)^3 p_1^4}{m_1^5 m_2^3} 
+ \frac{7}{32} \frac{(n_{12} p_2)^4 p_1^4}{m_1^4 m_2^4} 
+ \frac{3}{16} \frac{(n_{12} p_1)(n_{12} p_2) p_1^6}{m_1^7 m_2} 
 - \frac{3}{16} \frac{(n_{12} p_2)^2 p_1^4 p_2^2}{m_1^4 m_2^4} 
 \right.
 \nn\\
& \left. 
 \qquad- \frac{3}{16} \frac{p_1^6 (p_1 p_2)}{m_1^7 m_2} 
+ \frac{15}{64} \frac{(n_{12} p_2)^2 p_1^6}{m_1^6 m_2^2} 
 - \frac{45}{128} \frac{(n_{12} p_1)^2 (n_{12} p_2)^2 p_1^2 (p_1 p_2)}{m_1^5 m_2^3} 
- \frac{85}{256} \frac{(n_{12} p_1)^4 (n_{12} p_2)^2 (p_1 p_2)}{m_1^5 m_2^3} 
\right.
 \nn\\
& \left. 
 \qquad
- \frac{25}{256} \frac{(n_{12} p_1)(n_{12} p_2) p_1^4 p_2^2}{m_1^5 m_2^3} 
- \frac{1}{2} \frac{(n_{12} p_1)(n_{12} p_2)^3 p_1^2 (p_1 p_2)}{m_1^4 m_2^4} 
- \frac{23}{256} \frac{(n_{12} p_2)^4 p_1^2 (p_1 p_2)}{m_1^3 m_2^5} 
\right.
 \nn\\
& \left. 
 \qquad
- \frac{9}{16} \frac{(n_{12} p_1)(n_{12} p_2) p_1^4 (p_1 p_2)}{m_1^6 m_2^2} 
+ \frac{1}{16} \frac{(n_{12} p_1)(n_{12} p_2) p_1^2 p_2^2(p_1p_2)}{m_1^4 m_2^4} 
- \frac{1}{256} \frac{(n_{12} p_2)^2 p_1^4 (p_1 p_2)}{m_1^5 m_2^3}
\right.
 \nn\\
& \left. 
 \qquad
- \frac{3}{32} \frac{p_1^4 (p_1p_2)^2}{m_1^6 m_2^2}
+ \frac{7}{64} \frac{(n_{12} p_1)(n_{12} p_2) p_1^2 (p_1 p_2)^2}{m_1^5 m_2^3} 
+ \frac{1}{8} \frac{(n_{12} p_2)^2 p_1^2 (p_1 p_2)^2}{m_1^4 m_2^4} 
- \frac{3}{64} \frac{(n_{12} p_1)^2 (p_1 p_2)^3}{m_1^5 m_2^3} 
\right.
 \nn\\
& \left. 
 \qquad
+ \frac{3}{64} \frac{p_1^2 (p_1 p_2)^3}{m_1^5 m_2^3} 
 + \frac{3}{64} \frac{(n_{12} p_1)^2 p_1^4 p_2^2}{m_1^6 m_2^2} 
- \frac{7}{128} \frac{(n_{12} p_1)^3 (n_{12} p_2) p_1^2 p_2^2}{m_1^5 m_2^3}
+ \frac{7}{128} \frac{(n_{12} p_1)^2 p_1^2 (p_1 p_2) p_2^2}{m_1^5 m_2^3}
\right.
 \nn\\
& \left. 
 \qquad
+ \frac{21}{64} \frac{(n_{12} p_1)^2 (n_{12} p_2)^2 p_1^2 p_2^2}{m_1^4 m_2^4}
- \frac{7}{256} \frac{p_1^4 (p_1 p_2) p_2^2}{m_1^5 m_2^3} 
- \frac{1}{32} \frac{p_1^2 (p_1 p_2)^2 p_2^2}{m_1^4 m_2^4} 
    \right]
    + 1 \leftrightarrow 2\,,\\
H^{\rm 4PN(2)}_{\rm EFT}&=  \frac{G^2 m_1 m_2}{r_{12}^2}\left[
    \frac{369}{160} \frac{(n_{12} p_1)^6}{m_1^5} 
- \frac{549}{128} \frac{(n_{12} p_1)^5 (n_{12} p_2)}{m_1^4 m_2} 
+ \frac{3263}{1280} \frac{(n_{12} p_1)^4 (n_{12} p_2)^2}{m_1^3 m_2^2} 
- \frac{7}{4} \frac{(n_{12} p_1)^4 (n_{12} p_2)^2}{m_1^4 m_2} 
\right.
 \nn\\
& \left. 
 \qquad
+ \frac{1}{4} \frac{(n_{12} p_1)^3 (n_{12} p_2)^3}{m_1^2 m_2^3} 
- \frac{889}{192} \frac{(n_{12} p_1)^4 p_1^2}{m_1^5}   + \frac{67}{16} \frac{(n_{12} p_1)^3 (n_{12} p_2) p_1^2}{m_1^4 m_2} + \frac{797}{480} \frac{(n_{12} p_1)^2 (n_{12} p_2)^2 p_1^2}{m_1^3 m_2^2} 
\right.
 \nn\\
& \left. 
 \qquad
 + \frac{2}{3} \frac{(n_{12} p_1)^2 (n_{12} p_2)^2 p_1^2}{m_1^4 m_2}
- \frac{73}{192} \frac{(n_{12} p_1)(n_{12} p_2)^3 p_1^2}{m_1^2 m_2^3} + \frac{17}{96} \frac{(n_{12} p_1)(n_{12} p_2)^3 p_1^2}{m_1^3 m_2^2} + \frac{7}{4} \frac{(n_{12} p_2)^4 p_1^2}{m_1 m_2^4} 
\right.
 \nn\\
& \left. 
 \qquad
 + \frac{1673}{1920} \frac{(n_{12} p_2)^4 p_1^2}{m_1^2 m_2^3}
+ \frac{49}{16} \frac{(n_{12} p_1)^2 p_1^4}{m_1^5} - \frac{3}{16} \frac{m_2 (n_{12} p_1)^2 p_1^4}{m_1^6} + \frac{5}{8} \frac{(n_{12} p_1)(n_{12} p_2) p_1^4}{m_1^5} 
\right.
 \nn\\
& \left. 
 \qquad
 - \frac{191}{128} \frac{(n_{12} p_1)(n_{12} p_2) p_1^4}{m_1^4 m_2}
- \frac{487}{3840} \frac{(n_{12} p_2)^2 p_1^4}{m_1^3 m_2^2}  
+ \frac{209}{96} \frac{(n_{12} p_2)^2 p_1^4}{m_1^4 m_2} 
- \frac{43}{64} \frac{p_1^6}{m_1^5}  + \frac{55}{32} \frac{m_2 p_1^6}{m_1^6}
\right.
 \nn\\
& \left. 
 \qquad
 + \frac{1547}{256} \frac{(n_{12} p_1)^4 (p_1 p_2)}{m_1^4 m_2} 
- \frac{3571}{320} \frac{(n_{12} p_1)^3 (n_{12} p_2) (p_1 p_2)}{m_1^3 m_2^2} 
- \frac{7}{2} \frac{(n_{12} p_1)^3 (n_{12} p_2) (p_1 p_2)}{m_1^4 m_2} 
 \right.
 \nn\\
& \left. 
 \qquad
 + \frac{1421}{384} \frac{(n_{12} p_1)^2 (n_{12} p_2)^2 (p_1 p_2)}{m_1^2 m_2^3} 
 - \frac{851}{128} \frac{(n_{12} p_1)^2 p_1^2 (p_1 p_2)}{m_1^4 m_2}
 + \frac{2713}{480} \frac{(n_{12} p_1)(n_{12} p_2) p_1^2 (p_1 p_2)}{m_1^3 m_2^2} 
\right.
 \nn\\
& \left. 
 \qquad
 - \frac{133}{24} \frac{(n_{12} p_1)(n_{12} p_2) p_1^2 (p_1 p_2)}{m_1^4 m_2}
+ \frac{91}{384} \frac{(n_{12} p_2)^2 p_1^2 (p_1 p_2)}{m_1^2 m_2^3} 
- \frac{527}{384} \frac{(n_{12} p_2)^2 p_1^2 (p_1 p_2)}{m_1^3 m_2^2}
+\frac{77}{96} \frac{ (p_1 p_2)^3}{m_1^2 m_2^3}
\right.
 \nn\\
& \left. 
 \qquad
 - \frac{5}{4} \frac{p_1^4 (p_1 p_2)}{m_1^5} + \frac{1011}{256} \frac{p_1^4 (p_1 p_2)}{m_1^4 m_2}
 + \frac{4349}{1280} \frac{(n_{12} p_1)^2 (p_1 p_2)^2}{m_1^3 m_2^2} 
+ \frac{21}{16} \frac{(n_{12} p_1)^2 (p_1 p_2)^2}{m_1^4 m_2} 
+\frac{55}{384} \frac{ p_1^2(p_1 p_2)p_2^2}{m_1^2 m_2^3}
\right.
 \nn\\
& \left. 
 \qquad
- \frac{229}{192} \frac{(n_{12} p_1)(n_{12} p_2) (p_1 p_2)^2}{m_1^2 m_2^3}
- \frac{4901}{3840} \frac{p_1^2 (p_1 p_2)^2}{m_1^3 m_2^2} - \frac{53}{96} \frac{p_1^2 (p_1 p_2)^2}{m_1^4 m_2}
+ \frac{641}{3840} \frac{(n_{12} p_1)^2 p_1^2 p_2^2}{m_1^3 m_2^2} 
\right.
 \nn\\
& \left. 
 \qquad
 - \frac{11}{48} \frac{(n_{12} p_1)^2 p_1^2 p_2^2}{m_1^4 m_2} 
+ \frac{17}{32} \frac{(n_{12} p_1)(n_{12} p_2) p_1^2 p_2^2}{m_1^2 m_2^3} 
- \frac{23}{12} \frac{p_1^4 p_2^2}{m_1^4 m_2} - \frac{2479}{3840} \frac{p_1^4 p_2^2}{m_1^3 m_2^2} 
    \right]+ 1 \leftrightarrow 2\,,
\\
H^{\rm 4PN(3)}_{\rm EFT}&=  \frac{G^3 m_1 m_2}{r_{12}^3}\left[\frac{5027}{384} \frac{(n_{12} p_1)^4}{m_1^2} 
-\frac{23533}{1280} \frac{m_2 (n_{12} p_1)^4}{m_1^3} 
+\frac{9581}{240} \frac{(n_{12} p_1)^3 (n_{12} p_2)}{m_1^2} 
-\frac{3191}{640} \frac{(n_{12} p_1)^3 (n_{12} p_2)}{m_1 m_2} 
\right.
 \nn\\
& \left. 
 \qquad
+\frac{3}{4} \frac{m_2 (n_{12} p_1)^3 (n_{12} p_2)}{m_1^3} 
-\frac{13151}{960} \frac{(n_{12} p_1)^2 (n_{12} p_2)^2}{m_1^2} 
-\frac{27263}{960} \frac{(n_{12} p_1)^2 (n_{12} p_2)^2}{m_1 m_2}
-\frac{22993}{960} \frac{(n_{12} p_1)^2 p_1^2}{m_1^2} 
\right.
 \nn\\
& \left. 
 \qquad
+\frac{62847}{1600} \frac{m_2 (n_{12} p_1)^2 p_1^2}{m_1^3} 
-\frac{5}{4} \frac{m_2^2 (n_{12} p_1)^2 p_1^2}{m_1^4} 
-\frac{430911}{6400} \frac{(n_{12} p_1)(n_{12} p_2) p_1^2}{m_1^2} 
+\frac{16681}{1920} \frac{(n_{12} p_1)(n_{12} p_2) p_1^2}{m_1 m_2} 
\right.
 \nn\\
& \left. 
 \qquad
-\frac{23}{8} \frac{m_2 (n_{12} p_1)(n_{12} p_2) p_1^2}{m_1^3} 
+\frac{20377}{1600} \frac{(n_{12} p_2)^2 p_1^2}{m_1^2} 
+\frac{84383}{4800} \frac{(n_{12} p_2)^2 p_1^2}{m_2^2} 
+\frac{1243429}{19200} \frac{(n_{12} p_2)^2 p_1^2}{m_1 m_2} 
-\frac{1331}{1152} \frac{p_1^4}{m_1^2} 
\right.
 \nn\\
& \left. 
 \qquad
-\frac{159589}{19200} \frac{m_2 p_1^4}{m_1^3} 
+\frac{65}{16} \frac{m_2^2 p_1^4}{m_1^4} 
-\frac{1672183}{19200} \frac{(n_{12} p_1)^2 (p_1 p_2)}{m_1^2} 
+\frac{8777}{384} \frac{(n_{12} p_1)^2 (p_1 p_2)}{m_1 m_2} 
+\frac{m_2 (n_{12} p_1)^2 (p_1 p_2)}{4 m_1^3} 
\right.
 \nn\\
& \left. 
 \qquad
-\frac{73057}{2400} \frac{(n_{12} p_1)(n_{12} p_2) (p_1 p_2)}{m_1^2} 
+\frac{329791}{6400} \frac{(n_{12} p_1)(n_{12} p_2) (p_1 p_2)}{m_1 m_2}
+\frac{1029517}{14400} \frac{p_1^2 (p_1 p_2)}{m_1^2} 
+\frac{452369}{28800} \frac{p_1^2 (p_1 p_2)}{m_1 m_2} 
\right.
 \nn\\
& \left. 
 \qquad
-\frac{23}{8} \frac{m_2 p_1^2(p_1p_2)}{m_1^3}
+\frac{179}{100} \frac{(p_1p_2)^2}{m_1^2}
-\frac{1911149}{57600} \frac{(p_1p_2)^2}{m_1 m_2}
-\frac{79711}{4800} \frac{p_1^2 p_2^2}{m_1^2} 
-\frac{1772417}{57600} \frac{p_1^2 p_2^2}{m_1 m_2}
-\frac{40483 \pi^2}{16384} \frac{p_1^2 (p_1p_2)}{m_1^2} 
\right.
 \nn\\
& \left. 
 \qquad
+\frac{10631 \pi^2}{8192} \frac{(p_1 p_2)^2}{m_1 m_2}
+\frac{375 \pi^2}{8192} \frac{m_2 (n_{12} p_1)^4}{m_1^3} 
+\frac{2749 \pi^2}{8192} \frac{m_2 p_1^4}{m_1^3} 
+\frac{35655 \pi^2}{16384} \frac{(n_{12} p_1)^3 (n_{12} p_2)}{m_1^2} 
\right.
 \nn\\
& \left. 
 \qquad
-\frac{36405 \pi^2}{16384} \frac{(n_{12} p_1)^2 (n_{12} p_2)^2}{m_1 m_2}
+\frac{43101 \pi^2}{16384} \frac{(n_{12} p_1)(n_{12} p_2) p_1^2}{m_1^2} 
-\frac{1059 \pi^2}{1024} \frac{m_2 (n_{12} p_1)^2 p_1^2}{m_1^3} +\frac{13723 \pi^2}{16384} \frac{p_1^2 p_2^2}{m_1 m_2}
\right.
 \nn\\
& \left. 
 \qquad
-\frac{1059 \pi^2}{512} \frac{(n_{12} p_2)^2 p_1^2}{m_1 m_2} 
-\frac{6153 \pi^2}{2048} \frac{(n_{12} p_1)(n_{12} p_2) (p_1 p_2)}{m_1 m_2}
+\frac{56955 \pi^2}{16384} \frac{(n_{12} p_1)^2 (p_1 p_2)}{m_1^2} 
\right]+ 1 \leftrightarrow 2\,, \\
 H^{\rm 4PN(4)}_{\rm EFT}&=  \frac{G^4 m_1 m_2}{r_{12}^4}\left[-\frac{9841}{1600} m_1 (n_{12} p_1)^2  + \frac{626723}{19200} m_2 (n_{12} p_1)^2  + \frac{2732179}{57600} \frac{m_2^2 (n_{12} p_1)^2}{m_1} 
- \frac{21}{8} \frac{m_2^3 (n_{12} p_1)^2}{m_1^2} 
\right.
 \nn\\
& \left. 
 \qquad\qquad\qquad
- \frac{2421613}{19200} m_1 (n_{12} p_1)(n_{12} p_2) 
+ \frac{7}{8} \frac{m_1^2 (n_{12} p_1)(n_{12} p_2)}{m_2}  + \frac{72961}{4800} m_1 p_1^2 + \frac{2003633}{57600} m_2 p_1^2
\right.
 \nn\\
& \left. 
 \qquad\qquad\qquad
+ \frac{390961}{19200} \frac{m_2^2 p_1^2}{m_1} 
+ \frac{189}{32} \frac{m_2^3 p_1^2}{m_1^2} - \frac{3367117}{57600} m_1 (p_1 p_2) 
- \frac{529}{32} \frac{m_1^2 (p_1 p_2)}{m_2}  + \frac{21745 \pi^2}{16384} m_2 (n_{12} p_1)^2 
\right.
 \nn\\
 & \left.
 \qquad\qquad\qquad
- \frac{28691 \pi^2}{24576} \frac{m_2^2 (n_{12} p_1)^2}{m_1} 
+ \frac{63641 \pi^2}{24576} m_1 (n_{12} p_1)(n_{12} p_2) 
- \frac{199177 \pi^2}{49152} m_2 p_1^2 
 -\frac{21837 \pi^2}{8192} \frac{m_2^2 p_1^2}{m_1} 
 \right.
 \nn\\
 & \left.
 \qquad\qquad\qquad
+ \frac{176033 \pi^2}{24576} m_1 (p_1 p_2) 
 \right]+ 1 \leftrightarrow 2\,, \\
 H^{\rm 4PN(5)}_{\rm EFT}&=  \frac{G^5 m_1 m_2}{r_{12}^5}\left[    -\frac{11}{32} m_1^4  - \frac{172049}{2400} m_1^3 m_2  - \frac{152863}{1800} m_1^2 m_2^2 + \frac{6237 \pi^2}{1024} m_1^3 m_2  + \frac{44825 \pi^2}{6144} m_1^2 m_2^2\right]+ 1 \leftrightarrow 2\,.
\end{align}
\end{subequations}
Finally, the non-local 4PN tail Hamiltonian is given by
\begin{align}
\label{eq:Htail non CM}
    H_{\rm tail}^{\rm 4PN} = -\frac{G^2 m }{5 c^8 }Q_{ij}^{(3)}(t)\mathop{\mathrm{Pf}}\limits_{2r_{12}/c}\int_{-\infty}^{+\infty} \frac{ Q_{ij}^{(3)}(t')}{\lvert t-t' \rvert}\, \dd t'\,,
\end{align}
where $Q_{ij}$ is the quadrupole moment of the binary (which is Newtonian at this order), $Q_{ij}^{(\alpha)}$ denotes the $\alpha$-th derivative of $Q_{ij}$ and 
where Pf is the Hadamard \emph{Partie finie}. In general, for any well-behaved function $f(t)$ going to zero sufficiently fast when $t \rightarrow \pm \infty$, this \emph{Partie finie} is defined by
\begin{equation}
\mathop{\mathrm{Pf}}\limits_{\tau_0}\int_{-\infty}^{+\infty} \frac{f(t')}{\lvert t-t' \rvert}\, \mathrm{d}t' \equiv \int_{0}^{+\infty} \mathrm{d}\tau \, \mathrm{ln}\left( \frac{\tau}{\tau_0}\right) \left[ f^{(1)}(t-\tau)-f^{(1)}(t+\tau) \right]\,,
\end{equation}
where $f^{(1)}(t)\equiv\dd f/\dd t$.\footnote{In the present formalism, we assume that the quadrupole moment is constant in the remote past (for $t\leqslant -\mathcal{T}$ where $-\mathcal{T}$ is the instant of formation of the compact binary) and we extend this assumption to the remote future as well (for $t\geqslant +\mathcal{T}$), so that the integral is well-defined at infinity.} Note that, since we are working up to 4PN order, $ H_{\rm tail}^{\rm 4PN}$ takes the same expression in any gauges, for instance EFT or ADM gauges. 

In the frame of the center-of-mass (CM), using the notation $r \equiv r_{12}$, $\bm{n}\equiv \bm{n}_{12}$, $\bm{p}\equiv \bm{p}_1 = -\bm{p}_2$,  $p \equiv \rvert \boldsymbol{p} \rvert $ and $p_r \equiv \bm{n} \cdot \bm{p}$, the EFT Hamiltonian is given by
\begin{subequations}
\begin{align}
H_{\rm EFT}^{\rm N} =&\; \frac{p^2}{2m \nu}-\frac{G m^2 \nu}{r}\,,\\
H_{\rm EFT}^{\rm 1PN} =&\; \frac{p^4}{8 m^3 \nu^3} \Big[-1 + 3\nu\Big]  - \frac{G}{ \nu\,r} \left[ \frac{p^2}{2 } (3 + \nu) + \frac{p_r^2}{2}\,\nu \right]  + \frac{G^2 m^3 \nu}{2 r^2}\,,\\
H_{\rm EFT}^{\rm 2PN} =&\; \frac{p^6}{16 m^5 \nu^5} \Big[ 1 - 5\nu + 5\nu^2\Big]  + \frac{G}{8 m^2 \nu^3\,  r} \Big[ p^4 (5 - 16\nu - 3\nu^2) - p^2 p_r^2 (4 \nu + 2\nu^2) -3 p_r^4 \nu^2 \Big]  \nn\\
& + \frac{G^2 m}{2 \nu r^2} \left[ p^2 \left(\frac{11}{2} + 6\nu\right) +  p_r^2 (-1 + 3\nu) \right]  - \frac{G^3 m^4 }{4 r^3} \Big[2\nu + \nu^2\Big] \,,\\
H_{\rm EFT}^{\rm 3PN} =&\; - \frac{p^8}{128 m^7 \nu^7} \Big[5 - 35\nu + 70\nu^2 - 35\nu^3\Big] \nn\\
& + \frac{G}{m^4 \nu^5 r} \left[ - \frac{p^6}{16 } (7 - 38\nu + 41\nu^2 + 5\nu^3) + \frac{p^4 p_r^2}{16} \nu (4 - 10\nu - 3\nu^2) + \frac{3 p^2 p_r^4}{16} \nu^2 (1 - \nu) - \frac{5 p_r^6}{16}\nu^3 \right] \nn\\
& + \frac{G^2}{m \nu^3 r^2} \left[ - \frac{p^4}{16} (29 - 110\nu - 85\nu^2) + \frac{p^2 p_r^2}{16} (4 - 3\nu + 66\nu^2) + \frac{5 p_r^4}{12}\nu (1 + 5\nu) \right] \nn\\
& + \frac{G^3 m^2}{\nu r^3} \left[ - p^2 \left( \frac{17}{4} + \left(\frac{221}{48} - \frac{\pi^2}{64}\right)\nu + \frac{17}{8}\,\nu^2 \right) + p_r^2 \left( \frac{3}{2} - \left(\frac{107}{16} + \frac{3 \pi^2}{64}\right)\nu - 3 \nu^2 \right) \right] \nn \\
& + \frac{G^4 m^5 \nu}{r^4} \left[ \frac{3}{8} + \frac{103 \nu}{12} - \frac{21 \pi^2 \nu}{32} \right]\,,\\
H_{\rm EFT}^{\rm 4PN} =&\; \frac{p^{10}}{256 m^9 \nu^9} \Big[7 - 63\nu + 189\nu^2 - 210\nu^3 + 63\nu^4\Big] \nn\\
& + \frac{G}{m^6 \nu^7 r} \left[ \frac{p^8}{256} (90 - 672\nu + 1452\nu^2 - 773\nu^3 - 70\nu^4) - \frac{p^6 p_r^2}{64} \nu (12 - 54\nu + 37\nu^2 + 10\nu^3) \right. \nn \\ 
& \hspace{1.8cm} \left. - \frac{3 p^4 p_r^4}{128} \nu^2 (6 - 23\nu + 6\nu^2) - \frac{5 p^2 p_r^6}{64} \nu^3 (1 + 2\nu) + \frac{35 p_r^8}{256} \nu^3 (1 - 2\nu) \right] \nn \\
& + \frac{G^2}{m^3 \nu^5 r^2} \left[ \frac{p^6}{256} (440 - 2932\nu + 3777\nu^2 + 1855\nu^3) - \frac{p^4 p_r^2}{256} (48 - 960\nu + 2292\nu^2 - 1743\nu^3) \right. \nn \\ 
&  \left. \hspace{1.8cm}- \frac{p^2 p_r^4}{768} \nu (3556 - 9091\nu + 1025\nu^2) + \frac{p_r^6}{1280} \nu (2952 - 11510\nu + 7953\nu^2) \right] \nn \\
& + \frac{G^3}{\nu^3 r^3} \left[ p^4 \left( \frac{65}{16} - \left( \frac{572389}{19200} - \frac{2749 \pi^2}{8192} \right) \nu - \left( \frac{839689}{28800} - \frac{18491 \pi^2}{16384} \right) \nu^2 - \frac{433}{128} \nu^3 \right) \right. \nn \\
&  \left.\hspace{1.3cm} +\, p^2 p_r^2 \left( -\frac{5}{4} + \left( \frac{79047}{1600} - \frac{1059 \pi^2}{1024} \right) \nu - \left( \frac{2329}{48} + \frac{4035 \pi^2}{2048} \right) \nu^2 - \frac{313}{64} \nu^3 \right) \right. \nn \\
&  \left.\hspace{1.3cm} +\, p_r^4 \left( -\left( \frac{24493}{1280} - \frac{375 \pi^2}{8192} \right)\nu + \left( \frac{69143}{1920} - \frac{38655 \pi^2}{16384} \right) \nu^2 - \frac{517}{128} \nu^3 \right) \right] \nn \\
& + \frac{G^4 m^3}{\nu r^4} \left[ p^2 \left( \frac{189}{32} + \left( \frac{141361}{19200} - \frac{21837 \pi^2}{8192} \right) \nu + \left( \frac{524011}{19200} - \frac{158177 \pi^2}{49152} \right) \nu^2 \right) \right. \nn \\
& \hspace{1.6cm} \left. +\, p_r^2 \left( -\frac{21}{8} + \left( \frac{3437779}{57600} - \frac{28691 \pi^2}{24576} \right) \nu - \left( \frac{707}{3840} - \frac{110099 \pi^2}{49152} \right) \nu^2 \right) \right] \nn \\
& - \frac{G^5 m^6 \nu}{r^5} \left[ \frac{11}{32} + \left( \frac{168749}{2400} - \frac{6237 \pi^2}{1024} \right) \nu + \left( \frac{4889}{180} - \frac{7403 \pi^2}{3072} \right) \nu^2 \right]\,,\\
\label{eq:Htail}
    H_{\rm tail}^{\rm 4PN}&=-\frac{G^2 m  }{5 c^8 }Q_{ij}^{(3)}(t) \mathop{\mathrm{Pf}}\limits_{2r/c}\int_{-\infty}^{+\infty} \frac{ Q_{ij}^{(3)}(t')}{\lvert t-t' \rvert}\, \dd t'\,.
\end{align}
\end{subequations}

\section{Spin contributions in the EFT Hamiltonian up to 4PN order}
\label{app:spinEFT}

In this Appendix we present all the spin terms in the EFT Hamiltonian up to 4PN order, which were derived by Levi and Steinhoff~\cite{LS14,LS15a,LS15b,LS16a,LS16b,LS21}. In the formalism the non-spin part of the Hamiltonian $H_\text{NS}$ is given by the EFT Hamiltonian presented in the previous Appendix. To 4PN order we have to include beyond the usual spin-orbit ($SO$) and spin-spin ($SS$) terms higher order terms up to quartic order, denoted
\begin{align}
    H_S = H_{ SO} + H_{ SS} + H_{ SSS} + H_{SSSS} + \mathcal{O}\left(\frac{1}{c^9}\right)\,.
\end{align}
Furthermore we must take into account for the $SO$ and $SS$ terms the leading order (LO) followed by next-to-leading order (NLO) and next-to-next-to-leading order (NNLO) terms which gives the structure
\begin{subequations}
\begin{align}
    H_{SO} &= \frac{1}{c^3} H_{SO}^\text{LO} + \frac{1}{c^5} H_{SO}^\text{NLO} + \frac{1}{c^7} H_{SO}^\text{NNLO} + \mathcal{O}\left(\frac{1}{c^9}\right)\,,\\
    H_{SS} &= \frac{1}{c^4} H_{SS}^\text{LO} + \frac{1}{c^6} H_{SS}^\text{NLO} + \frac{1}{c^8} H_{SS}^\text{NNLO} + \mathcal{O}\left(\frac{1}{c^{10}}\right)\,,\\
    H_{SSS} &= \frac{1}{c^7} H_{SSS}^\text{LO} + \mathcal{O}\left(\frac{1}{c^9}\right)\,,\\
    H_{SSSS} &= \frac{1}{c^8} H_{SSSS}^\text{LO} + \mathcal{O}\left(\frac{1}{c^{10}}\right)\,,
\end{align}
\end{subequations}

Working in the CM frame, we introduce the following notations:\footnote{In~\cite{LS16a,LS21}, Levi and Steinhoff use the renormalised, dimensionless variables: $\tilde{r} \equiv r/(G m/c^2)$, $\bm{\tilde{p}} \equiv \bm{p}/(m \nu c)$,  $\bm{\tilde{L}}_p \equiv \bm{L}_p/(G m^2 \nu/c)$ and $\bm{\tilde{S}}_a \equiv \bm{S}_a/(G m^2 \nu)$.} $\bm{L}_p \equiv m \nu\, \bm{r} \times \bm{p}$, $L_p \equiv \rvert \bm{L}_p\rvert $, $S_1=\rvert\bm{S}_1\rvert$, $S_2=\rvert\bm{S}_2\rvert$, $S_{1n}=\bm{n} \cdot \bm{S}_1$, $S_{2n}=\bm{n} \cdot \bm{S}_2$, $(S_1 L_p) \equiv \bm{S}_1 \cdot \bm{L}_p$, $(S_2 L_p) \equiv \bm{S}_1 \cdot \bm{L}_p$,  $(p S_1) \equiv \bm{p} \cdot \bm{S}_1$, $(p S_2) \equiv \bm{p} \cdot \bm{S}_2$ and   $(S_1 S_2) \equiv \bm{S}_1 \cdot \bm{S}_2$. The $SO$ terms then read\footnote{These equations correspond to Eqs.~(4.26-4.28) of~\cite{LS16a}.} 
\begin{subequations}
\begin{align}
H_{SO}^{\rm LO} =&\; \frac{G}{4 r^3 \nu}(S_1 L_p) \left[ 3(1-\delta) + 2\nu \right]  + 1 \leftrightarrow 2\,,\\
H_{SO}^{\rm NLO} =&\; 
    (S_1 L_p) \Big\{ - \frac{G p_r^2}{16 m^2 \nu^3 r^3} \left[ 5(1-\delta) - 44\nu - 18\nu^2 + 34\delta \nu \right] -  \frac{G L_p^2}{16 m^2 \nu^3 r^5} \left[ 5(1-\delta) - 20\nu - 6\nu^2 +10\delta \nu \right]   \nn\\
    &- \frac{G^2 m}{8 \nu r^4} \left[ 20(1-\delta) + 13\nu - 5 \delta\nu \right] \Big\} + 1 \leftrightarrow 2\,,\\
 H_{SO}^{\rm NNLO} =&\; (S_1 L_p) \Big\{ \frac{G p_r^4}{32 m^4 \nu^5 r^3}  \left[ 7(1-\delta) - 70\nu + 175\nu^2 + 58\nu^3 + 56\delta\nu - 131\delta\nu^2 \right]   \nn\\     
 & + \frac{G L_p^2\, p_r^2}{32 m^4 \nu^5 r^5} \left[ 14(1-\delta) - 116\nu + 173\nu^2 + 38\nu^3 + 88\delta\nu - 73\delta\nu^2 \right]  \nn \\     
 & + \frac{G L_p^4}{32 m^4 \nu^5 r^7} \left[ 7(1-\delta) - 46\nu + 43\nu^2 + 10\nu^3 + 32\delta\nu - 17\delta\nu^2 \right]  \nn \\     
 & + \frac{G^2 p_r^2}{32 m \nu^3 r^4}  \left[ 90(1-\delta) - 459\nu - 81\nu^2 + 279\delta\nu + 63\delta\nu^2 \right]  \nn \\     & \hspace{1.3cm}+ \frac{G^2 L_p^2}{32 m \nu^3 r^6} \left[ 42(1-\delta) - 257\nu - 201\nu^2 + 173\delta\nu + 17\delta\nu^2 \right]  \nn \\    
 & + \frac{G^3 m^2}{16 \nu r^5} \left[  90(1-\delta) + 73\nu + 46\nu^2 - 61\delta\nu \right]  \Big\}  + 1 \leftrightarrow 2 \,.
\end{align}
\end{subequations}

Concerning the $SS$ part we further split it with obvious notation as
\begin{align}
H_{SS} = H_{S_1S_1} + H_{S_1S_2} + H_{S_2S_2}\,,
\end{align}
where $H_{S_1S_1}$ denotes the self interaction part depending on spin-induced deformability coefficients due to the internal structure of the compact objects (and $H_{S_2S_2}$ is obtained from $1 \leftrightarrow 2$), and where $H_{S_1S_2}$ is the interaction part between the two spins and is free of deformability coefficients. The interaction parts are given by\footnote{These explicit expressions are given in Eqs.~(4.29-4.30) of~\cite{LS16a} (LO and NLO) and (3.10) of~\cite{LS21} (NNLO).}
\begin{subequations}
\begin{align}
H_{S_1S_2}^{\rm LO} = & \frac{G}{r^3} \Big\{ 3 S_{1n} S_{2n} - (S_1 S_2) \Big\}\,,\\
    H_{S_1S_2}^{\rm NLO} =&  \frac{G}{4 m^2 \nu^2 r^3} \Bigg\{ 2(5 + \nu) (p S_1)(p S_2)  
    - 3 p_r \Big[ (5 + 3\nu+3\delta)  (p S_1) S_{2n}  + (5 + 3\nu-3\delta)  (p S_2) S_{1n})  \Big] \nn \\
    & \qquad\qquad\quad + p_r^2 \Big[ 7(2 - \nu) (S_1 S_2) +3(2+13\nu) S_{1n} S_{2n}\Big] \Bigg\} \nn \\
    &  -\frac{G L_p^2}{4 m^2 \nu^2 r^5} \Bigg\{ 2(5 + 2\nu) (S_1 S_2) - 3(2 + 3\nu) S_{1n} S_{2n} \Bigg\}+ \frac{G^2 m}{r^4}  \Bigg\{ 7 (S_1 S_2) - 13 S_{1n} S_{2n} \Bigg\}\,, 
    \\
   & \nn
    \\
H_{S_1S_2}^{\rm NNLO} = &  \frac{G}{16 m^4 \nu^4 r^3} \Bigg\{ 2 p_r^2 \left( -11 + 139\nu + 22\nu^2 \right) (p S_1)(p S_2) 
+3p_r^3 \Big[ \left( 11-115\nu -40 \nu^2 +5\delta (1-8\nu) \right) (p S_1) S_{2n}
\nn\\
& 
\qquad + \left( 11-115\nu -40 \nu^2 -5\delta (1-8\nu) \right) (p S_2) S_{1n} \Big]
+p_r^4 \Big[\left( -26 + 161\nu - 38\nu^2 \right) (S_1 S_2) 
\nn \\
& 
\qquad +  3\left( -6+73\nu +110\nu^2  \right) S_{1n} S_{2n} 
\Big]
\Bigg\} 
+ \frac{G}{r^5} \frac{L_p^2}{16 m^4 \nu^4} \Bigg\{ 2 \left( -11 + 25\nu + 4\nu^2 \right) (p S_1)(p S_2) \nn \\
& 
\qquad + 3 p_r \Big[ \left(11 - 25\nu - 10\nu^2 +  5 \delta (1 -2\nu) \right) (p S_1) S_{2n} + \left(11 - 25\nu - 10\nu^2 -5 \delta (1 - 2\nu) \right) (p S_2) S_{1n} \Big] \nn \\
& 
\qquad +p_r^2\Big[ -\left( 4+125\nu +52\nu^2 \right) (S_1 S_2) +6 \left( -6 + 33\nu + 25\nu^2 \right) S_{1n} S_{2n} \Big] 
\Bigg\}
\nn \\
& + \frac{G L_p^4}{16 m^4 \nu^4 r^7} \Bigg\{ 2 \left( 11 - 23\nu - 7\nu^2 \right) (S_1 S_2) + 3 \left( -6 - 7\nu + 10\nu^2 \right) S_{1n} S_{2n} \Bigg\}
\nn \\
& 
+ \frac{G^2}{8 m \nu^2 r^4} \Bigg\{ -\left( 124 + 19\nu \right) (p S_1)(p S_2) 
+ 2 p_r \Big[ \left( 90+23\nu+48\delta +7 \nu \delta \right) (p S_1) S_{2n} 
\nn \\
& 
\qquad
+ \left(90+23\nu-48\delta -7 \nu \delta  \right) (p S_2) S_{1n} \Big] 
+  p_r^2 \Big[6\left( -25 + 9\nu \right) (S_1 S_2) -  \left( 86 +259\nu \right) S_{1n} S_{2n} \Big]
\Bigg\}
\nn \\
& 
+ \frac{G^2 L_p^2}{8 m \nu^2 r^6} \Bigg[ 7 \left( 18 + 19\nu \right) (S_1 S_2) - \left( 86 + 169\nu \right) S_{1n} S_{2n} \Bigg]
\nn \\
& 
+\frac{G^3 m^2}{4 r^5} \Bigg\{ -\left( 78 + 23\nu \right) (S_1 S_2) + 3 \left( 42 + 29\nu \right) S_{1n} S_{2n} \Bigg\} \,. 
\end{align}
\end{subequations}
Next, the self parts are given by\footnote{These expressions can be found in Eqs.~(4.31-4.32) of~\cite{LS16a} (LO and NLO) and (3.11) of~\cite{LS21} (NNLO).}  
\begin{subequations}
\begin{align}
     H_{S_1S_1}^{\rm LO} &= \frac{G }{4 \nu r^3} \kappa_1(- 1 + 2\nu+\delta ) \left( S_1^2 - 3 S_{1n}^2 \right) \,,\\
H_{S_1S_1}^{\rm NLO} =&\;\frac{G}{16 m^2 \nu^3 r^3} \Big\{ 2 (p S_1)^2 \left[ -5 + 14\nu + 2\nu^2 + 2\kappa_1(1 - 3\nu) + \delta \left( 5 - 4\nu - 2\kappa_1(1 - \nu) \right) \right] \nn\\
& + 6 p_r (p S_1) S_{1n} \left[ 5 - 14\nu - 2\nu^2 - 2\kappa_1(1 - 2\nu - 2\nu^2) - \delta \left( 5 - 4\nu - 2\kappa_1 \right) \right]
\nn \\
& + p_r^2 S_1^2 \left[ 1 - 4\nu + 2\nu^2 + 2\kappa_1(1 - 10\nu + 8\nu^2) - \delta \left( 1 - 2\nu + 2\kappa_1(1 - 8\nu) \right) \right] \nn \\
& - 3 p_r^2 S_{1n}^2 \left[ 7 - 20\nu - 2\nu^2 - 6\kappa_1(1 - 4\nu^2) - \delta \left( 7 - 6\nu - 6\kappa_1(1 + 2\nu) \right) \right] \Big\} 
\nn \\
& -  \frac{G L_p^2}{16 m^2 \nu^3 r^5} \Big\{ 2 S_1^2 \left[ -5 + 14\nu + 2\nu^2 + \kappa_1(5 - 11\nu - 2\nu^2) - \delta \left( -5 + 4\nu + \kappa_1(5 - \nu) \right) \right] 
\nn \\
& + 3 S_{1n}^2 \left[ 7 - 20\nu - 2\nu^2 - 2\kappa_1(3 - 5\nu - 2\nu^2) - \delta \left( 7 - 6\nu - 2\kappa_1(3 + \nu) \right) \right] \Big\} \nn \\
& +  \frac{G^2 m}{4 \nu r^4} \Big\{ S_1^2 \left[ 2 - 2\nu + \kappa_1(4 - 11\nu) - \delta \left( 2 + 2\nu + \kappa_1(4 - 3\nu) \right) \right] \nn \\
& - S_{1n}^2 \left[ 2 + \kappa_1(10 - 27\nu) - \delta \left( 2 + 4\nu + \kappa_1(10 - 7\nu) \right) \right] \Big\} \,,\\
H_{S_1S_1}^{\rm NNLO} =&\; \frac{G}{32 m^4 \nu^5 r^3} \Big\{ p_r^2 (p S_1)^2 \Big[ -1 - 33\nu + 141\nu^2 + 32\nu^3 + 2\kappa_1(5 + 2\nu - 49\nu^2 - 6\nu^3) \nn \\
&  + \delta\left( 1 + 35\nu - 69\nu^2 - 2\kappa_1(5 + 12\nu - 15\nu^2) \right) \Big] \nn \\
& - 3 p_r^3 (p S_1) S_{1n} \left[ 7 - 59\nu + 141\nu^2 + 28\nu^3 + 2\kappa_1(3 + 10\nu - 39\nu^2 - 26\nu^3) - \delta\left( 7 - 45\nu + 65\nu^2 + 2\kappa_1(3 + 16\nu - \nu^2) \right) \right] \nn \\
& + p_r^4 S_1^2 \left[ 4 - 12\nu - 9\nu^2 + 10\nu^3 + \kappa_1(13 - 6\nu - 128\nu^2 + 48\nu^3) - \delta\left( 4 - 4\nu - 9\nu^2 + \kappa_1(13 + 20\nu - 72\nu^2) \right) \right] \nn \\
& - 3 p_r^4 S_{1n}^2 \left[ -6 + 44\nu - 97\nu^2 - 14\nu^3 + \kappa_1(5 - 46\nu + 24\nu^2 + 96\nu^3) + \delta\left( 6 - 32\nu + 45\nu^2 - \kappa_1(5 - 36\nu - 48\nu^2) \right) \right] \Big\} \nn \\
& + \frac{G L_p^2}{32 m^4 \nu^5 r^5} \Big\{ (p S_1)^2 \left[ 11 - 81\nu + 117\nu^2 + 8\nu^3 + 4\kappa_1(1 + \nu - 11\nu^2) - \delta\left( 11 - 59\nu + 21\nu^2 + 4\kappa_1(1 + 3\nu - 3\nu^2) \right) \right] \nn \\
& - 3 p_r (p S_1) S_{1n} \left[ 7 - 69\nu + 121\nu^2 + 8\nu^3 + 2\kappa_1(3 - 19\nu^2 - 6\nu^3) - \delta\left( 7 - 55\nu + 25\nu^2 + 2\kappa_1(3 + 6\nu - \nu^2) \right) \right] \nn \\
& + p_r^2 S_1^2 \left[ -7 + 54\nu - 126\nu^2 - 28\nu^3 + 2\kappa_1(7 - 48\nu + 70\nu^2 + 12\nu^3) + \delta\left( 7 - 40\nu + 60\nu^2 - 2\kappa_1(7 - 34\nu + 6\nu^2) \right) \right] \nn \\
& - 3 p_r^2 S_{1n}^2 \left[ -12 + 93\nu - 174\nu^2 - 18\nu^3 + 2\kappa_1(5 - 36\nu + 44\nu^2 + 16\nu^3) + \delta\left( 12 - 69\nu + 60\nu^2 - 2\kappa_1(5 - 26\nu - 8\nu^2) \right) \right] \Big\} \nn \\
& + \frac{G L_p^4}{32 m^4 \nu^5 r^7} \Big\{ S_1^2 \left[ -11 + 81\nu - 117\nu^2 - 8\nu^3 + \kappa_1(1 - 30\nu + 73\nu^2 + 6\nu^3) + \delta\left( 11 - 59\nu + 21\nu^2 - \kappa_1(1 - 28\nu + 9\nu^2) \right) \right] \nn \\
& - 3 S_{1n}^2 \left[ -6 + 49\nu - 77\nu^2 - 4\nu^3 + \kappa_1(5 - 26\nu + 29\nu^2 + 6\nu^3) + \delta\left( 6 - 37\nu + 15\nu^2 - \kappa_1(5 - 16\nu - 3\nu^2) \right) \right] \Big\} \nn \\
& + \frac{G^2}{16 m \nu^3 r^4} \Big\{ (p S_1)^2 \left[ 67 - 171\nu - 119\nu^2 + \kappa_1(4 - 105\nu + 198\nu^2) - \delta\left( 67 - 37\nu - 11\nu^2 + \kappa_1(4 - 97\nu + 4\nu^2) \right) \right] \nn \\
& + p_r (p S_1) S_{1n} \left[ -203 + 489\nu + 431\nu^2 - 2\kappa_1(20 - 293\nu + 505\nu^2) + \delta\left( 203 - 83\nu - 47\nu^2 + 2\kappa_1(20 - 253\nu + 15\nu^2) \right) \right] \nn \\
& + \frac{1}{2} p_r^2 S_1^2 \left[ 3 + 183\nu - 387\nu^2 - 2\kappa_1(141\nu - 292\nu^2) - \delta\left( 3 + 189\nu + 45\nu^2 - 2\kappa_1(141\nu + 26\nu^2) \right) \right] \nn \\
& + \frac{1}{2} p_r^2 S_{1n}^2 \left[ 269 - 723\nu - 517\nu^2 - 4\kappa_1(36 + 95\nu - 418\nu^2) - \delta\left( 269 - 185\nu - 157\nu^2 - 4\kappa_1(36 + 167\nu - 46\nu^2) \right) \right] \Big\} \nn \\
& + \frac{G^2 L_p^2}{32 m \nu^3 r^6} \Big\{ S_1^2 \left[ -139 + 371\nu + 245\nu^2 + 2\kappa_1(64 - 55\nu - 244\nu^2) + \delta\left( 139 - 93\nu - 33\nu^2 - 2\kappa_1(64 + 73\nu - 10\nu^2) \right) \right] \nn \\
& + S_{1n}^2 \left[ 269 - 795\nu - 205\nu^2 - 4\kappa_1(60 - 109\nu - 85\nu^2) - \delta\left( 269 - 257\nu - 61\nu^2 - 4\kappa_1(60 + 11\nu - 7\nu^2) \right) \right] \Big\} \nn \\
& + \frac{G^3 m^2}{56 \nu r^5} \Big\{ S_1^2 \left[ -98 + 168\nu - 136\nu^2 - \kappa_1(133 - 77\nu - 598\nu^2) + \delta\left( 98 + 28\nu + \kappa_1(133 + 189\nu) \right) \right] \nn \\
& + S_{1n}^2 \left[ 98 + 56\nu - 208\nu^2 + \kappa_1(259 - 371\nu - 674\nu^2) - \delta\left( 98 + 252\nu + \kappa_1(259 + 147\nu) \right) \right] \Big\} \,,
\end{align}
\end{subequations}
together with $H_{S_2S_2}$ obtained by label exchange $1 \leftrightarrow 2$. Finally the $SSS$ terms and $SSSS$ terms (both LO, respectively of 3.5 PN and 4PN order) read\footnote{These expressions can be found in Eq.~(4.33) of~\cite{LS16a} ($SSS$) and in Eq.~(3.12) of~\cite{LS21} ($SSSS$).}
\begin{subequations}
\begin{align}
    H_{SSS}^{\rm LO} =&\; (S_1 L_p) \Bigg\{ \frac{3 G}{4 m^2 \nu^2 r^3 L_p^2} (p S_2 - p_r S_{2n}) \Big[ (p S_1) \big( (1-\delta)(1+4\kappa_1) - 2\nu(1+\kappa_1) \big) 
    \nn
    \\
    &\qquad \qquad \qquad \qquad \qquad \qquad \qquad \quad 
    - (p S_2) \big( (1+\delta)(1+4\kappa_2) - 2\nu(1+\kappa_2) \big) \Big] \nn \\ 
    &+ \frac{G}{8 m^2 \nu^3 r^5}\Bigg[  \Big[ \kappa_1 \big( 3(1-\delta) - 6\nu(2-\delta) + 6\nu^2 \big) - 4\lambda_1 \big( 1-\delta - \nu(3-\delta) \big) \Big] (5 S_{1n}^2 - S_1^2)
    \nn
    \\ &
    \qquad\qquad \qquad
    + 6\nu\Big[(1+6\kappa_2)(1+\delta)-(2+3\kappa_2)\nu\Big]S_2^2
    -12\nu\Big[(1+2\kappa_1)(1-\delta)-(2+\kappa_1)\nu\Big](S_1S_2)
    \nn\\
    & \qquad \qquad \qquad + 
    30\nu(1-\delta-2\nu) S_{1n} S_{2n}-30
    \kappa_2 \nu (2 + 2\delta - \nu) S_{2n}^2 \Bigg] \Bigg\} + 1 \leftrightarrow 2 \,, 
    \\
 H^{\rm LO}_{SSSS} =&\; \frac{G}{16 m^2 \nu^3 r^5} \Bigg\{ 6\nu^2 \kappa_1 \kappa_2 \Big[20 S_{1n} (S_1 S_2) S_{2n} -  S_1^2 S_2^2 + 10 S_1^2 S_{2n}^2 - 2 (S_1 S_2)^2 - 35 S_{1n}^2 S_{2n}^2 \Big] 
 \nn
 \\ & + 4\nu \lambda_1 \Big[ 1 - \delta - 2\nu \Big] \Big[ 15 S_{1n} S_1^2 S_{2n} - 3 S_1^2 (S_1 S_2) - 35 S_{2n} S_{1n}^3 + 15 (S_1 S_2) S_{1n}^2 \Big] \nn\\ & - \iota_1 \Big[ (1-\delta) - 2\nu(2-\delta) + 2\nu^2 \Big] \Big[ 35 S_{1n}^4 - 30 S_1^2 S_{1n}^2 + 3 S_1^4 \Big] \Bigg\} + 1 \leftrightarrow 2\,.
\end{align}
\end{subequations}

\section{Spin contributions in the EFT equations of motion to 4PN order}
\label{app:EoM formulae}

The EFT equations of motion are obtained by varying the EFT Hamiltonian (in the CM frame), after consistent replacement of the canonical momenta in terms of position, ordinary velocity and spins, with use of the precession equations of the spins and in the final stage the order reduction of the accelerations. The EFT acceleration is made of a non-spin part and a spin part. Considering only the conservative dynamics (\textit{e.g.} neglecting the radiation reaction terms) the NS part reads
\begin{equation}
    \bm{a}_{\rm EFT} =\bm{a}_{\rm EFT}^{\rm N} +\frac{1}{c^2}\,\bm{a}_{\rm EFT}^{\rm 1PN}+\frac{1}{c^4}\,\bm{a}_{\rm EFT}^{\rm 2PN}+\frac{1}{c^6}\,\bm{a}_{\rm EFT}^{\rm 3PN}+\frac{1}{c^8}\left(\bm{a}_{\rm EFT}^{\rm 4PN} + \bm{a}_{\rm tail}^{\rm 4PN} \right) + \mathcal{O}\left(\frac{1}{c^9}\right)\,,
\end{equation}
with, singling out the 4PN tail term which is the same as in harmonic coordinates,
\begin{subequations}
\begin{align}
    \bm{a}_{\rm EFT}^{\rm N} &= -\frac{G m}{r^2}\,\bm{n}\,,\\
\bm{a}_{\rm EFT}^{\rm 1PN} &= \left[\frac{G m}{r^2}\left( -v^2 (1+3\nu) +\frac{3}{2}\,\nu\,u^2\right) +\frac{G^2 m^2}{r^3}\left( 4+2\nu\right) \right] \bm{n}+\frac{G m u}{r^2}\left(4-2\nu \right) \bm{v}\,,\\
\bm{a}_{\rm EFT}^{\rm 2PN} &= \left\{  \frac{Gm}{r^2} \left[ \frac{45}{8} u^4 \nu^2+ v^4 \left(- \frac{21}{8}\nu+4\nu^2\right)  +u^2 v^2  \left(\frac{9}{4}\nu - 6\nu^2\right) \right] \right. \\
&\quad \left. +\frac{G^2 m^2}{r^3} \left[ u^2\left(2 + \frac{7}{2}\nu + 2\nu^2\right) + v^2(7\nu - 2\nu^2) \right] -\frac{G^3 m^3}{r^4} \left(9 + \frac{47\nu}{4} \right) \right\} \bm{n} \nn\\
&\quad +  u \left\{ \frac{Gm}{r^2} \left[ u^2 \left(\frac{9}{2} \nu - 3\nu^2\right) +  v^2 \left(- \frac{3}{2}\nu+2\nu^2 \right) \right]- \frac{G^2 m^2 }{r^3} 2\left(1 + 2\nu^2\right)   \right\} \bm{v}\,,\nn\\
\bm{a}_{\rm EFT}^{\rm 3PN} &= \left\{ \frac{Gm}{r^2} \left[ \frac{175}{16} u^6 \nu^3 + u^4 v^2 \left(\frac{45}{8}\nu^2 - \frac{255}{8}\nu^3\right) + u^2 v^4 \left(\frac{33}{16}\nu - \frac{195}{16}\nu^2 + \frac{45}{2}\nu^3\right) + v^6\left(-\frac{17}{8}\nu + \frac{21}{2}\nu^2 - 13\nu^3\right) \right] \right. \nn\\
&\quad \left. + \frac{G^2 m^2}{r^3} \left[ u^4\left(-\frac{15}{2}\nu + \frac{111}{4}\nu^2 + 30\nu^3\right) + u^2 v^2\left(24\nu - \frac{75}{2}\nu^2 - 20\nu^3\right) + v^4\left(-8\nu + \frac{21}{2}\nu^2 + 10\nu^3\right) \right] \right. \nn\\
&\quad \left. + \frac{G^3 m^3}{r^4} \left[ u^2\left( -1 + \left(\frac{469}{16} + \frac{15\pi^2}{64}\right)\nu + \frac{67}{4}\nu^2 + 7\nu^3 \right) - v^2\left( \left(\frac{55}{16} + \frac{3\pi^2}{64}\right)\nu + \frac{27}{2}\nu^2 + \nu^3 \right) \right] \right. \nn\\
&\quad \left. + \frac{G^4 m^4}{r^5} \left[ 16 + \left(\frac{427}{6} - \frac{41\pi^2}{16}\right)\nu + \frac{47}{4}\nu^2 \right] \right\} \bm{n} \nn\\
&\quad + u \left\{ \frac{Gm}{r^2} \left[ -\frac{15}{4} u^4 \nu^3 + u^2 v^2 \left(\frac{15}{4}\nu - \frac{57}{4}\nu^2 + 12\nu^3\right) + v^4 \left(-\frac{5}{2}\nu + \frac{19}{2}\nu^2 - 6\nu^3\right) \right] \right. \nn\\
&\quad \left. + \frac{G^2 m^2}{r^3} \left[ u^2\left(30\nu - \frac{7}{2}\nu^2 - 18\nu^3\right) + v^2\left(-\frac{89}{4}\nu + 3\nu^2 + 10\nu^3\right) \right] \right. \nn\\
&\quad \left. - \frac{G^3 m^3}{r^4} \left[ -4 + \left(\frac{183}{8} + \frac{3\pi^2}{32}\right)\nu + \frac{7}{2}\nu^2 + 8\nu^3 \right] \right\} \bm{v}\,, \\
\bm{a}_{\rm EFT}^{\rm 4PN} &= \left\{ \frac{Gm}{r^2} \left[ u^8\left(-\frac{2205}{256}\nu^3 + \frac{2205}{128}\nu^4\right) + u^6 v^2\left(\frac{2415}{64}\nu^3 - \frac{735}{8}\nu^4\right) \right.\right. \nn\\
&\quad \left.\left. + u^4 v^4\left(\frac{45}{8}\nu^2 - \frac{9645}{128}\nu^3 + \frac{1335}{8}\nu^4\right) + u^2 v^6\left(\frac{63}{32}\nu - \frac{621}{32}\nu^2 + \frac{4725}{64}\nu^3 - 99\nu^4\right) \right.\right. \nn\\
&\quad \left.\left. + v^8\left(-\frac{239}{128}\nu + \frac{2193}{128}\nu^2 - \frac{13613}{256}\nu^3 + 54\nu^4\right) \right] \right. \nn \\
&\quad \left. + \frac{G^2 m^2}{r^3} \left[ u^6\left(-\frac{369}{4}\nu + \frac{5755}{16}\nu^2 - \frac{1431}{32}\nu^3 + 86\nu^4\right) + u^4 v^2\left(\frac{4521}{32}\nu - \frac{51051}{128}\nu^2 - \frac{3417}{64}\nu^3 - 228\nu^4\right) \right.\right. \nn\\
&\quad \left.\left. + u^2 v^4\left(-36\nu + \frac{855}{64}\nu^2 + \frac{2437}{16}\nu^3 + 104\nu^4\right) + v^6\left(-\frac{339}{32}\nu + \frac{7573}{128}\nu^2 - \frac{3493}{64}\nu^3 - 58\nu^4\right) \right] \right. \nn \\
&\quad \left. + \frac{G^3 m^3}{r^4} \left[ u^4\left( \left(\frac{410713}{1280} - \frac{7875\pi^2}{8192}\right)\nu + \left(-\frac{12933}{20} + \frac{811755\pi^2}{16384}\right)\nu^2 + \frac{24679}{384}\nu^3 + 126\nu^4 \right) \right.\right. \nn\\
&\quad \left.\left. + u^2 v^2\left( \left(-\frac{13237}{40} + \frac{11955\pi^2}{2048}\right)\nu + \left(\frac{480343}{960} - \frac{77055\pi^2}{4096}\right)\nu^2 - \frac{6053}{64}\nu^3 - 63\nu^4 \right) \right.\right. \nn\\
&\quad \left.\left. + v^4\left( \left(\frac{275787}{6400} - \frac{8889\pi^2}{8192}\right)\nu + \left(-\frac{256507}{4800} - \frac{7935\pi^2}{16384}\right)\nu^2 + \frac{5521}{128}\nu^3 + 16\nu^4 \right) \right] \right. \nn \\
&\quad \left. + \frac{G^4 m^4}{r^5} \left[ u^2\left( \frac{5}{4} + \left(-\frac{2874901}{9600} + \frac{55505\pi^2}{4096}\right)\nu + \left(-\frac{34965}{128} + \frac{217135\pi^2}{8192}\right)\nu^2 + \frac{243}{16}\nu^3 + 24\nu^4 \right) \right.\right. \nn\\
&\quad \left.\left. + v^2\left( \frac{1}{8} + \left(\frac{1533187}{28800} - \frac{134471\pi^2}{12288}\right)\nu + \left(\frac{6007417}{28800} - \frac{314825\pi^2}{24576}\right)\nu^2 - \frac{201}{16}\nu^3 \right) \right] \right. \nn \\
&\quad \left. + \frac{G^5 m^5}{r^6} \left[ -\frac{801}{32} + \left(-\frac{9088001}{14400} + \frac{282211\pi^2}{6144}\right)\nu + \left(-\frac{1984957}{7200} + \frac{236611\pi^2}{12288}\right)\nu^2 \right] \right\} \bm{n}  \nn\\
&\quad + u \left\{ \frac{Gm}{r^2} \left[ u^6\left(\frac{105}{32}\nu^3 - \frac{35}{8}\nu^4\right) + u^4 v^2\left(-\frac{345}{32}\nu^3 + \frac{105}{4}\nu^4\right) \right.\right. \nn\\
&\quad \left.\left. + u^2 v^4\left(\frac{57}{16}\nu - \frac{435}{16}\nu^2 + \frac{2229}{32}\nu^3 - 54\nu^4\right) + v^6\left(-\frac{45}{16}\nu + \frac{331}{16}\nu^2 - \frac{1473}{32}\nu^3 + 24\nu^4\right) \right] \right. \nn\\
&\quad \left. + \frac{G^2 m^2}{r^3} \left[ u^4\left(\frac{849}{16}\nu - \frac{8451}{64}\nu^2 + \frac{137}{64}\nu^3 - 30\nu^4\right) + u^2 v^2\left(-\frac{401}{8}\nu + \frac{3369}{32}\nu^2 - \frac{899}{96}\nu^3 + 98\nu^4\right) \right.\right. \nn\\
&\quad \left.\left. + v^4\left(\frac{119}{16}\nu - \frac{923}{64}\nu^2 + \frac{841}{64}\nu^3 - 44\nu^4\right) \right] \right. \nn \\
&\quad \left. + \frac{G^3 m^3}{r^4} \left[ u^2\left( \left(-\frac{50497}{160} + \frac{5415\pi^2}{512}\right)\nu + \left(\frac{21575}{96} + \frac{19455\pi^2}{1024}\right)\nu^2 - \frac{4477}{96}\nu^3 - 74\nu^4 \right) \right.\right. \nn\\
&\quad \left.\left. + v^2\left( \left(\frac{339777}{1600} - \frac{17103\pi^2}{2048}\right)\nu - \left(\frac{22367}{300} + \frac{85449\pi^2}{4096}\right)\nu^2 + \frac{961}{32}\nu^3 + 34\nu^4 \right) \right] \right. \nn \\
&\quad \left. + \frac{G^4 m^4}{r^5} \left[ -\frac{7}{4} + \left(-\frac{151877}{1200} + \frac{6415\pi^2}{256}\right)\nu + \left(\frac{120169}{1440} + \frac{25349\pi^2}{1536}\right)\nu^2 - \frac{157}{8}\nu^3 - 16\nu^4 \right] \right\} \bm{v}\,, \\
\label{eq:tail acceleration}\bm{a}_{\rm tail}^{{\rm 4PN}\,i} =&\; \frac{8 G^2 m}{5} \left\{-  x^{j}\int_{0}^{\infty} \mathrm{d}\tau \, \ln\left( \frac{c\tau}{2r}\right) Q_{ij}^{(7)}(t-\tau)  + x^j\left[\left(Q_{ij}^{(3)}\ln r\right)^{(3)}-Q_{ij}^{(6)}\ln
    r\right] -\frac{1}{4 m \nu r}\,Q_{kl}^{(3)}\,Q_{kl}^{(3)}\, n^i\right\} \,. 
\end{align}
\end{subequations}

The spin-dependent terms up to 4PN order have the structure
\begin{subequations}
\begin{align}
\bm{a}_{SO} &= \frac{1}{c^3} \bm{a}_{SO}^\text{LO} + \frac{1}{c^5} \bm{a}_{SO}^\text{NLO} + \frac{1}{c^7} \bm{a}_{SO}^\text{NNLO} + \mathcal{O}\left(\frac{1}{c^9}\right)\,,\\
\bm{a}_{SS} &= \frac{1}{c^4} \bm{a}_{SS}^\text{LO} + \frac{1}{c^6} \bm{a}_{SS}^\text{NLO} + \frac{1}{c^8} \bm{a}_{SS}^\text{NNLO} + \mathcal{O}\left(\frac{1}{c^{10}}\right)\,,\\
\bm{a}_{SSS} &= \frac{1}{c^7} \bm{a}_{SSS}^\text{LO} + \mathcal{O}\left(\frac{1}{c^9}\right)\,,\\
\bm{a}_{SSSS} &= \frac{1}{c^8} \bm{a}_{SSSS}^\text{LO} + \mathcal{O}\left(\frac{1}{c^{10}}\right)\,.
\end{align}
\end{subequations}
In details, we have (recall our notation $\kappa_\pm = \kappa_1\pm\kappa_2$, $\lambda_\pm = \lambda_1\pm\lambda_2$ and $\iota_\pm = \iota_1\pm\iota_2$)
\begin{subequations}
	\begin{align}
		\bm{a}_{SO}^\text{LO} &= \frac{G}{2 r^3} \bigg\{ \Big[ 21 (S n v) + 9 \delta (\Sigma n v)\Big]\bm{n} + 21 u\, \bm{n} \times \bm{S} - 14\,\bm{v} \times \bm{S} - 6 \delta\, \bm{v} \times \bm{\Sigma} \bigg\}\,,\\
		\bm{a}_{SO}^\text{NLO} &= 
		\left\{ \frac{G}{8 r^3} \Big[ 3 \left(- 90\nu u^2+ \left( - 5+59\nu  \right) v^2  \right) (S n v) + 3 \delta \left(- 40\nu u^2+ \left(  - 5 +26\nu \right) v^2  \right) (\Sigma n v) \Big] \right. \nn \\
		& \left. - \frac{G^2 m}{4 r^4} \Big[ \left( 143 + 75\nu \right) (S n v) + 9\delta \left( 7 + 4\nu \right) (\Sigma n v) \Big] \right\} \bm{n}  + \frac{G u}{4 r^3} \Big[ \left(-27+ 57\nu \right) (S n v) + \delta \left( -3+ 24\nu \right) (\Sigma n v) \Big] \bm{v} \nn  \\
		& + u \left\{ -\frac{3G}{8 r^3} \Big[ -90\nu u^2 + v^2\left( 5 + 13\nu \right)  \Big] + \frac{G^2 m}{4 r^4} \left( -125 + 9\nu \right) \right\} \bm{n} \times \bm{S} \nn \\
		& + \left\{ \frac{G}{4r^3} \Big[ - 54\nu u^2+ v^2\left( 5 - 23\nu \right) \Big] + \frac{ G^2 m}{r^4} \left( 22 + 5\nu \right) \right\} \bm{v} \times \bm{S} \nn \\
		& + \delta \left\{ \frac{G}{r^3} \Big[ -6\nu u^2+\frac{5}{4}\,v^2 \left( 1 - 2\nu \right)  \Big] + \frac{ G^2 m}{ r^4} \left( 10 +\frac{5}{2} \nu \right) \right\} \bm{v} \times \bm{\Sigma} \nn \\
		& - \frac{G^2}{4 \nu r^5} \Big[ \left( 9+13\nu \right) S_n + 3 \delta \left(  3+2\nu \right) \Sigma_n \Big] \bm{L}\,, \\
		\bm{a}_{SO}^\text{NNLO} &= 
		\bigg\{ \frac{G}{16 r^3} \Big[ \left( 315\nu \left( 1 - 11\nu \right) u^4+ \left( -9 + 243\nu - 771\nu^2 \right) v^4 - 15\nu \left( 40 - 211\nu \right) u^2 v^2  \right) (S n v) \nn \\
		&  + \delta \left(- 1575\nu^2 u^4+ \left( -9 + 102\nu - 339\nu^2 \right) v^4 - 15\nu \left( 8 - 95\nu \right) u^2 v^2  \right) (\Sigma n v) \Big] \nn \\
		&  + \frac{G^2 m}{8 r^4} \Big[ \left(- 6 \left( 24 + 257\nu + 117\nu^2 \right) u^2+ 2 \left( 9 + 38\nu + 59\nu^2 \right) v^2  \right) (S n v) \nn \\
		&  + \delta \left(- 3 \left( 48 + 175\nu + 103\nu^2 \right) u^2+ \left( 18 - 5\nu + 45\nu^2 \right) v^2  \right) (\Sigma n v) \Big] \nn \\
		&  + \frac{G^3 m^2}{8 r^5} \Big[ 2 \left( 300 + 193\nu - 20\nu^2 \right) (S n v) + \delta \left( 264 + 190\nu - 13\nu^2 \right) (\Sigma n v) \Big] \bigg\} \bm{n} \nn \\
		& + u \bigg\{ \frac{G}{16 r^3} \Big[ - 105\nu \left( 1 - 11\nu \right) u^4+\left( -9 - 93\nu + 177\nu^2 \right) v^4 + 15\nu \left( 32 - 123\nu \right) u^2 v^2  \Big] \nn \\
		&  + \frac{G^2 m}{8 r^4} \Big[ 6 \left( 24 + 246\nu + 197\nu^2 \right) u^2 - \left( 80 + 749\nu + 187\nu^2 \right) v^2 \Big] + \frac{G^3 m^2}{4 r^5} \left( 296 + 271\nu + 55\nu^2 \right) \bigg\} \bm{n} \times \bm{S} \nn \\
		& + u \bigg\{ \frac{G}{8 r^3} \Big[ \left(- 75\nu \left( 2 - 5\nu \right) u^2+ 3 \left( 1 + 50\nu - 89\nu^2 \right) v^2  \right) (S n v)  + \delta \left( 165\nu^2 u^2 +3 \left( 1 + 7\nu - 38\nu^2 \right) v^2 \right) (\Sigma n v) \Big] \nn \\
		& + \frac{G^2 m}{4 r^4} \Big[ \left( 25 + 188\nu + 202\nu^2 \right) (S n v) + \delta \left( 25 + 73\nu + 83\nu^2 \right) (\Sigma n v) \Big] \bigg\} \bm{v} \nn \\
		& + \bigg\{ \frac{G}{8 r^3} \Big[  15\nu \left( 1 - 11\nu \right) u^4+\left( 3 - 25\nu + 99\nu^2 \right) v^4 - 3\nu \left( 28 - 79\nu \right) u^2 v^2  \Big] \nn \\
		&  + \frac{G^2 m}{8 r^4} \Big[- \left( 48 + 499\nu + 432\nu^2 \right) u^2+ 2 \left( 9 + 91\nu - 29\nu^2 \right) v^2  \Big]  - \frac{G^3 m^2}{4 r^5} \left( 186 + 145\nu \right) \bigg\} \bm{v} \times \bm{S} \nn \\
		& + \delta \bigg\{ \frac{G}{8 r^3} \Big[ - 75\nu^2 u^4 +\left( 3 - 18\nu + 43\nu^2 \right) v^4 - 3 \left( 8\nu - 35\nu^2 \right) u^2 v^2 \Big] \nn \\
		& + \frac{G^2 m}{8 r^4} \Big[ - 6 \left( 8 + 27\nu + 31\nu^2 \right) u^2+\left( 18 + 73\nu - 23\nu^2 \right) v^2  \Big] - \frac{G^3 m^2}{4 r^5} \left( 90 + 61\nu \right) \bigg\} \bm{v} \times \bm{\Sigma} \nn \\
		& + \bigg\{ \frac{G^2}{8 \nu r^5} \Big[ \left(-u^2 (72 \nu+90\nu^2)+v^2 (6-41\nu-2\nu^2)   \right) S_n \nn \\
		&  + \delta \left(-u^2 (72 \nu+42\nu^2)+v^2 (6-19\nu-\nu^2) \right) \Sigma_n \Big] \nn \\
		&  + \frac{G^3 m}{4 \nu r^6} \Big[ \left( 33 + 35\nu - 3\nu^2 \right) S_n + \delta \left( 33 + 20\nu - \nu^2 \right) \Sigma_n \Big] \bigg\} \bm{L}\,,\nn \\
		\bm{a}_{SS}^\text{LO} &= \frac{3 G}{4 m r^4} \bigg\{ \Big[ 2(2+\kappa_{+}) (5 S_n^2 - S^2) + 2(2\delta - \kappa_{-} + \delta \kappa_{+}) (5 S_n \Sigma_n - (S\Sigma)) \nn \\
		&  + (\kappa_{+}(1-2\nu) - \delta \kappa_{-} - 4\nu) (5 \Sigma_n^2 - \Sigma^2) \Big] \bm{n} - 2 \Big[ 2(2+\kappa_{+})S_n + (2\delta - \kappa_{-} + \delta \kappa_{+})\Sigma_n \Big] \bm{S} \nn \\
		&  - 2 \Big[ (2\delta - \kappa_{-} + \delta \kappa_{+})S_n + (\kappa_{+}(1-2\nu) - \delta \kappa_{-} - 4\nu)\Sigma_n \Big] \bm{\Sigma} \bigg\}\,,\nn \\
		\bm{a}_{SS}^\text{NLO} &= \text{given in the ancillary file~\cite{ancillaryfile}}\,,\\
		\bm{a}_{SS}^\text{NNLO} &= \text{given in the ancillary file~\cite{ancillaryfile}}\,,\\
		\bm{a}_{SSS}^\text{LO} &= \text{given in the ancillary file~\cite{ancillaryfile}}\,,\\
		\bm{a}_{SSSS}^\text{LO} &= \frac{5 G}{16 m^3 r^6} \Bigg\{ 3 \bigg[ -\left( 2\iota_{+} - 3(\kappa_{-}^2 - \kappa_{+}^2) + 8\lambda_{+} \right) \left( 21 S_n^4 - 14 S_n^2 S^2 + S^4 \right) \nn \\
		& + 2 \left( 2(\iota_{-} - \delta\iota_{+}) + 3\delta(\kappa_{-}^2 - \kappa_{+}^2) + 4(\lambda_{-} - 2\delta\lambda_{+}) \right) \left( 21 S_n^3 \Sigma_n - 7 S_n^2 (S\Sigma) - 7 S_n \Sigma_n S^2 + S^2 (S\Sigma) \right) \nn \\
		& + \left( 2\big(\delta\iota_{-} - (1-2\nu)\iota_{+}\big) + (1-6\nu)(\kappa_{-}^2 - \kappa_{+}^2) + 4\big(\delta\lambda_{-} - (1-4\nu)\lambda_{+}\big) \right) \nn \\
		& \times \left( 63 S_n^2 \Sigma_n^2 - 7 S_n^2 \Sigma^2 - 7 \Sigma_n^2 S^2 + S^2 \Sigma^2 + 2 (S\Sigma)^2 - 28 S_n \Sigma_n (S\Sigma) \right) \nn \\
		& + 2 \left( 2(1-3\nu)\iota_{-} - 2\delta(1-\nu)\iota_{+} - 3\nu\delta(\kappa_{-}^2 - \kappa_{+}^2) + 2(1-6\nu)\lambda_{-} - 2\delta(1-4\nu)\lambda_{+} \right) \nn \\
		& \times \left( 21 S_n \Sigma_n^3 - 7 S_n \Sigma_n \Sigma^2 - 7 \Sigma_n^2 (S\Sigma) + (S\Sigma)\Sigma^2 \right) \nn \\
		& + \left( \delta(1-2\nu)\iota_{-} - (1-4\nu+2\nu^2)\iota_{+} + 3\nu^2(\kappa_{-}^2 - \kappa_{+}^2) - 4\nu\delta\lambda_{-} + 4\nu(1-2\nu)\lambda_{+} \right) \left( 21 \Sigma_n^4 - 14 \Sigma_n^2 \Sigma^2 + \Sigma^4 \right) \bigg] \bm{n} \nn \\
		& + 2 \bigg[ 2 \left( 2\iota_{+} - 3(\kappa_{-}^2 - \kappa_{+}^2) + 8\lambda_{+} \right) \left( 7 S_n^3 - 3 S_n S^2 \right) \nn \\
		& - 3 \left( 2(\iota_{-} - \delta\iota_{+}) + 3\delta(\kappa_{-}^2 - \kappa_{+}^2) + 4(\lambda_{-} - 2\delta\lambda_{+}) \right) \left( 7 S_n^2 \Sigma_n - 2 S_n (S\Sigma) - \Sigma_n S^2 \right) \nn \\
		& - 3 \left( 2\big(\delta\iota_{-} - (1-2\nu)\iota_{+}\big) + (1-6\nu)(\kappa_{-}^2 - \kappa_{+}^2) + 4\big(\delta\lambda_{-} - (1-4\nu)\lambda_{+}\big) \right)  \left( 7 S_n \Sigma_n^2 - S_n \Sigma^2 - 2 \Sigma_n (S\Sigma) \right) \nn \\
		& - \left( 2(1-3\nu)\iota_{-} - 2\delta(1-\nu)\iota_{+} - 3\nu\delta(\kappa_{-}^2 - \kappa_{+}^2) + 2(1-6\nu)\lambda_{-} - 2\delta(1-4\nu)\lambda_{+} \right) \nn \\
		& \times \left( 7 \Sigma_n^3 - 3 \Sigma_n \Sigma^2 \right) \bigg] \bm{S} - 2 \bigg[ \left( 2(\iota_{-} - \delta\iota_{+}) + 3\delta(\kappa_{-}^2 - \kappa_{+}^2) + 4(\lambda_{-} - 2\delta\lambda_{+}) \right) \left( 7 S_n^3 - 3 S_n S^2 \right) \nn \\
		& + 3 \left( 2\big(\delta\iota_{-} - (1-2\nu)\iota_{+}\big) + (1-6\nu)(\kappa_{-}^2 - \kappa_{+}^2) + 4\big(\delta\lambda_{-} - (1-4\nu)\lambda_{+}\big) \right) \left( 7 S_n^2 \Sigma_n - 2 S_n (S\Sigma) - \Sigma_n S^2 \right)\nn \\
		& + 3 \left( 2(1-3\nu)\iota_{-} - 2\delta(1-\nu)\iota_{+} - 3\nu\delta(\kappa_{-}^2 - \kappa_{+}^2) + 2(1-6\nu)\lambda_{-} - 2\delta(1-4\nu)\lambda_{+} \right) 
		\left( 7 S_n \Sigma_n^2 - S_n \Sigma^2 - 2 \Sigma_n (S\Sigma) \right) \nn \\
		& + 2 \left( \delta(1-2\nu)\iota_{-} - (1-4\nu+2\nu^2)\iota_{+} + 3\nu^2(\kappa_{-}^2 - \kappa_{+}^2) - 4\nu\delta\lambda_{-} + 4\nu(1-2\nu)\lambda_{+} \right) \left( 7 \Sigma_n^3 - 3 \Sigma_n \Sigma^2 \right) \bigg] \bm{\Sigma} \Bigg\}\,.
	\end{align}
\end{subequations}
As an important consistency check, we have verified that the complete EFT CM Hamiltonian including all spin terms up to 4PN order reduces on-shell, after expressing the momenta in terms of velocities and applying the order-reduction of accelerations, to the total 4PN energy of the system, which is conserved as a consequence of the previous EFT equations of motion including spin terms (and taking into account the spin precession equations).

\bibliography{ListeRef_BLL26.bib}

\end{document}